\documentclass[a4paper]{article}
\usepackage[dvips]{epsfig}
\usepackage{amssymb}
\usepackage{bm}
\usepackage{dcolumn}
\usepackage{amsmath}
\usepackage{lmodern,amsmath,amssymb,microtype,booktabs,tabularx,enumitem}
\usepackage{caption}
\usepackage{mathrsfs}

\catcode`\@=11
\def\marginnote#1{}
\newcount\hour
\newcount\minute
\newtoks\amorpm
\hour=\time\divide\hour by60
\minute=\time{\multiply\hour by60 \global\advance\minute
by-\hour}\edef\standardtime{{\ifnum\hour<12
\global\amorpm={am}%
        \else\global\amorpm={pm}\advance\hour by-12 \fi
        \ifnum\hour=0 \hour=12 \fi
        \number\hour:\ifnum\minute<10
0\fi\number\minute\the\amorpm}}
\edef\militarytime{\number\hour:\ifnum\minute<10
0\fi\number\minute}

\def\draftlabel#1{{\@bsphack\if@filesw {\let\thepage\relax
   \xdef\@gtempa{\write\@auxout{\string
      \newlabel{#1}{{\@currentlabel}{\thepage}}}}}\@gtempa
   \if@nobreak \ifvmode\nobreak\fi\fi\fi\@esphack}
        \gdef\@eqnlabel{#1}}
\def\@eqnlabel{}
\def\@vacuum{}
\def\draftmarginnote#1{\marginpar{\raggedright\scriptsize\tt#1}}
\def\draft{\oddsidemargin -.5truein
        \def\@oddfoot{\sl preliminary draft \hfil
        \rm\thepage\hfil\sl\today\quad\militarytime}
        \let\@evenfoot\@oddfoot \overfullrule 3pt
        \let\label=\draftlabel
        \let\marginnote=\draftmarginnote

\def\@eqnnum{(\theequation)\rlap{\kern\marginparsep\tt\@eqnlabel}%
\global\let\@eqnlabel\@vacuum}  }

\def\numberbysection{\@addtoreset{equation}{section}
        \def\theequation{\thesection.\arabic{equation}}}

\def\underline#1{\relax\ifmmode\@@underline#1\else
 $\@@underline{\hbox{#1}}$\relax\fi}

\catcode`@=12
\relax

\numberbysection

\advance \topmargin by -\headheight
\advance \topmargin by -\headsep

\evensidemargin \oddsidemargin
\newcommand{\dd}{\mathrm d}
\newcommand{\ii}{\mathrm i}

\newcommand{\sgn}{\operatorname{sgn}}
\newcommand{\sech}{\operatorname{sech}}

\def\br{\begin{eqnarray}}
\def\er{\end{eqnarray}}
\def\be{\begin{equation}}
\def\ee{\end{equation}}

\def\({\left(}
\def\){\right)}

\relax

\def\bv{\bigm\vert}

\def\a{\alpha}

\def\bw{\bar{w}}
\def\r{\rho}
\def\bG{\bar{\G}}
\def\bJ{\bar{J}}

\def\b{\beta}

\def\D{\Delta}
\def\eps{\epsilon}

\def\G{\Gamma}

\def\l{\lambda}
\def\L{\Lambda}

\def\pa{\partial}

\def\s{\sigma}

\def\tp0{\Theta_{+}^{(0)}}
\def\tm0{\Theta_{-}^{(0)}}

\def\bv{{\bar{v}}}
\def\bM{{\bar{M}}}
\def\bN{{\bar{N}}}
\def\bH{{\bold H}}
\def\Ep{{\bold E}_{+}}
\def\Em{{\bold E}_{-}}

\def\f#1#2#3 {f^{#1#2}_{#3}}

\def\win1{{\sf w_{1+\infty}}}

\def\Win1{{\sf W_{1+\infty}}}

\def\rlx{\relax\leavevmode}
\def\inbar{\vrule height1.5ex width.4pt depth0pt}
\def\IZ{\rlx\hbox{\sf Z\kern-.4em Z}}
\def\IR{\rlx\hbox{\rm I\kern-.18em R}}
\def\IC{\rlx\hbox{\,$\inbar\kern-.3em{\rm C}$}}
\def\IN{\rlx\hbox{\rm I\kern-.18em N}}
\def\IO{\rlx\hbox{\,$\inbar\kern-.3em{\rm O}$}}
\def\IP{\rlx\hbox{\rm I\kern-.18em P}}
\def\IQ{\rlx\hbox{\,$\inbar\kern-.3em{\rm Q}$}}
\def\IF{\rlx\hbox{\rm I\kern-.18em F}}
\def\IG{\rlx\hbox{\,$\inbar\kern-.3em{\rm G}$}}
\def\IH{\rlx\hbox{\rm I\kern-.18em H}}
\def\II{\rlx\hbox{\rm I\kern-.18em I}}
\def\IK{\rlx\hbox{\rm I\kern-.18em K}}
\def\IL{\rlx\hbox{\rm I\kern-.18em L}}
\def\one{\hbox{{1}\kern-.25em\hbox{l}}}
\def\0#1{\relax\ifmmode\mathaccent"7017{#1}%
B        \else\accent23#1\relax\fi}

\begin{document}
\begin{titlepage}

\vskip .6in

\begin{center}
{\large {\bf Variable-mass sine-Gordon with point defects: integrability, soliton transmission, and quasi-conservation}}
\end{center}

\normalsize

\begin{center}

A.R. Aguirre$^{(a,b)}$, H. Blas$^{(c)}$, H.F. Callisaya$^{(d)}$ and M.C. de Oliveira$^{(c)}$

\par \vskip .1in \noindent

$^{(a)}$ Instituto de Física e Química\\
 Universidade Federal de Itajub\'a—IFQ (UNIFEI)\\
 Av. BPS 1303, CEP 37500-903,
 Itajub\'a, MG, Brazil\\
$^{(b)}$ Instituto de Física de São Carlos\\
 Universidade de São Paulo, IFSC - USP \\
CEP 13566-590, São Carlos, SP, Brasil.
\\
$^{(c)}$Instituto de F\'{\i}sica\\
Universidade Federal de Mato Grosso\\
Av. Fernando Correa, $N^{0}$ \, 2367\\
Bairro Boa Esperan\c ca, Cep 78060-900, Cuiab\'a - MT - Brazil. \\ 
$^{(d)}$ Departamento de Matem\'atica\\
Universidade Federal de Mato Grosso\\
Av. Fernando Correa, $N^{0}$ \, 2367\\
Bairro Boa Esperan\c ca, Cep 78060-900, Cuiab\'a - MT - Brazil.  
\\
Corresponding author: M.C. de Oliveira
\end{center}

We study integrable variable-mass sine-Gordon model (vmSG) with point defects. Using the Lax and B\"acklund-gauge formulations, we construct type-I and type-II defect matrices, derive the corresponding sewing conditions, and generate the bulk and defect contributions to an infinite hierarchy of conserved charges. The lowest members reduce to the standard sine-Gordon energy and momentum in the homogeneous limit, while for inhomogeneous backgrounds they define integrability-generated energy- and momentum-type quantities. Defect compatibility imposes matching conditions on the variable-mass functions across the defect. An analytical transmission factor $z$ characterizes soliton transmission, topological conversion, and absorption/emission processes. We then deform the type-I sewing conditions by parameters $\a$ and $\b$ and derive the associated defect anomalies. Consistent one-soliton transmission is recovered for $\a \b =1$, whereas for $\a \b  \neq 1$ parity-centered kink–kink and kink–antikink transmissions  display vanishing lowest order integrated anomalies but no generic vanishing at higher order. Numerical simulations reproduce the integrable transmission/conversion regimes and show that non-integrable defects generate small residual disturbances and lowest order quasi-conserved charges, while quasi-conservation of the higher order charges remains unestablished. These results elucidate how exact defect integrability deforms into charge-dependent quasi-conservation and establish a framework for defect-controlled soliton transport in inhomogeneous media, with potential applications to nonuniform Josephson junctions, magnetic and nonlinear-optical systems, and effective molecular and DNA models.
\end{titlepage}

\section{Introduction}
\label{sec:introduction}

The sine-Gordon model is one of the paradigmatic nonlinear field theories in which exact
soliton dynamics, integrability, topology, and physical applications can be studied within
a common framework. Its zero-curvature representation generates an infinite hierarchy of
conserved quantities, while its spectrum contains kinks, antikinks, and breathers undergoing
elastic scattering \cite{rajaraman, babelon}. The model and its variants arise in several contexts, including long
Josephson junctions, nonlinear optical media, magnetic systems, and effective descriptions
of molecular excitations. In particular, sine-Gordon solitons in Josephson devices have also
been proposed as sensitive probes of weak localized perturbations
\cite{Ustinov1998, Soloviev2015}.

Realistic propagation media are generally inhomogeneous, so the coefficients controlling
the nonlinear dynamics may depend on space and time. An integrable realization of this
idea is the variable-mass sine-Gordon equation (vmSG)
\begin{equation}
 \partial_{\xi}\partial_{\eta}w
 +2m_0^2\rho_0(\xi)\rho_1(\eta)\sin(2w)=0,
 \qquad
 \xi=\frac{t+x}{2},
 \quad
 \eta=\frac{t-x}{2},
 \label{eq:intro-vmsg}
\end{equation}
which admits accelerating and shape-changing solitons
\cite{Kundu2007}. Equation (\ref{eq:intro-vmsg}) also defines an integrable relativistic generalization of sine-Gordon, obtained as a reduction of a conformal affine Toda model \cite{ferreira1, ferreira2, bonora1}.  

Locally, when $\rho_0\rho_1\neq0$, the variable coefficient can be
removed through the transformation
\begin{equation}
\label{tr01}
 X=\int^\xi \rho_0(s)\,ds,
 \qquad
 Y=\int^\eta \rho_1(s)\,ds.
\end{equation}
Nevertheless, this mapping does not trivialize the defect problem. The coordinate transformation removes the factorized variable mass locally from the bulk equation, but does not trivialize the defect theory: a defect static at $x=0$ in the physical spacetime becomes, in general, a moving trajectory in the transformed sine-Gordon coordinates $(X,Y)$, while its sewing conditions, field discontinuity, topological accounting, and defect charges remain determined by the original inhomogeneous geometry. See a discussion on the defect trajectory in  Appendix \ref{app:defectrajetory}.

The variable-mass sine-Gordon framework has also been proposed as an effective description of inhomogeneous molecular systems, including DNA-promoter dynamics \cite{Kundu2007}. We do not attempt here to construct a microscopic DNA model; rather, such applications provide part of the motivation for studying soliton propagation and localized defects in spatially and temporally nonuniform media.

Our setting is related to the fluxon-scattering model of Soloviev et al. \cite{Soloviev2015}, where sine-Gordon solitons probe localized inhomogeneities. Here, however, the interaction is described by B\"acklund-type point defects in a vmSG background, yielding purely transmitting integrable regimes and, away from closure, localized anomalies with weak reflection and radiation.

Point defects provide a controlled way of introducing a localized discontinuity into an
integrable field theory. Early studies showed that certain defects can be purely transmitting
and can preserve integrability by contributing localized terms to the conserved charges
\cite{BowcockCorriganZambon2004a,
BowcockCorriganZambon2004b}. At the classical level, their sewing conditions can be
interpreted as B\"acklund transformations frozen at the defect position. The inverse-scattering
and defect-matrix formulation subsequently provided a systematic construction of the complete
hierarchy of modified charges \cite{Caudrelier2008}. In a type-I defect, the bulk fields on the
two half-lines interact directly at the defect, whereas a type-II defect contains an additional
localized degree of freedom. The latter class was developed through defect fusion, Lagrangian
methods, and B\"acklund--gauge transformations
\cite{CorriganZambon2009,CorriganZambon2010,AguirreEtAl2011}.
Complementary Hamiltonian, $r$-matrix, and multisymplectic formulations have further clarified
the canonical and Liouville-integrable structure of these systems
\cite{AvanDoikou2012,CaudrelierKundu2015}.

For readers unfamiliar with the defect formalism, the basic picture is as follows. A point defect divides space into two half-lines and allows the bulk field to have independent limiting values at $x=0^-$ and $x=0^+$. For a type-I integrable defect these two fields are related directly by first-order sewing conditions which can be viewed as a B\"acklund transformation frozen at the defect position; there is no independent dynamical field living at the defect. A type-II defect is more general and introduces an additional localized variable whose evolution forms part of the sewing conditions. In either case, integrability requires that the flux of each bulk conserved quantity through $x=0$ be exactly compensated by an appropriate localized defect contribution. The B\"acklund-gauge matrices introduced below provide a systematic algebraic realization of these conditions and generate the corresponding defect contributions to the hierarchy of conserved charges.

Integrable defects are subject to stringent constraints, since their contributions must exactly compensate the flux of every conserved bulk charge. This raises the question of whether a quasi-integrable theory can be generated by deforming the integrable sewing conditions themselves. The idea is closely related to bulk quasi-integrability, where deformations of an integrable model introduce anomalies that need not vanish pointwise but may have zero time integral for soliton configurations with suitable parity. This mechanism was first developed for deformed sine-Gordon theories and later shown to produce towers of exact or asymptotically conserved charges \cite{FerreiraZakrzewski2011,FerreiraZakrzewski2014,BlasCallisaya2018}.

Here, however, we will perform a deformation at the localized integrable defect: by modifying the integrable matching conditions and the associated defect contributions, one can obtain a distinct model with lowest order quasi-conserved charges. In the present work, by quasi-conservation we mean the following. If a bulk-plus-defect charge $\mathcal Q=Q+Q_D$ satisfies $d\mathcal Q/dt=\mathcal A(t)$, exact conservation corresponds to $\mathcal A(t)\equiv0$. We call $\mathcal Q$ asymptotically quasi-conserved for a given scattering process when $\mathcal A(t)$ is nonzero during the interaction but $\int_{-\infty}^{+\infty}\mathcal A(t)\,dt=0$, so that $\mathcal Q(+\infty)=\mathcal Q(-\infty)$. Oddness of the anomaly about the collision time provides a sufficient mechanism for this cancellation. This property need not hold uniformly throughout the hierarchy; accordingly, we distinguish charge-dependent quasi-conservation from quasi-integrability of an infinite hierarchy.

The main purpose of this work is to construct integrable point defects in the variable-mass
sine-Gordon model. Starting from the Lax representation of Eq.~\eqref{eq:intro-vmsg},
we derive the type-I and type-II B\"acklund--gauge matrices and the corresponding sewing
conditions. The matrices retain the algebraic form of their constant-mass counterparts,
while the B\"acklund equations are dressed by the background functions $\rho_0$ and
$\rho_1$. The relevant compatibility conditions arise for the type-I and type-II constructions:
preservation of the B\"acklund-gauge structure requires 
\begin{equation}
 \bar{\rho}_j=   \rho_j,\,\,\, j=0, 1, 
 \label{eq:intro-background-compatibility}
\end{equation}
where $\rho_j$ and $\bar{\rho}_j$ denote the background functions on each side of the defect, respectively. Both type B\"acklund--gauge matrices are compatible 
only for equal variable-mass backgrounds. 

We will solve the auxiliary Riccati equations recursively and construct the bulk and defect
contributions to the hierarchy of integrability-generated charges. The first nontrivial
combinations define momentum- and energy-type quantities, denoted by ${\cal P}$ and
${\cal E}$. For general $\rho_0$ and $\rho_1$, these are not ordinary Noether energy and
momentum because spacetime translation invariance is absent. They reduce, however, to the
canonical sine-Gordon charges in the homogeneous limit $\rho_0=\rho_1=1$. The lowest order type-I and type-II defect contributions retain the standard homogeneous form, whereas the
background dependence enters explicitly in the higher order Riccati charges.

The analytical transmission factor $z$  separates two branches. Its positive-sign sector
describes kink--kink transmission, whereas its negative-sign sector describes
kink--antikink conversion, with the change in bulk topological charge exchanged with the
defect. The critical limits in which the transmission factor vanishes or diverges correspond
to absorption or emission configurations rather than to a regular transmitted soliton.

In order to investigate departures from exact integrability of the vmSG model with defect, we deform the type-I sewing conditions
by two parameters $\a$ and $\b$. The resulting balance equations contain
momentum-type and energy-type anomalies localized at the defect
\br
\label{int:ano1}
\frac{d}{dt}
\left(
{\cal P} + {\cal P}_D\right) &=& (\a \b -1)
{\cal A}_1,\\
\frac{d}{dt}
\left(
{\cal E} + {\cal E}_D\right) &=& (\a \b -1) \label{int:ano2}
{\cal A}_2,
\er
where ${\cal P}_D$ and ${\cal E}_D$ are the defect contributions, and ${\cal A}_j,\, j=1,2$ are the field dependent anomalies. Exact integrability and a consistent one-soliton transmission ansatz are simultaneously recovered when
\begin{equation}
 \a \b =1.
 \label{eq:intro-closure}
\end{equation}
Under this closure condition, the deformation parameters can be absorbed into a single effective B\"acklund parameter. When Eq.~\eqref{eq:intro-closure} is violated, the two sewing
conditions cannot be reduced to the same frozen B\"acklund transformation, and the defect
becomes genuinely non-integrable.

Moreover, for time-symmetric backgrounds $\rho_j(0,-t)= \rho_j(0,t),\, j=0,1$ and appropriately centered soliton phases, both lowest order anomalies ${\cal A}_j, j=1,2$ are odd under $t\rightarrow -t$. However, both third-order anomalies are generally asymmetric. 

We test these predictions with a second-order finite-difference scheme on the two-half-line geometry. In the closure-compatible regime (\ref{eq:intro-closure}), the simulations reproduce the predicted kink--kink and kink--antikink transmission branches. Outside this regime, transmission remains dominant, with weak reflected and small residual disturbances. The first-order anomalies ${\cal A}_j, j=1,2$, remain localized near the defect, while their time integrals vanish within numerical accuracy, supporting asymptotic conservation of the momentum- and energy-type charges ${\cal P}$ and ${\cal E}$. Moreover, both third-order anomalies have non-vanishing time integrals, indicating that the corresponding third-order charges are not asymptotically conserved.

Our setting is physically related to the fluxon-scattering framework of Soloviev et al. \cite{Soloviev2015}, where localized external perturbations of the sine-Gordon equation affect the propagation of fluxons and are investigated using collective-coordinate methods. The present approach is mathematically distinct: the localized interaction is introduced through B\"acklund-type sewing conditions connecting two variable-mass sine-Gordon fields across a point defect. This construction admits exactly transmitting integrable configurations, whereas deformations of the sewing conditions generate localized anomalies in the conservation laws. In the non-integrable regime, weak reflection and residual disturbances may also occur. Thus, although both models address soliton interactions with localized structures, our emphasis is on defect integrability and the hierarchy-dependent quasi-conservation of bulk-plus-defect charges, rather than on the perturbative dynamics of fluxons under external forcing.

The paper is organized as follows. The introduction section defines  the variable-mass
sine-Gordon model, reviews the motivation for integrable and deformed point
defects, and summarizes the main results. Section~2 reviews the
sine-Gordon model with point defect, including its Lax representation, type-I and type-II
B\"acklund--gauge transformations, Riccati conservation laws, and
defect-modified charges. Section~3 extends these constructions to the
variable-mass model, derives the corresponding type-I and type-II defect
matrices and compatibility conditions, and develops the Riccati hierarchy,
including the first-, second-, and third-order bulk and defect contributions.
Section~4 introduces the deformed type-I sewing conditions, derives the
first- and third-order defect anomalies, establishes the closure condition
for exact transmission, classifies the kink--kink and kink--antikink branches
together with their absorption and emission limits, examines the relevant
parity properties. Section \ref{sec:background} presents the numerical simulations of the integrable
and non-integrable regimes. Section~6 summarizes the conclusions, clarifies
the hierarchy-dependent character of the observed quasi-conservation, and
discusses possible extensions. The defect trajectory is discussed in  Appendix \ref{app:defectrajetory}. Appendix \ref{app:thirdorder} provides the third order charge, defect contribution and anomalies. In Appendix \ref{app:reality} we present a real reduction for nonzero parameters of the type-II B\"acklund transformation. In Appendix \ref{app:emission-absorption} we present a more complete
demonstration of absorption and emission processes. 
 Appendix \ref{app:numerical-scheme} provides the finite-difference scheme for the bulk and defect evolution, and the initialization of the
kink--kink and kink--antikink configurations.

\section{The sine-Gordon model}

Before introducing the variable-mass sine-Gordon model, its
defect and modified defect extensions, it is useful to recall the standard constant-mass sine-Gordon theory in a form adapted to the subsequent generalization. The ordinary sine-Gordon model provides the
undeformed integrable reference system: its equation of motion
follows from a zero-curvature condition, its B\"acklund
transformations can be realized as gauge transformations relating
two auxiliary linear problems, and its infinite hierarchy of
conserved quantities can be generated systematically from the
associated Riccati equations. These structures will be the basic
building blocks for the variable-mass construction discussed in the
following sections.

We also revisit the treatment of ordinary sine-Gordon model with integrable defects. In the constant-mass theory, type-I and type-II defects can be encoded in defect matrices
obtained from the same gauge/B\"acklund framework. The defect
contributions to the conserved charges then follow directly from
the expansion of the defect matrix together with the auxiliary
Riccati fields. This provides a systematic method for constructing
modified conserved quantities, rather than deriving each defect
energy or momentum contribution independently.

The goal of this section is therefore to establish the algebraic and
analytical framework in the simplest setting. We first introduce
the standard $sl(2)$ Lax representation of the sine-Gordon
equation, then derive the corresponding B\"acklund-gauge
transformations, and finally construct the generating functions for
the conserved charges and their defect modifications. In the
subsequent sections the same logic will be applied to the
variable-mass model, allowing one to identify precisely how the
functions $\rho_0$ and $\rho_1$ deform the Lax pair, the Riccati
recursion relations, and the conserved quantities.

A Lax representation consists of a spectral-parameter-dependent auxiliary linear system 
whose compatibility reproduces the nonlinear field equation. For the present system, 
this compatibility is expressed by the zero-curvature condition (\ref{zceq}). A B\"acklund–gauge 
transformation is a matrix $K$ satisfying (\ref{gauge1}), which relates two auxiliary systems and
induces first-order relations between their nonlinear fields. Conserved integrals are 
subsequently obtained from the associated current or Riccati expansions.

\subsection{The Lax Pair}
The usual sine-Gordon equation 
\br
 \pa_{\xi}\pa_{\eta} w + 2m_0^2 \sin 2w =0,\label{sge}
\er
can be derived as a compatibility condition for the associated linear problem
\br
 \pa_\xi v &=& M v,  \qquad \pa_\eta v = N v, \label{auxsg1} 
\er
i.e., the compatibility condition $\pa_{\xi} \pa_{\eta} v = \pa_{\eta} \pa_{\xi} v$ gives rise to the zero curvature equation
\br
\label{zceq}
 \pa_\eta M - \pa_\xi N + [ M, N] =0,
\er
where $v = (v_1, v_2)^T$ is a two-component vector, and the Lax pair $M$ and $N$ are given by\footnote{Here $\{\bH, \Ep, \Em\}$ are the generators of the $sl(2)$ Lie algebra, which satisy the following commutation relations,
\br 
 \left[\bH, {\bf E}_\pm \right] &=& \pm 2\,{\bf E}_\pm,\qquad \left[{\bf E}_+ ,{\bf E}_-\right] = \bH,
\er
and the standard two-dimensional representation given by,
\br  
 \bH &=&  \left[
   \begin{array}{lc}  1  &  0\\
                             0  &  -1
    \end{array}
 \right],\qquad
 \Ep \,=\, \left[\begin{array}{cc}  0 \mbox{\,\,\,}&  1\\
                                          0\mbox{\,\,\,}  &  0
    \end{array}
    \right],\qquad
    \Em \,=\, \left[\begin{array}{cc}  0 \mbox{\,\,\,}&  0\\[0.3cm]
                                          1\mbox{\,\,\,}  &  0
    \end{array}
    \right]. 
\er
We have also introduced
the light cone coordinates $\xi=\frac{1}{2}(t+x)$, $\eta=\frac{1}{2}(t-x)$,  and the derivatives $\pa_\xi = \pa_t+\pa_x$, and $\pa_\eta = \pa_t -\pa_x$.} 
\br 
M &=& -\frac{i}{2}(\pa_\xi w)\bH -\l m_0 \(e^{iw}\,\Ep - e^{-iw}\, \Em\) \,=\, 
 \left[
   \begin{array}{cc} 
	-\frac{i}{2}(\pa_\xi w)\quad   & -m_0 \l\,  e^{iw} \\
    m_0\l \, e^{-iw}  &  \frac{i}{2}(\pa_\xi w)
   \end{array}\right], \label{laxsine1}\\
 N &=& + \frac{i}{2}(\pa_\eta w)\bH - \frac{m_0}{\l}\(e^{-iw}\,\Ep -e^{iw}\,\Em\)\,\,=\,\,
 \left[
   \begin{array}{cc} \frac{i}{2}(\pa_\eta w)\quad   & - \frac{m_0}{\l}\, e^{-iw}  \\
    \frac{m_0}{\l}\,e^{iw}  \quad & -\frac{i}{2}(\pa_\eta w) 
   \end{array}
 \right],\qquad \mbox{}\label{laxsine2}
\er
with $m_0$ a mass parameter, and $\l$ the spectral parameter.

\subsection{The B\"acklund-gauge transformation}

Let us consider two different configurations $v$ and $\bar{v}$ corresponding to solutions of the auxiliary linear problem (\ref{auxsg1}) described by 	Lax pairs $(M,N)$ and $(\bar{M},\bar{N})$\footnote{The Lax pair $(\bM,\bN)$ is obtained directly from $(M,N)$ by replacing the field $w$ for another solution of the sine-Gordon equation $\bw$.}, respectively. Let us now introduce a matrix $K$ connecting the two configurations, namely,
\br
 \bv = K v, \label{eq2.6}
\er
where $K $ satisfies differential equations, namely
\br
 \pa_\xi K  = \bM K  - K M, \qquad \pa_\eta K  = \bN K  - K  N.\label{gauge1}
\er
which correspond to a gauge transformation. Solutions of the above gauge equations for the matrix $K$ will generate some relations between different solutions of the sine-Gordon equation, i.e. the B\"acklund transformations for the sine-Gordon equation. 

By considering the ansatz $K_{I} = K_{0} + \l^{-1} K_{1}$ for the $K$ matrix in (\ref{gauge1}), where $K_{0}$ and $K_{1}$ are also $2 \times 2 $ matrices, and solving grade by grade the gauge equations (\ref{gauge1}), we find that the $K_I$ matrix takes the following form
\br
 K_{I}  &=&  e^{-\frac{iw}{2}\bH}\left[\mathbb{I} +\frac{\s}{\l}(\Ep -\Em)\right] e^{\frac{i\bw}{2}\bH}\,=\, 
 \left[
   \begin{array}{r c} e^{-\frac{i}{2}w_-} \,\,\,\mbox{}& \frac{\s}{\l}\, e^{-\frac{i}{2}w_+}\\[0.3cm]
-\frac{\s}{\l}\, e^{\frac{i}{2}w_+} \,\,\,\mbox{}&  e^{\frac{i}{2}w_-}
   \end{array}
 \right], \label{eq2.10}\quad \mbox{}
\er
where $w_\pm = w \pm \bw$. This $K_I$ matrix generates the well-known B\"acklund transformations for the sine-Gordon equations, namely
\br 
\label{B10}
 \pa_\xi w^{-} &=& -{2m_0\s} \sin w^{+},\\
 \pa_\eta w^{+} &=& +\frac{2m_0}{\s}\sin w^{-}, \label{B20}
\er
where $\s$ is the B\"acklund parameter. 

A second solution can be derived by considering a generalized form of the $K$ matrix \cite{AguirreEtAl2011}, as follows
\br 
 K_{II}  = T_0 + \frac{1}{\l} T_1 + \frac{1}{\l^2} T_2,
\er 
where $T_k$ are $2 \times 2$ matrices. Solving the gauge transformations grade by grade, the $K_{II}$ matrix will take the following explicit form,
\br
 K_{II} &=& \(\begin{array}{cc}
a  e^{-\frac{i}{2} w^{-}} +  \frac{c}{\l^2} e^{\frac{i}{2} w^{-}} & \frac{a c}{\l b} e^{i(\L - \frac{1}{2} w^{+})} (e^{i w^{-}} + e^{-i w^{-}}  +c_0)\\
\frac{b}{\l} e^{-i(\L - \frac{1}{2} w^{+})}&  a e^{\frac{i}{2} w^{-}} + \frac{c}{\l^2} e^{-\frac{i}{2} w^{-}} \end{array}\), \label{eq2.14}
\er
which generates what is referred to as type-II B\"acklund transformations for the sine-Gordon model. 

Note that (\ref{eq2.14}) is a solution of (\ref{gauge1}) together with the next relationships
\br
\label{back11}
i\pa_{\xi} w^{-} &=& \frac{m_0}{a} \, (b\, e^{-i(\Lambda - w^{+})} + \b_{12} \, e^{-\frac{i}{2} w^{+}} )\\
\label{back12}
i\pa_{\eta} w^{-} &=& -\frac{m_0 }{c} \, (\b_{12} \, e^{\frac{i}{2} w^{+}}  + b \, e^{-i\Lambda} )\\
\label{back13}
i \pa_{\xi} \Lambda &=& -\frac{m_0 c}{b} \, e^{i(\Lambda - w^{+})}( e^{i w^{-}} - e^{-i w^{-}} )\\
\label{back14}
i\pa_{\eta}(\Lambda - w^{+}) &=& \frac{m_0 a}{b}  \, e^{i\Lambda} ( e^{i w^{-}} - e^{-i w^{-}})
\er
and 
\br
\pa_{\xi} \b_{12} &=& -\frac{i}{2} \pa_{\xi} w^{+} \b_{12} + c m_0 e^{\frac{i}{2} w^{+}} ( e^{i w^{-}} - e^{-i w^{-}})\\
\pa_{\eta} \b_{12} &=& \frac{i}{2} \pa_{\eta} w^{+} \b_{12} - a m_0   e^{-\frac{i}{2} w^{+}} (e^{i w^{-}} - e^{-i w^{-}}). \label{b12}
\er
One can show that $\b_{12}$ becomes
\br
\b_{12} = \frac{a c}{b}  e^{i(\Lambda -\frac{1}{2} w^{+})} ( e^{i w^{-}} + e^{-i w^{-}} + c_{0})
\er
with $c_0$ a constant. Moreover, one has

\br
\pa_{\xi} \pa_{\eta} \Lambda &=& \frac{im_0^2 a c}{b^2} e^{i(\Lambda -w^{+})} (e^{i w^{-}}-e^{-i w^{-}}) \times\\
&&[\frac{b^2}{a c} + e^{2 i\Lambda} +e^{i \Lambda} (e^{i w^{-}}-e^{-i w^{-}})+ e^{2i\Lambda}(e^{i w^{-}}+e^{-i w^{-}}+ c_0)], 
\er
where $a,b,c,$ are some B\"acklund parameters, and $\L$ is an auxiliary field. By cross-differentiating Eqs.~(\ref{back11}) and (\ref{back12}), one verifies that if $w$ satisfies the sine-Gordon equation, then $\bw$ does as well. In Appendix \ref{app:reality} we present a real reduction $\L = \ell + i \chi $ for nonzero parameters $a, b, c$, with $c_0 \in \IR$.

\subsection{Conservation laws}
Now, let us construct explicitly the conserved quantities for the sine-Gordon model. From the auxiliary linear problem (\ref{auxsg1}), and from (\ref{laxsine1})-(\ref{laxsine2}) we can directly derive two conservation laws, namely
\br 
 \pa_\xi J_k^{\xi} + \pa_\eta J_k^{\eta}=0,  \qquad k=1,2,\label{cl1}
\er
with,
\br
J_1^{\xi} &=& \,\frac{i}{2}(\pa_\eta w) -\frac{m_0}{\l} e^{-iw}\G_{21}, \quad \qquad J_1^{\eta}= \,\frac{i}{2}(\pa_\xi w) +{m_0}{\l} e^{iw}\G_{21},\label{eq2.20a}\\
J_2^{\xi} &=& -\frac{i}{2}(\pa_\eta w) +\frac{m_0}{\l} e^{iw}\G_{12}, \quad \qquad J_2^{\eta}= -\frac{i}{2}(\pa_\xi w) -{m_0}{\l} e^{-iw}\G_{12},\label{eq2.20}
\er
where we have introduced the auxiliary functions $\G_{21} = v_2 v_1^{-1}$ and  $\G_{12} = v_1 v_2^{-1}$. From these, we obtain that the associated generating functions of the conserved quantities are given by,
\br
 Q_k = \frac{1}{2}\int_{-\infty}^{\infty} dx\, ( J_k^{\xi}+J_k^{\eta}), \qquad k=1,2.\label{eq2.22}
\er
Now, the auxiliary functions $\G_{12}$  and $\G_{21}$ satisfy the following set of Riccati equations, 
\br
\pa_\xi \G_{21} &=& i(\pa_\xi w) \G_{21} + m_0\l \(e^{-iw}+e^{iw} \G_{21}^2\),\label{eq2.23}\\
   \pa_\eta \G_{21} &=& -i(\pa_\eta w) \G_{21} + \frac{m_0}{\l} \(e^{iw}+e^{-iw} \G_{21}^2\),\label{eq2.24} \\
 \pa_\xi \G_{12} &=& -i(\pa_\xi w) \G_{12} - m_0\l \(e^{iw}+e^{-iw} \G_{12}^2\)\label{eq2.25}\\
 \pa_\eta \G_{12} &=& i(\pa_\eta w) \G_{12} - \frac{m_0}{\l} \(e^{-iw}+e^{iw} \G_{12}^2\).\label{eq2.26} 
\er 
Since these equations are not coupled we can solve them separately. In order to determine the auxiliary function $\G_{21}$, let us consider firstly its expansion as $\l\to\infty$, namely
\br
\G_{21} = \sum_{n=0}^{\infty}  \frac{\G_{21}^{(-n)}}{\l^n}.
\er
and substitute into eqs. (\ref{eq2.23}) and (\ref{eq2.24}). Then, we obtain the following  coefficients,
\br
\G_{21}^{(-0)} &=& ie^{-iw},\nonumber\\
  \G_{21}^{(-1)} &=& -\frac{i}{m_0}(\pa_\xi w) e^{-iw},\\
 \G_{21}^{(-2)} &=& \frac{i}{2m_0^2}\left[(\pa_\xi w)^2 + i\pa_\xi^2 w\right] e^{-iw}, \nonumber\\
 \G_{21}^{(-3)} &=& \frac{i}{4m_0^3}\left[\pa_\xi^3 w - 2i(\pa_\xi w)(\pa_\xi^2 w)\right] e^{-iw}, \nonumber\\&&\dots\dots \nonumber \mbox{}
\er
Thus, we can get the first infinite set of conserved quantities generated from,
\br
Q_1&=&\int_{-\infty}^{\infty} dx \left[ \frac{i}{2}(\pa_t w) + \frac{m_0}{2}\(\l e^{iw} - \l^{-1} e^{-iw}   \)\G_{21}\right]=\sum_{n=0}^{\infty}  \l^{-n} Q_1^{(-n)}.\label{gf1}
\er
From the coefficients  of the expansion of $\G_{21}$, we can obtain directly the first two non-trivial conserved charges, namely
\br
Q_1^{(-1)}&=& \frac{i}{4m_0}\int_{-\infty}^{\infty} dx \left[(\pa_\xi w)^2+i(\pa_\xi^2 w) - 2m_0^2 e^{-2iw} \right],\label{eq2.38}\\
 Q_1^{(-2)} &=& \frac{i}{8m_0^2}\int_{-\infty}^{\infty} dx \left[\pa_\xi^3 w -2i(\pa_\xi w)(\pa_\xi^2 w) +4m_0^2(\pa_\xi w) e^{-2iw}\right].
\er
Now, if we consider the series expansion of the auxiliary function $\G_{21}$ as $\l\to 0$, namely,
\br
 \G_{21} = \sum_{n=0}^{\infty} {\l^n}\, {\G_{21}^{(n)}}.
\er
we obtain the following coefficients,
\br
\G_{21}^{(+0)}&=&ie^{iw}\nonumber\\
\G_{21}^{(+1)}&=&\frac{i}{m_0}(\pa_\eta w) e^{iw},\\
\G_{21}^{(+2)} &=&  \frac{i}{2m_0^2}\left[(\pa_\eta w)^2 - i\pa_\eta^2 w\right] e^{iw}, \quad \mbox{}\nonumber\\
 \G_{21}^{(+3)} &=& -\frac{i}{4m_0^3}\left[\pa_\eta^3 w+ 2i(\pa_\eta w)(\pa_\eta^2 w)\right] e^{iw}.\quad\nonumber \mbox{}
\er
Then, from eq. (\ref{gf1}) we obtain the following non-trivial charges,
\br
 Q_1^{(+1)}&=& -\frac{i}{4m_0}\int_{-\infty}^{\infty} dx \left[(\pa_\eta w)^2-i(\pa_\eta^2 w) - 2m_0^2 e^{2iw} \right],\\
 Q_1^{(+2)} &=& \frac{i}{8m_0^2}\int_{-\infty}^{\infty} dx \left[\pa_\eta^3 w +2i(\pa_\eta w)(\pa_\eta^2 w) +4m_0^2(\pa_\eta w) e^{2iw}\right].
\er 
Similarly, we can solve the Riccati equations (\ref{eq2.25}) and (\ref{eq2.26})  recursively to obtain the following coefficients of the auxiliary function $\G_{12}$,
\br 
\G_{12}^{(-0)} &=& ie^{iw},\qquad \qquad \qquad\qquad \qquad \qquad \quad\,\,\,\, \G_{12}^{(+0)} \,=\, ie^{-iw},    \nonumber\\
 \G_{12}^{(-1)} &=& -\frac{i}{m_0}(\pa_\xi w) e^{iw}, \qquad \qquad \qquad \qquad \quad \,\, \G_{12}^{(+1)} \,=\, \frac{i}{m_0}(\pa_\eta w) e^{-iw},  \\
 \G_{12}^{(-2)} &=& \frac{i}{2m_0^2}\left[ (\pa_\xi w)^2 -i\pa_\xi^2 w\right]e^{iw}, \qquad \qquad \,\,\,\,   \G_{12}^{(+2)} \,=\, \frac{i}{2m_0^2}\left[(\pa_\eta w)^2+i\pa_\eta^2 w\right]e^{-iw}, \nonumber\\
 \G_{13}^{(-3)} &=& \frac{i}{4m_0^3}\left[(\pa_\xi^3 w) + 2i(\pa_\xi w)(\pa_\xi^2 w)\right]e^{iw},\quad\,\, \G_{13}^{(+3)} \,=\, -\frac{i}{4m_0^3}\left[\pa_\eta^3 w - 2i(\pa_\eta w)(\pa_\eta^2 w)\right]e^{-iw}. \nonumber
\er
Therefore, from the second generating function of the infinite conserved quantities,
\br
 Q_2 &=&\int_{-\infty}^{\infty} dx \left[ -\frac{i}{2}(\pa_t w) + \frac{m_0}{2}\(\l^{-1} e^{iw} - \l e^{-iw}   \)\G_{12}\right],\label{gf2}
\er
we obtain the following non-vanishing charges,
\br
 Q_2^{(-1)} &=& -\frac{i}{4m_0}\int_{-\infty}^{\infty} dx \left[(\pa_\xi w)^2-i(\pa_\xi^2 w) - 2m_0^2 e^{2iw} \right],\\
 Q_2^{(-2)} &=& -\frac{i}{8m_0^2}\int_{-\infty}^{\infty} dx \left[\pa_\xi^3 w +2i(\pa_\xi w)(\pa_\xi^2 w) +4m_0^2(\pa_\xi w)e^{2iw} \right],\\
  Q_2^{(+1)} &=& \frac{i}{4m_0}\int_{-\infty}^{\infty} dx \left[(\pa_\eta w)^2+i(\pa_\eta^2 w) - 2m_0^2 e^{-2iw} \right],\\
 Q_2^{(+2)} &=& -\frac{i}{8m_0^2}\int_{-\infty}^{\infty} dx \left[\pa_\eta^3 w -2i(\pa_\eta w)(\pa_\eta^2 w) +4m_0^2(\pa_\eta w) e^{-2iw}\right].\label{eq2.41}
\er
Note that the Riccati system of equations (\ref{eq2.23})-(\ref{eq2.26}) is invariant under the  transformation $\G_{12}^* = \G_{21},\,\, \l^{*}  = - \l$, then one can show $Q_2^\dagger = Q_1$, which allows us to obtain the expressions for the canonical energy and momentum for the sine-Gordon system by considering the following simple combinations of the first order conserved charges $Q_1^{(\pm 1)}$ and $Q_2^{(\pm 1)}$,
{\small
\br
 P&=& \frac{im_0}{2}\left[(Q_2^{(-1)}-Q_1^{(-1)}) + (Q_2^{(+1)}-Q_1^{(+1)}) \right] =\int_{-\infty}^{\infty} dx\,  (\pa_t w)(\pa_x w),\\[0.1cm]
  E&=& \frac{im_0}{2}\left[(Q_2^{(-1)}-Q_1^{(-1)}) - (Q_2^{(+1)}-Q_1^{(+1)}) \right]=\int_{-\infty}^{\infty} dx \left[\frac{1}{2}(\pa_t w)^2+\frac{1}{2}(\pa_x w)^2 - m_0^2 \cos 2w\right]\!\!.\qquad \,\,\mbox{}
\er}
It is worth pointing out that dual sequences of charges can be constructed by interchanging $\pa_\eta \leftrightarrow \pm \pa_\xi$ (P- and T- symmetry), and simultaneously performing the transformations $w \to \mp w$ and $\l \to \pm \l^{-1}$ in the above equations. In general, these transformations, which leave to dual Lax pairs, are symmetries of the equation of motion (\ref{sge}), and the associated conserved charges can be determined from eqs. (\ref{eq2.38}) -- (\ref{eq2.41}) straightforwardly.


\subsection{Defect conserved charges}

Let us now introduce a dynamical defect in the model as an internal boundary condition at a fixed point, say $x=0$. First of all,  let us consider $v$ and $\bv$ as the auxiliary functions corresponding to the associated linear problems for $x < 0$ described by the Lax pair $M$ and $N$, and for $x > 0$ by $\bM$ and $\bN$, respectively. These two configurations will be connected at the defect point $x=0$ by the $K$-matrix, which will be referred to as \emph{defect matrix}, by equation (\ref{eq2.6}), where now the fields $w$ and $\bw$ on both sides of the defect, and their derivatives,  will take values on the defect point $x=0$.

Now, the generating functions $Q_k$ of the infinite set of conserved charges (\ref{eq2.22}) will be modified in the presence of such defect in the following way,
\br
 Q_k &=& \frac{1}{2}\int_{-\infty}^{0} dx \,(J_k^{\xi}+J_k^{\eta})+\frac{1}{2}\int_{0}^{\infty} dx\,  (\bar{J}^{\xi}_k+\bar{J}_k^{\eta}).
\er
By taking the time-derivative, and using respectively the conservation law (\ref{cl1}), we get
\br
 \frac{dQ_k}{dt} &=& \frac{1}{2}\left(J_k^{\eta}-J_k^{\xi}\right)\Big|_{x=0}- \frac{1}{2}\left(\bJ_k^{\eta}-\bJ_k^{\xi}\right)\Big|_{x=0}.\label{eq2.47}
\er
From the relation between the two auxiliary linear problems (\ref{eq2.6}), we find that
\br
 \bG_{12} = \frac{K_{12} + K_{11}\G_{12}}{K_{22}+K_{21} \G_{12}}, \qquad \bG_{21} = \frac{K_{21} + K_{22}\G_{21}}{K_{11}+K_{12} \G_{21}},
\er
where $K_{ij}$ are the respective components of the defect matrix. Therefore, by introducing these relations in (\ref{eq2.47}), we obtain that
\br
 \frac{d}{dt}\(Q_k+D_k\)= 0,
\er
where $D_k$ are the defect contributions to the  generating functions of the infinite modified conserved charges ${\cal Q}_k=Q_k+D_k$, and given by
\br
 D_k &=& - \ln\Big[K_{kk}+K_{kl}\G_{lk}\Big]\Big|_{x=0}, \qquad k =1,2, \qquad \mbox{and} \quad l\neq k.
\er
Notice that expansions in powers of $\l$ provides the defect contributions to the modified conserved charges at all orders. Then, taking into account the type I defect matrix $K$ in (\ref{eq2.10}) and using the above formula to compute the respective first order coefficients of the defect contributions, we get
\br  
 D_1^{(-1)} &=& -i\s \,e^{-i(w+\bw)}, \qquad \quad  D_1^{(+1)} \,=\, \frac{i}{\s}e^{-i(w-\bw)} -\frac{1}{m_0}\(\pa_t w - \pa_x w\), \\
 D_2^{(-1)} &=& i\s\, e^{i(w+\bw)},\qquad \qquad \,\, D_2^{(+1)}\,=\,-\frac{i}{\s}e^{i(w-\bw)} -\frac{1}{m_0}\(\pa_t w - \pa_x w\).
\er
Then, from these results it is possible to obtain the corresponding defect contributions to the canonical energy and momentum for the sine-Gordon model by performing the following linear combinations,
\br
  P^{(1)}_{\mbox{\tiny D,I}} &=& \frac{im_0}{2} \left[(D_2^{(-1)}-D_1^{(-1)})+(D_2^{(+1)}-D_1^{(+1)}) \right] =
 \! -m_0\!\left[\s\cos(w+\bw) -\frac{1}{\s}\cos(w-\bw)\right],\label{PI} \\
 E^{(1)}_{\mbox{\tiny D, I}} &=& \frac{im_0}{2}  \left[(D_2^{(-1)}-D_1^{(-1)})-(D_2^{(+1)}-D_1^{(+1)}) \right]= -m_0\!\left[\s\cos(w+\bw) +\frac{1}{\s}\cos(w-\bw)\right].\label{EI}
\er
These results correspond to the energy and momentum of the type I defect sine-Gordon model previously derived in \cite{kundu2, CorriganZambon2009}. It is worth pointing out that this systematic way allows us to compute defect contributions to the conserved quantities at all orders, by considering the corresponding coefficients of the auxiliary functions.\\

The defect contributions associated with the type-II defect matrix (\ref{eq2.14}) can also be derived. The first order defect contribution are then given by
\br
  P^{(1)}_{\mbox{\tiny D,II}} &=& \frac{m_0}{2}[\frac{b}{a} e^{i(w^{+}-\L)} + \frac{b}{c} e^{-i\L}-(\frac{c}{b} e^{-i(w^{+}-\L)} +\frac{a}{b} e^{i \L})(e^{iw^{-}} + e^{-iw^{-}} + c_0)]  \label{PII}\\
 E^{(1)}_{\mbox{\tiny D,II}} &=& \frac{m_0}{2}[\frac{b}{a} e^{i(w^{+}-\L)} - \frac{b}{c} e^{-i \L} - (\frac{c}{b} e^{-i(w^{+}-\L)} -\frac{a}{b} e^{i \L})(e^{iw^{-}} + e^{-iw^{-}} + c_0)]  \label{EII}
\er

Thus the modified conserved quantities incorporating the type $I$ and $II$ defect contributions are, respectively
\br
{\cal P}^{(1)}_a = P^{(1)}_a + P_{D,a}^{(1)},\,\,\,
{\cal E}^{(1)} = E^{(1)}_a + E_{D,a}^{(1)},\,\,\,\,\,\,\,\,\, a= I,II,
\er  
where the $I$-type and $II$-type defect contributions are provided in (\ref{PI})-(\ref{EI}) and (\ref{PII})-(\ref{EII}), respectively.  

\section{Variable mass sine-Gordon model}

The variable-mass sine-Gordon model (vmSG) (\ref{eq:intro-vmsg}) is an inhomogeneous extension of the
standard sine-Gordon theory in which the constant mass scale is dressed by light-cone dependent functions. Such a deformation is physically motivated by soliton propagation in nonuniform media, where spatial or temporal variations of the effective coupling may induce acceleration, distortion, or interaction with external backgrounds. At the same time, for the special factorized dependence, the model still admits a Lax representation, making it possible to investigate how B\"acklund transformations, integrable defects, and conserved quantities are deformed by the functions $\rho_0(\xi)$ and $\rho_1(\eta)$. Even though the bulk equation can be locally mapped to the constant-mass sine-Gordon equation by a light-cone coordinate transformation (\ref{tr01}), fixed defects and physical boundary conditions are defined in the original coordinates. Hence the variable-mass formulation
remains nontrivial and provides the appropriate framework for analyzing
soliton-defect interactions in inhomogeneous media.
 
The variable mass sine-Gordon model (\ref{eq:intro-vmsg}) can be associated to the Lax pair $M$ and $N$ given by 
\br 
M &=& -\frac{i}{2}(\pa_\xi w){\bH} -\l m_0 \rho_0(\xi) \(e^{iw}\,\Ep - e^{-iw}\, \Em\) = 
 \left[
   \begin{array}{cc} -\frac{i}{2}(\pa_\xi w)   & -\l m_0 \rho_0(\xi)e^{iw} \\
    \l m_0\rho_0(\xi) e^{-iw}  &  \frac{i}{2}(\pa_\xi w)
   \end{array}
 \right] , \,\,\,\,\,\quad\mbox{}\label{laxvm1}\\
 N &=&+ \frac{i}{2}(\pa_\eta w)\bH - \frac{m_0\rho_1(\eta) }{\l}\(e^{-iw}\,\Ep -e^{iw}\,\Em\)\,=\,
 \left[
   \begin{array}{cc} \frac{i}{2}(\pa_\eta w)\quad   & - \frac{m_0\rho_1(\eta)}{\l}\, e^{-iw}  \\
    \frac{m_0\rho_1(\eta)}{\l}\,e^{iw}  \quad & -\frac{i}{2}(\pa_\eta w) 
   \end{array}
 \right],\qquad \label{laxvm2}
\er
where $m_0$ is the mass parameter, $\l$ is the spectral parameter, the fields $\rho_0(\xi)$ and $\rho_1(\eta)$ are some functions depending on the light-cone coordinates and $w$ is the field of the variable mass SG model.  
It is important to note that the standard sine-Gordon equation can be recovered when the functions satisfy the condition $\rho_0 =\rho_1 =1 $.

\subsection{The B\"acklund-Gauge transformation}

Let us consider two configurations, $v$ and $\bar{v}$, satisfying the auxiliary linear problem (\ref{auxsg1}) with Lax pairs $(M,N)$ and $(\bar{M},\bar{N})$, respectively. The first pair is defined in (\ref{laxvm1})–(\ref{laxvm2}), while the second is obtained by replacing $w$ with $\bw$ and the background functions $(\rho_0,\rho_1)$ with $(\bar{\rho}_0,\bar{\rho}_1)$. We introduce a matrix $K$ relating these configurations through (\ref{eq2.6}) and satisfying the gauge transformation equations (\ref{gauge1}). Solutions of these equations for $K$ yield B\"acklund transformations for the variable-mass sine-Gordon equation.

The background compatibility conditions will be shown below. For the type-I and type-II matrices considered below, the B\"acklund-gauge equations admit transformations between the two field configurations with identical background functions, $\bar{\rho}_j=\rho_j$, $j=0,1$. In the Table 1 we present a compatibility summary for the type I and II defects.  

\begin{center}
\small
\begin{tabularx}{\textwidth}{p{3.5cm}XX}
\toprule
Requirement & Type I & Type II\\
\midrule
Backgrounds & $\bar\rho_j=\rho_j$, $j=0,1$ & $\bar\rho_j=\rho_j$, $j=0,1$\\[4pt]
Field relations & Eqs. (\ref{bacvm1})--(\ref{bacvm2}), one B\"acklund parameter & Eqs. (\ref{back11v})--(\ref{L11}), auxiliary field $\L$\\[4pt]
Auxiliary field & Absent & Present\\[4pt]
Real physical sector & Real fields and real nonzero $\sigma$ & Additional reality restrictions; Appendix C provides a sufficient reduction\\
\bottomrule
\end{tabularx}
\captionof{table}{Compatibility summary for the type I and II defects.  
}
\end{center} 

\subsection{Type I B\"acklund transformation}

By considering the ansatz $K(\l)= K_0 + \l^{-1} K_1$ for the $K$ matrix, where $K_0$ and $K_1$ are also $2 \times 2 $ matrices, and solving grade by grade the gauge equations (\ref{gauge1}) with the relevant Lax pair $(M,N)$ in (\ref{laxvm1})-(\ref{laxvm2}), we find that the $K$ matrix takes the same form as in (\ref{eq2.10}), i.e.
\br
 K_{I}^{vm}  &=&  e^{-\frac{iw}{2}\bH}\left[\mathbb{I} +\frac{\s}{\l}(\Ep -\Em)\right] e^{\frac{i\bw}{2}\bH}\,=\, 
 \left[
   \begin{array}{r c} e^{-\frac{i}{2}w_-} \,\,\,\mbox{}& \frac{\s}{\l}\, e^{-\frac{i}{2}w_+}\\
-\frac{\s}{\l}\, e^{\frac{i}{2}w_+} \,\,\,\mbox{}&  e^{\frac{i}{2}w_-}
   \end{array}
 \right]. \label{eq2.10vm}
\er
This $K^{vm}_I$ matrix generates the following relations
\br 
 \pa_\xi w_- &=& -\epsilon \, 2m_0\s \rho_0 \,\sin w_+,\label{bacvm1}\\ 
 \pa_\eta w_+ &=& \epsilon \,\frac{2m_0 }{\s}\rho_1 \, \sin w_-,\,\,\,\, \epsilon = \pm 1,\,\,\,\, \bar{\rho}_j=\rho_j,\, j=0,1,\label{bacvm2}
\er
where $\s$ is a B\"acklund parameter. It can be directly verified from the above equations that  if $w$ satisfies the vmSG equation (\ref{eq:intro-vmsg}), then $\bw$ also does, i.e., eqs (\ref{bacvm1}) and (\ref{bacvm2}) constitute the B\"acklund transformations for the variable mass sine-Gordon equation. Note that with the redefinition $\sigma\to\epsilon\sigma$, the sign $\epsilon$ is incorporated into the B\"acklund parameter, and we  recover the usual sine-Gordon B\"acklund transformation (\ref{B10})-(\ref{B20}) for $\rho_0=\rho_1=1$.

Therefore, the B\"acklund-gauge matrix (\ref{eq2.10}) of the type I gauge transformation for the usual SG model preserves the algebraic structure of the variable mass sine-Gordon case, while the corresponding B\"acklund equations  (\ref{bacvm1})-(\ref{bacvm2}) acquire explicit dependence on the background functions $\rho_0(\xi)$ and $\rho_1(\eta)$. In the limit
$\rho_0=\rho_1=1$, one recovers the standard transformations
(\ref{B10})-(\ref{B20}). For nontrivial variable-mass backgrounds, these
 relations provide a constructive tool for relating different solutions of
the vmSG equation and, in particular, for obtaining multi-soliton configurations
by a method complementary to the Hirota construction used in Ref.\cite{Kundu2007}.

\subsection{Type II B\"acklund transformation}

The type II B\"acklund transformation for the variable mass sine-Gordon (\ref{eq:intro-vmsg}) can be derived by considering a generalized form of the $K$ matrix, as follows
\br 
\label{type2v}
 K_{II}^{vm}  = K_0 + \frac{1}{\l} K_1+ \frac{1}{\l^2} K_2,
\er 
where $K_0, K_1, K_2$ are $2 \times 2$ matrices. Solving the gauge transformations grade by grade, the $K_{II}^{vm}$ matrix takes the same form as in (\ref{eq2.14}), i.e
\br
 K_{II}^{vm} &=& \(\begin{array}{cc}
a  e^{-\frac{i}{2} w^{-}} +  \frac{c}{\l^2} e^{\frac{i}{2} w^{-}} & \frac{a c}{\l b} e^{i(\L - \frac{1}{2} w^{+})} (e^{i w^{-}} + e^{-i w^{-}}  +c_0)\\
\frac{b}{\l} e^{-i(\L - \frac{1}{2} w^{+})}&  a e^{\frac{i}{2} w^{-}} + \frac{c}{\l^2} e^{-\frac{i}{2} w^{-}} \end{array}\), \label{eq2.14vm}
\er
which will generate the type-II B\"acklund transformation for the variable mass sine-Gordon model provided that 
\br
\label{rho12c}
\bar{\rho}_0 = \rho_0,\,\,\,\,\bar{\rho}_1 = \rho_1.
\er 
Note that (\ref{eq2.14vm}) solves the gauge equation (\ref{gauge1}), with the matrices $M$ and $N$ defined in (\ref{laxvm1})-(\ref{laxvm2}), provided that the following relations are satisfied
\br
\label{back11v}
i\pa_{\xi} w^{-} &=& \frac{m_0}{a } \rho_0\, (b  e^{-i(\Lambda - w^{+})} + \b_{12} e^{-\frac{i}{2} w^{+}} ),\\
\label{back12v}
i\pa_{\eta} w^{-} &=& -\frac{m_0 }{c }  \rho_1\, (\b_{12} e^{\frac{i}{2} w^{+}}  + b  e^{-i\Lambda} ),\\
\label{back13v}
i \pa_{\xi} \Lambda &=& -\frac{m_0 c }{b } \rho_0 \, e^{i(\Lambda - w^{+})}( e^{i w^{-}} - e^{-i w^{-}} ),\\
\label{back14v}
i\pa_{\eta}(\Lambda - w^{+}) &=& \frac{m_0 a }{b }  \rho_1 \, e^{i\Lambda} ( e^{i w^{-}} - e^{-i w^{-}}),
\er
and 
\br
\label{b12r}
\pa_{\xi} \b_{12} &=& -\frac{i}{2} \pa_{\xi} w^{+} \b_{12} + c  m_0 \rho_0 e^{\frac{i}{2} w^{+}} ( e^{i w^{-}} - e^{-i w^{-}}),\\
\label{b12rr}
\pa_{\eta} \b_{12} &=& \frac{i}{2} \pa_{\eta} w^{+} \b_{12} - a m_0 \rho_1 e^{-\frac{i}{2} w^{+}} (e^{i w^{-}} - e^{-i w^{-}}), \label{b12v}\\
\b_{12} &=& \frac{a  c }{b }  e^{i(\Lambda -\frac{1}{2} w^{+})} ( e^{i w^{-}} + e^{-i w^{-}} + c_{0})\label{L11}
,\er
with $c_0$ a constant. Moreover, one has
\br
\nonumber
\pa_{\xi} \pa_{\eta} \Lambda &=& \frac{im_0^2 a  c \rho_0 \rho_1}{b ^2} e^{i(\Lambda -w^{+})} (e^{i w^{-}}-e^{-i w^{-}}) \times\\
&&(\frac{b^2}{a c } + e^{2i\Lambda} +e^{i \Lambda} (e^{i w^{-}}-e^{-i w^{-}})+ e^{2i\Lambda} (e^{i w^{-}}+e^{-i w^{-}}+ c_0). 
\er

Therefore, the type-II B\"acklund-gauge matrix $K^{\rm vm}_{II}$ in
(\ref{eq2.14vm}) preserves the algebraic structure of the ordinary sine-Gordon type-II construction, while the corresponding B\"acklund equations are dressed
by the variable-mass background functions $\rho_0(\xi)$ and
$\rho_1(\eta)$. The auxiliary field $\Lambda$, together with
$\beta_{12}$, encodes the additional degree of freedom characteristic of
type-II B\"acklund transformations and type-II defects.

In the homogeneous limit $\rho_0=\rho_1=1$, the relations (\ref{back11v})-(\ref{b12v}) reduce to the
standard type-II B\"acklund transformations of the sine-Gordon model  (\ref{back11})-(\ref{b12}), whereas
for nontrivial variable-mass backgrounds they provide a constructive
framework for relating vmSG solutions in the presence of an auxiliary defect degree of freedom.

The background-matching conditions are identical for the type-I and type-II matrices considered here. Their field compatibility requirements differ: type I involves the two first-order relations for the bulk fields, whereas type II additionally requires the auxiliary-field equations and the relation defining $\beta_{12}$.

For real matched backgrounds and real nonzero parameters $a,b,c$, with $c_0\in\mathbb R$, the type-II B\"acklund equations admit the real reduction $\Lambda=\ell+\ii\chi$, where $e^{2\chi}=-(ac/b^2)(2\cos w^-+c_0)>0$. Under this constraint, equations (\ref{back11v})--(\ref{back14v}) become a real first-order system for $w^\pm$ and $\ell$. The imaginary parts of the auxiliary-field equations coincide with the differentiated constraint, establishing its local preservation. The same reduction makes the lowest-order type-II defect energy- and momentum-type contributions real. The details of these results are presented in Appendix \ref{app:reality}.

\subsection{Conservation laws}

Now, let us construct explicitly the conserved quantities for the variable mass sine-Gordon model. From an auxiliary linear problem similar to (\ref{auxsg1}) for the Lax pair (\ref{laxvm1}) and (\ref{laxvm2}) we can directly derive two conservation laws, namely
\br 
 \pa_\xi J_k^{\xi} + \pa_\eta J_k^{\eta}=0,  \qquad k=1,2,\label{cl2}
\er
with
\br
J_1^{\xi} &=& \,\frac{i}{2}(\pa_\eta w) +\frac{im_0}{\l}\r_1\(1+i e^{-iw}\G_{21}\),  \qquad J_1^{\eta}= \,\frac{i}{2}(\pa_\xi w) -i{\l} m_0\r_0\(1+i e^{iw}\G_{21}\),\label{eq3.15}\qquad \mbox{}\\
J_2^{\xi} &=& -\frac{i}{2}(\pa_\eta w) -\frac{im_0}{\l} \r_1\(1+i e^{iw}\G_{12}\), \qquad J_2^{\eta}= -\frac{i}{2}(\pa_\xi w) +i{\l}m_0 \r_0 \(1+i e^{-iw}\G_{12}\)\!\!,\qquad\,\,\mbox{}\label{eq3.16}
\er
where we have introduced the auxiliary functions $\G_{21} = v_2 v_1^{-1}$ and  $\G_{12} = v_1 v_2^{-1}$. From these, we can obtain  the associated generating functions of the conserved quantities
\br
 Q_k = \frac{1}{2}\int_{-\infty}^{\infty} dx\, ( J_k^{\xi}+J_k^{\eta}), \qquad k=1,2.\label{eq3.17}
\er
The auxiliary functions $\G_{12}$  and $\G_{21}$ satisfy the following set of Riccati equations, 
\br
\pa_\xi \G_{21} &=& i(\pa_\xi w) \G_{21} + \l m_0 \r_0\(e^{-iw}+e^{iw} \G_{21}^2\),\label{eq3.18}\\
   \pa_\eta \G_{21} &=& -i(\pa_\eta w) \G_{21} + \frac{m_0}{\l}\r_1 \(e^{iw}+e^{-iw} \G_{21}^2\),\label{eq3.19} \\
 \pa_\xi \G_{12} &=& -i(\pa_\xi w) \G_{12} - \l m_0 \r_0\(e^{iw}+e^{-iw} \G_{12}^2\)\label{eq3.20}\\
 \pa_\eta \G_{12} &=& i(\pa_\eta w) \G_{12} - \frac{m_0}{\l}\r_1 \(e^{-iw}+e^{iw} \G_{12}^2\).\label{eq3.21} 
\er 
Since these equations are not coupled we can solve them separately. In order to determine the auxiliary function $\G_{21}$, let us consider firstly the series expansion as $\l\to\infty$, namely
\br
\G_{21} = \sum_{n=0}^{\infty}  \frac{\G_{21}^{(-n)}}{\l^n}.
\er
By substituting into eqs. (\ref{eq3.18}) and (\ref{eq3.19}), we obtain the following coefficients,
\br
\G_{21}^{(-0)} &=& ie^{-iw},\nonumber\\
  \G_{21}^{(-1)} &=&  \frac{1}{m_0 \r_0}(\pa_\xi  e^{-iw}) = -\frac{i}{m_0\r_0}(\pa_\xi w) e^{-iw},\nonumber\\
 \G_{21}^{(-2)} &=& -\frac{i}{2m_0^2\r_0}\pa_\xi\left[\frac{1}{\r_0}(\pa_\xi  e^{-iw})\right] = \frac{i}{2m_0^2\r_0^2}\left[(\pa_\xi w)^2 + i\pa_\xi^2 w -\frac{i}{\r_0}(\pa_\xi\r_0)(\pa_\xi w)\right] e^{-iw}, \nonumber\\[0.2cm]
 \G_{21}^{(-3)} &=& -\frac{1}{4m_0^3\r_0}\left\{\pa_\xi\left[ \frac{1}{\r_0}\pa_\xi\left[\frac{1}{\r_0}(\pa_\xi  e^{-iw})\right] \right] +\frac{i}{\r_0}(\pa_\xi w)\pa_\xi\left[\frac{1}{\r_0}(\pa_\xi e^{-iw})\right]\right\}\nonumber \\
&=& \frac{i}{4m_0^3\r_0^3}\left[\pa_\xi^3 w - 2i(\pa_\xi w)(\pa_\xi^2 w) -\frac{1}{\r_0}(\pa_\xi^2\r_0)(\pa_\xi w) +\frac{2i}{\r_0}(\pa_\xi\r_0)(\pa_\xi w )^2 -\frac{3}{\r_0}(\pa_\xi \r_0)(\pa_\xi^2 w) \right. \nonumber \\
&& \left. \qquad \quad\,\, +\frac{3}{\r_0^2}(\pa_\xi \r_0)^2 (\pa_\xi w) \right] e^{-iw},\\&&\dots\dots \nonumber \mbox{}
\er
Thus, we can get the first infinite set of conserved quantities generated from,
\br
Q_1&=&\int_{-\infty}^{\infty} dx \left[ \frac{i}{2}(\pa_t w) -\frac{im_0}{2}\(\l\r_0-\l^{-1}\r_1\)+ \frac{m_0}{2}\(\l \r_0 e^{iw} - \l^{-1} \r_1 e^{-iw}   \)\G_{21}\right].\label{vgf1}\qquad\mbox{}
\er
From the coefficients  of the expansion of $\G_{21}$, it is not difficult to see that 
the zero grade term gives rise to a topological charge. Then, the first non-trivial conserved charges are given by,
\br
Q_1^{(-1)}&=& \frac{i}{4m_0}\int_{-\infty}^{\infty} dx\, \frac{1}{\r_0}\left[(\pa_\xi w)^2+i(\pa_\xi^2 w)- \frac{i}{\r_0}(\pa_\xi \r_0)(\pa_\xi w)- 2m_0^2\r_0\r_1 (e^{-2iw}-1) \right],\label{eq3.26}\\
 Q_1^{(-2)} &=& \frac{i}{8m_0^2}\int_{-\infty}^{\infty} dx \,\frac{1}{\r_0^2}\left[\pa_\xi^3 w -2i(\pa_\xi w)(\pa_\xi^2 w) -\frac{1}{\r_0}(\pa_\xi^2\r_0)(\pa_\xi w) +\frac{2i}{\r_0}(\pa_\xi\r_0)(\pa_\xi w )^2 \right.\nonumber \\&&\left. \qquad \qquad \qquad \,\,\,\,\,\, -\frac{3}{\r_0}(\pa_\xi \r_0)(\pa_\xi^2 w) +\frac{3}{\r_0^2}(\pa_\xi \r_0)^2 (\pa_\xi w)+4m_0^2\r_0\r_1(\pa_\xi w) e^{-2iw} \right]\!\!.\qquad\,\, \mbox{}
\er
Now, if we consider the series expansion of the auxiliary function $\G_{21}$ as $\l\to 0$, namely,
\br
 \G_{21} = \sum_{n=0}^{\infty} {\l^n}\, {\G_{21}^{(n)}}.
\er
we obtain the following coefficients,
\br
\G_{21}^{(0)}&=& ie^{iw}\nonumber\\
\G_{21}^{(1)}&=&\frac{1}{m_0\r_1} (\pa_\eta e^{iw}) = \frac{i}{m_0\r_1}(\pa_\eta w) e^{iw},\nonumber\\
\G_{21}^{(2)} &=& -\frac{i}{2m_0^2\r_1}\pa_\eta\left[\frac{1}{\r_1} (\pa_\eta e^{iw}) \right] =  \frac{i}{2m_0^2\r_1^2}\left[(\pa_\eta w)^2 - i\pa_\eta^2 w +\frac{i}{\r_1}(\pa_\eta\r_1)(\pa_\eta w)\right] e^{iw}, \quad \mbox{}\nonumber\\[0.2cm]
 \G_{21}^{(3)} &=&-\frac{1}{4m_0^3\r_1}\left\{\pa_\eta\left[ \frac{1}{\r_1}\pa_\eta\left[\frac{1}{\r_1}(\pa_\eta  e^{iw})\right] \right] -\frac{i}{\r_1}(\pa_\eta w)\pa_\eta\left[\frac{1}{\r_1}(\pa_\eta e^{iw})\right]\right\}\nonumber \\[0.1cm]
 &=&-\frac{i}{4m_0^3\r_1^3}\left[\pa_\eta^3 w + 2i(\pa_\eta w)(\pa_\eta^2 w) -\frac{1}{\r_1}(\pa_\eta^2\r_1)(\pa_\eta w) -\frac{2i}{\r_1}(\pa_\eta\r_1)(\pa_\eta w )^2 \right. \nonumber \\
&& \left. \qquad \qquad \,-\frac{3}{\r_1}(\pa_\eta \r_1)(\pa_\eta^2 w) +\frac{3}{\r_1^2}(\pa_\eta \r_1)^2 (\pa_\eta w) \right] e^{iw}.
\er
Again, by substituting these coefficients in eq. (\ref{vgf1}) we obtain a topological charge from the zero grade term, and then the first corresponding non-trivial charges are given by,
\br
 Q_1^{(+1)}&=& -\frac{i}{4m_0}\int_{-\infty}^{\infty} dx\,\frac{1}{\r_1} \left[(\pa_\eta w)^2-i(\pa_\eta^2 w) +\frac{i}{\r_1}(\pa_\eta \r_1)(\pa_\eta w) - 2m_0^2\r_0\r_1 (e^{2iw}-1) \right]\!\!,\qquad \,\,\mbox{}\\
 Q_1^{(+2)} &=& \frac{i}{8m_0^2}\int_{-\infty}^{\infty} dx \,\frac{1}{\r_1^2} \left[\pa_\eta^3 w +2i(\pa_\eta w)(\pa_\eta^2 w) -\frac{1}{\r_1}(\pa_\eta^2\r_1)(\pa_\eta w) -\frac{2i}{\r_1}(\pa_\eta\r_1)(\pa_\eta w )^2 \right.\nonumber \\&&\left. \qquad \qquad \qquad \,\,\,\,\,\,\, -\frac{3}{\r_1}(\pa_\eta \r_1)(\pa_\eta^2 w) +\frac{3}{\r_1^2}(\pa_\eta \r_1)^2 (\pa_\eta w) +4m_0^2\r_0\r_1(\pa_\eta w) e^{2iw} \right]\!\!.\qquad \mbox{}
\er 
Analogously, the Riccati equations (\ref{eq3.20}) and (\ref{eq3.21}) can be solved recursively to obtain the coefficients of the $\G_{12}$ auxiliary function, namely,
\br
\G_{12}^{(-0)} &=& ie^{iw},\nonumber\\
  \G_{12}^{(-1)} &=&  -\frac{1}{m_0\r_0}(\pa_\xi  e^{iw}) = -\frac{i}{m_0\r_0}(\pa_\xi w) e^{iw},\nonumber\\
 \G_{12}^{(-2)} &=& -\frac{i}{2m_0^2\r_0}\pa_\xi\left[\frac{1}{\r_0}(\pa_\xi  e^{iw})\right] = \frac{i}{2m_0^2\r_0^2}\left[(\pa_\xi w)^2 - i\pa_\xi^2 w +\frac{i}{\r_0}(\pa_\xi\r_0)(\pa_\xi w)\right] e^{iw}, \nonumber\\[0.2cm]
 \G_{12}^{(-3)} &=& \frac{1}{4m_0^3\r_0}\left\{\pa_\xi\left[ \frac{1}{\r_0}\pa_\xi\left[\frac{1}{\r_0}(\pa_\xi  e^{iw})\right] \right] -\frac{i}{\r_0}(\pa_\xi w)\pa_\xi\left[\frac{1}{\r_0}(\pa_\xi e^{iw})\right]\right\}\nonumber \\
&=& \frac{i}{4m_0^3\r_0^3}\left[\pa_\xi^3 w + 2i(\pa_\xi w)(\pa_\xi^2 w) -\frac{1}{\r_0}(\pa_\xi^2\r_0)(\pa_\xi w) -\frac{2i}{\r_0}(\pa_\xi\r_0)(\pa_\xi w )^2 -\frac{3}{\r_0}(\pa_\xi \r_0)(\pa_\xi^2 w) \right. \nonumber \\
&& \left. \qquad \quad \,\,+\frac{3}{\r_0^2}(\pa_\xi \r_0)^2 (\pa_\xi w) \right] e^{iw}, 
\er
and, 
\br
\G_{12}^{(0)}&=& ie^{-iw},\nonumber\\
\G_{12}^{(1)}&=&-\frac{1}{m_0\r_1} (\pa_\eta e^{-iw}) = \frac{i}{m_0\r_1}(\pa_\eta w) e^{-iw},\nonumber\\
\G_{12}^{(2)} &=& -\frac{i}{2m_0^2\r_1}\pa_\eta\left[\frac{1}{\r_1} (\pa_\eta e^{-iw}) \right] =  \frac{i}{2m_0^2\r_1^2}\left[(\pa_\eta w)^2 + i\pa_\eta^2 w -\frac{i}{\r_1}(\pa_\eta\r_1)(\pa_\eta w)\right] e^{-iw}, \quad \mbox{}\nonumber\\[0.2cm]
 \G_{12}^{(3)} &=&\frac{1}{4m_0^3\r_1}\left\{\pa_\eta\left[ \frac{1}{\r_1}\pa_\eta\left[\frac{1}{\r_1}(\pa_\eta  e^{-iw})\right] \right] +\frac{i}{\r_1}(\pa_\eta w)\pa_\eta\left[\frac{1}{\r_1}(\pa_\eta e^{-iw})\right]\right\}\nonumber \\[0.1cm]
 &=&-\frac{i}{4m_0^3\r_1^3}\left[\pa_\eta^3 w - 2i(\pa_\eta w)(\pa_\eta^2 w) -\frac{1}{\r_1}(\pa_\eta^2\r_1)(\pa_\eta w) +\frac{2i}{\r_1}(\pa_\eta\r_1)(\pa_\eta w )^2 -\frac{3}{\r_1}(\pa_\eta \r_1)(\pa_\eta^2 w) \right. \nonumber \\
&& \left. \qquad \qquad\,+\frac{3}{\r_1^2}(\pa_\eta \r_1)^2 (\pa_\eta w) \right] e^{-iw}.
\er
Therefore, from the second generating function of the infinite conserved quantities,
\br
Q_2 &=&\int_{-\infty}^{\infty} dx \left[ -\frac{i}{2}(\pa_t w) +\frac{im_0}{2}\(\l \r_0 - \l^{-1} \r_1\)+ \frac{m_0}{2}\(\l^{-1} \r_1e^{iw} - \l \r_0 e^{-iw}   \)\G_{12}\right],\qquad \mbox{}
\er
we obtain the following first non-trivial conserved charges,
\br
 Q_2^{(-1)} &=& -\frac{i}{4m_0}\int_{-\infty}^{\infty} dx \,\frac{1}{\r_0}\left[(\pa_\xi w)^2-i(\pa_\xi^2 w) +\frac{i}{\r_0}(\pa_\xi\r_0)(\pa_\xi w) - 2m_0^2\r_0\r_1 (e^{2iw}-1) \right]\!\!,\quad\,\,\mbox{}\\
 Q_2^{(-2)} &=& -\frac{i}{8m_0^2}\int_{-\infty}^{\infty} dx \,\frac{1}{\r_0^2}\left[\pa_\xi^3 w +2i(\pa_\xi w)(\pa_\xi^2 w) -\frac{1}{\r_0}(\pa_\xi^2\r_0)(\pa_\xi w) -\frac{2i}{\r_0}(\pa_\xi\r_0)(\pa_\xi w )^2 \right.\nonumber \\&&\left. \qquad \qquad \qquad \quad \,\,\,\,-\frac{3}{\r_0}(\pa_\xi \r_0)(\pa_\xi^2 w) +\frac{3}{\r_0^2}(\pa_\xi \r_0)^2 (\pa_\xi w)+4m_0^2\r_0\r_1(\pa_\xi w) e^{2iw} \right]\!\!,\qquad \mbox{}\\
  Q_2^{(+1)} &=& \frac{i}{4m_0}\int_{-\infty}^{\infty} dx\,\frac{1}{\r_1} \left[(\pa_\eta w)^2+i(\pa_\eta^2 w)-\frac{i}{\r_1}(\pa_\eta \r_1)(\pa_\eta w) - 2m_0^2\r_0\r_1(e^{-2iw}-1) \right]\!\!,\\
 Q_2^{(+2)} &=& -\frac{i}{8m_0^2}\int_{-\infty}^{\infty} dx\,\frac{1}{\r_1^2}\left[\pa_\eta^3 w - 2i(\pa_\eta w)(\pa_\eta^2 w) -\frac{1}{\r_1}(\pa_\eta^2\r_1)(\pa_\eta w) +\frac{2i}{\r_1}(\pa_\eta\r_1)(\pa_\eta w )^2  \right. \nonumber \\
&&\left. \qquad \qquad \qquad \quad  -\frac{3}{\r_1}(\pa_\eta \r_1)(\pa_\eta^2 w)+\frac{3}{\r_1^2}(\pa_\eta \r_1)^2 (\pa_\eta w) +4m_0^2\r_0\r_1(\pa_\eta w) e^{-2iw}\right]\!\!. \qquad \mbox{}
\er
\subsubsection{First order conserved charges}

The corresponding linear combinations of the first order conserved charges $Q_1^{(\pm 1)}$ and $Q_2^{(\pm 1)}$ become,
\br
 P^{(1)}_{\mbox{\tiny vm}}&=& \frac{im_0}{2}\left[(Q_2^{(-1)}-Q_1^{(-1)}) + (Q_2^{(+1)}-Q_1^{(+1)}) \right] \nonumber \\
 &=&\frac{1}{4}\int_{-\infty}^{\infty} dx \left\{\(\frac{1}{\r_0}-\frac{1}{\r_1}\)\Big[(\pa_t w)^2+(\pa_x w)^2\Big] +2\(\frac{1}{\r_0}+\frac{1}{\r_1}\)(\pa_t w)(\pa_x w)\nonumber\right. \\&& \qquad \qquad \quad + 2m_0^2(\r_0-\r_1) (\cos 2w-1)\Big\}.\label{p1}\\
 E^{(1)}_{\mbox{\tiny vm}}&=& \frac{im_0}{2}\left[(Q_2^{(-1)}-Q_1^{(-1)}) - (Q_2^{(+1)}-Q_1^{(+1)}) \right]\nonumber\\
 &=&\frac{1}{4}\int_{-\infty}^{\infty} dx \left\{\(\frac{1}{\r_0}+\frac{1}{\r_1}\)\Big[(\pa_t w)^2+(\pa_x w)^2\Big] +2\(\frac{1}{\r_0}-\frac{1}{\r_1}\)(\pa_t w)(\pa_x w)\nonumber\right. \\&& \qquad \qquad \quad - 2m_0^2(\r_0+\r_1)(\cos 2w-1)\Big\}. \label{e1}
\er

For general background fields $\rho_0$ and $\rho_1$, the vmSG model lacks space-time translation invariance and therefore does not possess the usual conserved energy and momentum. Nevertheless, the quantities defined in Eqs.~(\ref{p1})--(\ref{e1}) reduce to the canonical momentum and energy of the standard sine-Gordon model in the homogeneous limit $\rho_0=\rho_1=1$.

\subsubsection{Second order conserved charges: trivial on the bulk}
The second-order combinations simplify when written in terms of the
background-covariant derivatives
\br
\label{cov1}
 D_0\equiv \frac{1}{\rho_0}\partial_\xi,
 \qquad
 D_1\equiv \frac{1}{\rho_1}\partial_\eta .
\er
Indeed, the differences of the Riccati charges can be expressed as
\begin{equation}
 Q_2^{(-2)}-Q_1^{(-2)}=-\frac{i}{4m_0^2}\int_{-\infty}^{+\infty}dx\,F_\xi,
 \qquad
 Q_2^{(+2)}-Q_1^{(+2)}=-\frac{i}{4m_0^2}\int_{-\infty}^{+\infty}dx\,F_\eta,
 \label{eq:second_order_differences}
\end{equation}
where
\begin{equation}
 \begin{split}
 F_\xi&=\rho_0D_0^3w+4m_0^2\rho_1(D_0w)\cos(2w),\\
 F_\eta&=\rho_1D_1^3w+4m_0^2\rho_0(D_1w)\cos(2w).
 \end{split}
 \label{eq:second_order_F}
\end{equation}
Therefore, in analogy with Eqs.~(3.41)--(3.42),
\begin{equation}
 \begin{split}
 P^{(2)}_{\mbox{\tiny vm}}&\equiv \frac{im_0}{2}\Big[
 (Q_2^{(-2)}-Q_1^{(-2)})+(Q_2^{(+2)}-Q_1^{(+2)})\Big]
 =\frac{1}{8m_0}\int_{-\infty}^{+\infty}dx\,(F_\xi+F_\eta),\\
 E^{(2)}_{\mbox{\tiny vm}}&\equiv \frac{im_0}{2}\Big[
 (Q_2^{(-2)}-Q_1^{(-2)})-(Q_2^{(+2)}-Q_1^{(+2)})\Big]
 =\frac{1}{8m_0}\int_{-\infty}^{+\infty}dx\,(F_\xi-F_\eta).
 \end{split}
 \label{eq:second_order_PE}
\end{equation}
Using the field equation
$D_0D_1w+2m_0^2\sin(2w)=0$ together with
$2\partial_x=\rho_0D_0-\rho_1D_1$, one finds
\begin{equation}
 F_\xi=2\partial_x(D_0^2w),
 \qquad
 F_\eta=-2\partial_x(D_1^2w),
 \label{eq:second_order_total_derivatives}
\end{equation}
and hence
\begin{equation}
 P^{(2)}_{\mbox{\tiny vm}}=\frac{1}{4m_0}
 \Big[D_0^2w-D_1^2w\Big]_{-\infty}^{+\infty},
 \qquad
 E^{(2)}_{\mbox{\tiny vm}}=\frac{1}{4m_0}
 \Big[D_0^2w+D_1^2w\Big]_{-\infty}^{+\infty}.
 \label{eq:second_order_surface_charges}
\end{equation}
Thus, for localized kink, antikink, or breather configurations with vanishing
field derivatives at spatial infinity, $P^{(2)}_{\mbox{\tiny vm}}=E^{(2)}_{\mbox{\tiny vm}}=0$. These combinations
are therefore surface charges rather than new independent bulk charges. In the two-half-line defect problem, however, they reduce to boundary values at $x=0^\pm$ and must be supplemented by the corresponding order-$\lambda^{\pm2}$ defect contributions.

\subsubsection{Third order conserved charges}
\label{3order}
 To display explicitly the next odd member of the hierarchy, it is
convenient to introduce the background-covariant light-cone derivatives as in (\ref{cov1}) and then one can define
\begin{equation}
 u\equiv \mathcal{D}_{0}w,
 \qquad
 v\equiv \mathcal{D}_{1}w,
 \label{eq:vmSG-covariant-derivatives}
\end{equation}
and the recursive notation
\begin{equation}
 u_{n}\equiv \mathcal{D}_{0}^{\,n}u,
 \qquad
 v_{n}\equiv \mathcal{D}_{1}^{\,n}v,
 \qquad n=1,2,\ldots .
 \label{eq:vmSG-recursive-uv}
\end{equation}
In the original coordinates these definitions begin with
\begin{equation}
 u=\frac{w_{t}+w_{x}}{\rho_{0}},
 \qquad
 v=\frac{w_{t}-w_{x}}{\rho_{1}},
 \qquad
 u_{1}=\frac{1}{\rho_{0}}(\partial_{t}+\partial_{x})u,
 \qquad
 v_{1}=\frac{1}{\rho_{1}}(\partial_{t}-\partial_{x})v.
 \label{eq:vmSG-uv-tx}
\end{equation}
For later use, define
\begin{align}
 \mathcal{S}_{\xi}&\equiv
 u^{4}+2u u_{2}+u_{1}^{2},
 &
 \mathcal{T}_{\xi}&\equiv
 u^{2}\cos(2w)+u_{1}\sin(2w),
 \label{eq:vmSG-ST-xi}
 \\
 \mathcal{S}_{\eta}&\equiv
 v^{4}+2v v_{2}+v_{1}^{2},
 &
 \mathcal{T}_{\eta}&\equiv
 v^{2}\cos(2w)+v_{1}\sin(2w).
 \label{eq:vmSG-ST-eta}
\end{align}

The coefficients displayed in Eqs.~(3.26), (3.31), (3.34), and
(3.35) are sufficient for the first two orders.  Because the generating
functions contain explicit factors of $\lambda^{\pm 1}$, the charges
$Q_{k}^{(\pm3)}$ require the Riccati recursion one step further.  The
relevant grade-four coefficients are
\begin{align}
 \Gamma_{21}^{(-4)}
 &=\frac{e^{-iw}}{8m_{0}^{4}}
 \left[u_{3}+2u^{2}u_{1}-i\mathcal{S}_{\xi}\right],
 &
 \Gamma_{12}^{(-4)}
 &=\frac{e^{iw}}{8m_{0}^{4}}
 \left[-u_{3}-2u^{2}u_{1}-i\mathcal{S}_{\xi}\right],
 \label{eq:vmSG-Gamma-minus4}
 \\
 \Gamma_{21}^{(4)}
 &=\frac{e^{iw}}{8m_{0}^{4}}
 \left[-v_{3}-2v^{2}v_{1}-i\mathcal{S}_{\eta}\right],
 &
 \Gamma_{12}^{(4)}
 &=\frac{e^{-iw}}{8m_{0}^{4}}
 \left[v_{3}+2v^{2}v_{1}-i\mathcal{S}_{\eta}\right].
 \label{eq:vmSG-Gamma-plus4}
\end{align}
Substitution into the generating functions (3.27) and (3.36) gives
\begin{align}
 Q_{1}^{(-3)}
 &=\int_{-\infty}^{\infty}\!\mathrm{d}x\,
 \left\{
 \frac{\rho_{0}}{16m_{0}^{3}}
 \left[u_{3}+2u^{2}u_{1}-i\mathcal{S}_{\xi}\right]
 +\frac{\rho_{1}}{4m_{0}}
 \left(u_{1}-iu^{2}\right)e^{-2iw}
 \right\},
 \label{eq:vmSG-Q1-minus3}
 \\
 Q_{2}^{(-3)}
 &=\int_{-\infty}^{\infty}\!\mathrm{d}x\,
 \left\{
 \frac{\rho_{0}}{16m_{0}^{3}}
 \left[u_{3}+2u^{2}u_{1}+i\mathcal{S}_{\xi}\right]
 +\frac{\rho_{1}}{4m_{0}}
 \left(u_{1}+iu^{2}\right)e^{2iw}
 \right\},
 \label{eq:vmSG-Q2-minus3}
 \\
 Q_{1}^{(+3)}
 &=\int_{-\infty}^{\infty}\!\mathrm{d}x\,
 \left\{
 \frac{\rho_{1}}{16m_{0}^{3}}
 \left[v_{3}+2v^{2}v_{1}+i\mathcal{S}_{\eta}\right]
 +\frac{\rho_{0}}{4m_{0}}
 \left(v_{1}+iv^{2}\right)e^{2iw}
 \right\},
 \label{eq:vmSG-Q1-plus3}
 \\
 Q_{2}^{(+3)}
 &=\int_{-\infty}^{\infty}\!\mathrm{d}x\,
 \left\{
 \frac{\rho_{1}}{16m_{0}^{3}}
 \left[v_{3}+2v^{2}v_{1}-i\mathcal{S}_{\eta}\right]
 +\frac{\rho_{0}}{4m_{0}}
 \left(v_{1}-iv^{2}\right)e^{-2iw}
 \right\}.
 \label{eq:vmSG-Q2-plus3}
\end{align}
The real derivative terms proportional to
$u_{3}+2u^{2}u_{1}$ and $v_{3}+2v^{2}v_{1}$ cancel in the
combinations entering the momentum- and energy-type quantities.  In fact,
\begin{align}
 Q_{2}^{(-3)}-Q_{1}^{(-3)}
 &=i\int_{-\infty}^{\infty}\!\mathrm{d}x\,
 \left[
 \frac{\rho_{0}}{8m_{0}^{3}}\mathcal{S}_{\xi}
 +\frac{\rho_{1}}{2m_{0}}\mathcal{T}_{\xi}
 \right],
 \label{eq:vmSG-Qdiff-minus3}
 \\
 Q_{2}^{(+3)}-Q_{1}^{(+3)}
 &=-i\int_{-\infty}^{\infty}\!\mathrm{d}x\,
 \left[
 \frac{\rho_{1}}{8m_{0}^{3}}\mathcal{S}_{\eta}
 +\frac{\rho_{0}}{2m_{0}}\mathcal{T}_{\eta}
 \right].
 \label{eq:vmSG-Qdiff-plus3}
\end{align}
Accordingly, the third-order analogues of the combinations in
Eqs.~(3.41) and (3.42) are defined by
\begin{align}
 P^{(3)}_{\mbox{\tiny vm}}
 &\equiv \frac{im_{0}}{2}
 \left[
 \left(Q_{2}^{(-3)}-Q_{1}^{(-3)}\right)
 +\left(Q_{2}^{(+3)}-Q_{1}^{(+3)}\right)
 \right],
 \label{eq:vmSG-P3-definition}
 \\
 E^{(3)}_{\mbox{\tiny vm}}
 &\equiv \frac{im_{0}}{2}
 \left[
 \left(Q_{2}^{(-3)}-Q_{1}^{(-3)}\right)
 -\left(Q_{2}^{(+3)}-Q_{1}^{(+3)}\right)
 \right].
 \label{eq:vmSG-E3-definition}
\end{align}
Using Eqs.~\eqref{eq:vmSG-Qdiff-minus3} and
\eqref{eq:vmSG-Qdiff-plus3}, one obtains the explicitly real expressions
\begin{align}
 P^{(3)}_{\mbox{\tiny vm}}
 &=\int_{-\infty}^{\infty}\!\mathrm{d}x\,
 \left[
 \frac{\rho_{1}}{16m_{0}^{2}}\mathcal{S}_{\eta}
 -\frac{\rho_{0}}{16m_{0}^{2}}\mathcal{S}_{\xi}
 +\frac{\rho_{0}}{4}\mathcal{T}_{\eta}
 -\frac{\rho_{1}}{4}\mathcal{T}_{\xi}
 \right],
 \label{eq:vmSG-P3}
 \\
 E^{(3)}_{\mbox{\tiny vm}}
 &=-\int_{-\infty}^{\infty}\!\mathrm{d}x\,
 \left[
 \frac{\rho_{0}}{16m_{0}^{2}}\mathcal{S}_{\xi}
 +\frac{\rho_{1}}{16m_{0}^{2}}\mathcal{S}_{\eta}
 +\frac{\rho_{1}}{4}\mathcal{T}_{\xi}
 +\frac{\rho_{0}}{4}\mathcal{T}_{\eta}
 \right].
 \label{eq:vmSG-E3}
\end{align}
For real $w$, $\rho_{0}$, and $\rho_{1}$, both charges are
real.  As in the first-order case, for generic inhomogeneous backgrounds
they are integrability-generated third order charges. In the homogeneous limit $\rho_{0}=\rho_{1}=1$, one has
$u=w_{t}+w_{x}$ and $v=w_{t}-w_{x}$, and the preceding expressions
reduce to the corresponding third-order members of the ordinary
sine--Gordon hierarchy,
\begin{align}
 P^{(3)}_{\mbox{\tiny vm}}\big|_{\rho_{0}=\rho_{1}=1}
 &=\int_{-\infty}^{\infty}\!\mathrm{d}x\,
 \left[
 \frac{\mathcal{S}_{\eta}-\mathcal{S}_{\xi}}{16m_{0}^{2}}
 +\frac{\mathcal{T}_{\eta}-\mathcal{T}_{\xi}}{4}
 \right],
 \label{eq:vmSG-P3-homogeneous}
 \\
 E^{(3)}_{\mbox{\tiny vm}}\big|_{\rho_{0}=\rho_{1}=1}
 &=-\int_{-\infty}^{\infty}\!\mathrm{d}x\,
 \left[
 \frac{\mathcal{S}_{\xi}+\mathcal{S}_{\eta}}{16m_{0}^{2}}
 +\frac{\mathcal{T}_{\xi}+\mathcal{T}_{\eta}}{4}
 \right].
 \label{eq:vmSG-E3-homogeneous}
\end{align}

\subsection{Defect conserved charges}

As in the homogeneous case, we introduce a defect at $x=0$, where the auxiliary problems $(M,N)$ and $(\bM,\bN)$ are connected by the defect matrix $K$ in Eq.~(\ref{eq2.6}), relating $w$ and $\bw$ at the defect.

Now, the generating functions $Q_k$ of the infinite set of conserved charges (\ref{eq3.17}) will be modified in the presence of such defect in the following way,
\br
\label{dch}
 Q_k &=& \frac{1}{2}\int_{-\infty}^{0} dx \,(J_k^{\xi}+J_k^{\eta})+\frac{1}{2}\int_{0}^{\infty} dx\,  (\bar{J_k}^{\xi}+\bar{J}_k^{\eta}).
\er
By taking the time-derivative, and using respectively the conservation law (\ref{cl2}), we get
\br
 \frac{dQ_k}{dt} &=& \frac{1}{2}\left(J_k^{\eta}-J_k^{\xi}\right)\Big|_{x=0}- \frac{1}{2}\left(\bJ_k^{\eta}-\bJ_k^{\xi}\right)\Big|_{x=0}.\label{eq3.42}
\er
From the relation between the two auxiliary linear problems (\ref{eq2.6}), we find that
\br
 \bG_{12} = \frac{K_{12} + K_{11}\G_{12}}{K_{22}+K_{21} \G_{12}}, \qquad \bG_{21} = \frac{K_{21} + K_{22}\G_{21}}{K_{11}+K_{12} \G_{21}},
\er
where $K_{ij}$ are the respective components of the defect matrix. Therefore, by introducing these relations in (\ref{eq3.42}), we get that
\br
 \frac{d}{dt}\(Q_k+D_k\)= 0,
\er
where $D_k$ are the defect contributions to the  generating functions of the infinite modified conserved charges ${\cal Q}_k=Q_k+D_k$, and given by
\br
\label{Def12}
 D_k &=& - \ln\Big[K_{kk}+K_{kl}\G_{lk}\Big]\Big|_{x=0}, \qquad k =1,2, \qquad \mbox{and} \qquad l\neq k.
\er
Notice that expansions in powers of $\l$ of the auxiliary functions $\G_{ij}$ provides the defect contributions to the modified conserved charges at all orders. Then, taking into account the type I defect matrix $K$ in (\ref{eq2.10vm}) and using the above formula to compute the respective first order coefficients of the defect contributions, we get 
\br 
\label{dd11} 
 D_1^{(-1)} &=& -i\s \,e^{-i(w+\bw)}, \qquad \qquad  D_1^{(+1)} \,=\, \frac{i}{\s}e^{-i(w-\bw)} -\frac{1}{m_0 \r_1}\(\pa_t w - \pa_x w\), \\
 D_2^{(-1)} &=& i\s\, e^{i(w+\bw)},\qquad \qquad \,\, \,\,\,\,\,\,D_2^{(+1)}\,=\,-\frac{i}{\s}e^{i(w-\bw)} -\frac{1}{m_0 \r_1}\(\pa_t w - \pa_x w\).\label{dd22}
\er
Then, from these results it is possible to obtain the corresponding defect contributions to the $P^{(1)}$  (\ref{p1}) and $E^{(1)}$ (\ref{e1}) charges of the variable mass sine-Gordon model by performing the following linear combinations,
\br
\label{pd1}
P^{(1)}_{\mbox{\tiny D,I}}  &=&  \frac{im_0}{2} \left[(D_2^{(-1)}-D_1^{(-1)})+(D_2^{(+1)}-D_1^{(+1)}) \right] =
   -m_0 \left[\s\cos(w+\bw) -\frac{1}{\s}\cos(w-\bw)\right] ,\qquad \,\,\,\\
 \label{ed1}
E^{(1)}_{\mbox{\tiny D,I}}  &=&  \frac{im_0}{2} \left[(D_2^{(-1)}-D_1^{(-1)})-(D_2^{(+1)}-D_1^{(+1)}) \right] =   -m_0 \left[\s\cos(w+\bw) +\frac{1}{\s}\cos(w-\bw)\right] .
\er
So, these are defect contributions to the relevant charges defined above in (\ref{p1})-(\ref{e1}), i.e. $P^{(1)}$ and $E^{(1)}$, respectively. In fact,  these results remain the same as the ones corresponding  to the momentum  ($P$) and energy ($E$) defect contributions of the type I defect sine-Gordon model previously derived in (\ref{PI})-(\ref{EI}). 
 
Similarly, it is possible to obtain the corresponding defect contributions to the $P^{(1)}$  (\ref{p1}) and $E^{(1)}$ (\ref{e1}) charges for the type II defect
\br
\label{pd11}
P^{(1)}_{\mbox{\tiny D,II}}  &=&  \frac{im_0}{2} \left[(D_2^{(-1)}-D_1^{(-1)})+(D_2^{(+1)}-D_1^{(+1)}) \right] \\
&=&\frac{m_0}{2}[\frac{b}{a} e^{i(w^{+}-\L)} + \frac{b}{c} e^{-i\L}-(\frac{c}{b} e^{-i(w^{+}-\L)} +\frac{a}{b} e^{i \L})(e^{iw^{-}} + e^{-iw^{-}} + c_0)],  \label{P'II}\\
 \label{ed11}
E^{(1)}_{\mbox{\tiny D,II}}  &=&  \frac{im_0}{2} \left[(D_2^{(-1)}-D_1^{(-1)})-(D_2^{(+1)}-D_1^{(+1)}) \right] \\
&=&  \frac{m_0}{2}[\frac{b}{a} e^{i(w^{+}-\L)} - \frac{b}{c} e^{-i \L} - (\frac{c}{b} e^{-i(w^{+}-\L)} -\frac{a}{b} e^{i \L})(e^{iw^{-}} + e^{-iw^{-}} + c_0)]  \label{E'II} .
\er  
It is important to observe that, although the vmSG Riccati coefficients
depend explicitly on the inhomogeneous functions $\rho_0(\xi)$ and
$\rho_1(\eta)$, the first type-I and type-II defect contributions entering the
combinations $P^{(1)}_{D}$ and $E^{(1)}_{D}$ are determined only by the
leading Riccati coefficients. Consequently they retain the same algebraic
form as in the constant-mass sine-Gordon case. The dependence on
$\(\rho_0,\rho_1\)$ appears in the next order coefficients of the defect generating functions, such as $D_k^{(\pm2)}$, and hence in the higher
integrability-generated defect charges.

In fact, the $\rho_j-$dependence appears at the next order charges. For example
\br
D_1^{(-2)} &=& -\frac{c}{a} e^{i w^{-}} + \frac{i c}{b m_0 \rho_0} (\pa_{\xi}w) e^{i(\L-w^{+})} S - \frac{c^2}{2b^2} e^{2i(\L-w^{+})} S^2,\\
D_2^{(-2)} &=& -\frac{c}{a} e^{-i w^{-}} + \frac{i b}{a m_0 \rho_0} (\pa_{\xi}w) e^{-i(\L-w^{+})}  - \frac{b^2}{2a^2} e^{-2i(\L-w^{+})}, 
\er 
and
\br
D_1^{(+2)} &=& -\frac{a}{c} e^{-i w^{-}} - \frac{i a}{b m_0 \rho_1} (\pa_{\eta}w) e^{i \L} S - \frac{a^2}{2b^2} e^{2i\L} S^2,\\
D_2^{(+2)} &=& -\frac{a}{c} e^{i w^{-}} - \frac{i b}{c m_0 \rho_1} (\pa_{\eta}w) e^{-i \L}  - \frac{b^2}{2c^2} e^{-2i \L},
\er 
where $S\equiv e^{iw^{-}} + e^{-iw^{-}} + c_0$.
 
The inverse-background Riccati formulas are understood on regions where the real backgrounds 
are sufficiently smooth and nonzero. For the explicitly retained orders, we assume $C^3$ backgrounds 
with bounded derivatives through third order and $0 < r_{\star} \leq |\rho_j| \leq r^{\star}< \infty$. Smoothness is assumed 
when discussing the full formal hierarchy. The spatial charges are restricted to configurations 
with integrable vacuum-subtracted densities and vanishing asymptotic fluxes. These are sufficient 
hypotheses; more general local backgrounds may be considered on regular patches.

\section{Defect conditions, soliton transmission, and anomalies}

The previous section showed that the first integrability-generated charges $P^{(1)}_{\mbox{\tiny vm}}$ and $E^{(1)}_{\mbox{\tiny vm}}$, together with their defect contributions, $P^{(1)}_{\mbox{\tiny D, a}}$ and $E^{(1)}_{\mbox{\tiny D, a}}$, $\mbox{\tiny a=I,II}$, can be obtained
systematically from the Riccati expansion and the defect matrix. Although the
type-I and type-II defect terms keep the same algebraic form as in the
homogeneous sine-Gordon model, the bulk charges, $P^{(1)}_{\mbox{\tiny vm}}$ and $E^{(1)}_{\mbox{\tiny vm}}$, are explicitly dressed by the
variable-mass functions $\rho_0(\xi)$ and $\rho_1(\eta)$. Thus, for generic
inhomogeneous backgrounds, these quantities are not the canonical momentum and
energy, but modified charges generated by the integrable structure.

In this section, we examine the conservation of these quantities by differentiating the corresponding bulk charges with respect to time in the presence of the defect. Although distinct background functions may be considered on the two half-lines, $\{\rho_0,\rho_1\}\neq \{\bar\rho_0,\bar\rho_1\}$, the B\"acklund–gauge constructions derived above do not cover this case, and the existence of a local defect contribution cannot be assumed. We therefore restrict the following analysis to matching backgrounds  
\br
\label{rhodif}
   \rho_j = \bar{\rho}_j,\,\,\,\, j=0,1,
\er
and compute the time derivatives of $P^{(1)}_{\mbox{\tiny vm}}$ and $E^{(1)}_{\mbox{\tiny vm}}$. The aim is to determine
whether the boundary fluxes can be written as total time derivatives of local defect functionals,
\br
\label{quasi10}
   \frac{d {\cal P}}{dt} &=&{\cal A}_1^{(1)},\,\,\,\,\,\, {\cal P} \equiv  P^{(1)}_{\mbox{\tiny vm}} + P^{(1)}_D,\\
	\label{quasi20}
   \frac{d {\cal E}}{dt}&=&{\cal A}_2^{(1)},\,\,\,\,\,\, {\cal E} \equiv  E^{(1)}_{\mbox{\tiny vm}} + E^{(1)}_D,
\er
where ${\cal A}_1^{(1)}$ and ${\cal A}_2^{(1)}$ are the anomalies that prevent the charges ${\cal P}$ and ${\cal E}$ from being exactly conserved. 

When the anomalies ${\cal A}_1^{(1)}$ and ${\cal A}_2^{(1)}$ vanish, the corresponding modified charges are exactly conserved. In general, these anomalies arise from deformations of the defect sewing conditions defined by the frozen B\"acklund transformation. Matching the variable-mass backgrounds, $\bar{\rho}_j=\rho_j$ for $j=0,1$, does not by itself ensure vanishing anomalies: assigning different values of the B\"acklund parameter $\sigma$ to the two sewing equations can also generate nonzero contributions. These anomalies quantify the failure of exact conservation of the modified charges.

Firstly, let us examine the defect contributions to $P^{(1)}_{\mbox{\tiny vm}}$ and $E^{(1)}_{\mbox{\tiny vm}}$ by deriving the boundary fluxes and then impose the matching-background condition (\ref{rhodif}). So, one has
\br
 \frac{d}{dt}P^{(1)}_{\mbox{\tiny vm}}&=& \frac{1}{4}\int_{-\infty}^{0} dx \frac{\pa}{\pa_t}\Big\{\(\frac{1}{\r_0}-\frac{1}{\r_1}\)\Big[(\pa_t w)^2+(\pa_x w)^2\Big] +\nonumber\\
&& 2\(\frac{1}{\r_0}+\frac{1}{\r_1}\)(\pa_t w)(\pa_x w)  + 2m_0^2(\r_0-\r_1) (\cos 2w-1)\Big\}+ \nonumber \\
\nonumber
&& \frac{1}{4}\int^{+\infty}_{0} dx \frac{\pa}{\pa_{t}} \Big\{\(\frac{1}{\bar{\r}_0}-\frac{1}{\bar{\r}_1}\)\Big[(\pa_t  \bar{w})^2+(\pa_x  \bar{w})^2\Big] +\\
&&2\(\frac{1}{\bar{\r}_0}+\frac{1}{\bar{\r}_1}\)(\pa_t  \bar{w})(\pa_x  \bar{w}) + 2m_0^2(\bar{\r}_0-\bar{\r}_1) (\cos 2 \bar{w}-1)\Big\}\\
\nonumber
\frac{d}{dt}E^{(1)}_{\mbox{\tiny vm}}&=&  \frac{1}{4}\int_{-\infty}^{0} dx  \frac{\pa}{\pa_t} \Big\{\(\frac{1}{\r_0}+\frac{1}{\r_1}\)\Big[(\pa_t w)^2+(\pa_x w)^2\Big] + \\
\nonumber
&& 2\(\frac{1}{\r_0}-\frac{1}{\r_1}\)(\pa_t w)(\pa_x w)-  2m_0^2(\r_0+\r_1)(\cos 2w-1)\Big\}+\\
\nonumber
&&\frac{1}{4}\int_{0}^{+\infty} dx  \frac{\pa}{\pa_t} \Big\{\(\frac{1}{\bar{\r}_0}+\frac{1}{\bar{\r}_1}\)\Big[(\pa_t  \bar{w})^2+(\pa_x  \bar{w})^2\Big] + \\
&& 2\(\frac{1}{\bar{\r}_0}-\frac{1}{\bar{\r}_1}\)(\pa_t  \bar{w})(\pa_x  \bar{w}) - 2m_0^2(\bar{\r}_0+\bar{\r}_1)(\cos 2 \bar{w}-1)\Big\}
\er

Then one has 
\br \nonumber
\frac{d}{dt}P^{(1)}_{\mbox{\tiny vm}}&=& \frac{1}{4} \Big[ 2 \r^{-} \pa_x w\pa_t w + \r^{+} ((\pa_x w)^2 +(\pa_t w)^2) +2 m_0^2 \r^{+} \rho_{0} \rho_{1} (\cos{2 w} -1) \Big]_{x=0}-\\
&& \frac{1}{4}  \Big[ 2 \bar{\r}^{-} \pa_x \bar{w}\pa_t \bar{w} + \bar{\r}^{+} ((\pa_x \bar{w})^2 +(\pa_t \bar{w})^2) + 2 m_0^2 \bar{\r}^{+} \bar{\rho}_{0} \bar{\rho}_{1} (\cos{2 \bar{w}} -1) \Big]_{x=0},\nonumber\\
\label{P0}
\er
and
\br \nonumber
\frac{d}{dt}E^{(1)}_{\mbox{\tiny vm}}&=& \frac{1}{4} \Big[ 2 \r^{+} \pa_x w\pa_t w + \r^{-} ((\pa_x w)^2 +(\pa_t w)^2) +2 m_o^2 \r^{-} \rho_{0} \rho_{1} (\cos{2 w} -1) \Big]_{x=0}-\\
&& \frac{1}{4}  \Big[ 2 \bar{\r}^{+} \pa_x \bar{w}\pa_t \bar{w} + \bar{\r}^{-} ((\pa_x \bar{w})^2 +(\pa_t \bar{w})^2) + 2 m_o^2 \bar{\r}^{-} \bar{\rho}_{0} \bar{\rho}_{1} (\cos{2 \bar{w}} -1) \Big]_{x=0}\nonumber\\
\label{E0}
\er

with 
\br
\rho^{\pm} \equiv \frac{\rho_1 \pm \rho_0}{\rho_1 \rho_0},\,\,\,\,
\bar{\rho}^{\pm} \equiv \frac{\bar{\rho}_1 \pm \bar{\rho}_0}{\bar{\rho}_1 \bar{\rho}_0}.
\er
Then, assuming the matching backgrounds (\ref{rhodif}) one has 
\br
\label{rhopm}
\rho^{\pm} = \bar{\rho}^{\pm}.
\er
We seek the relevant terms in the quasi-conservation equations (\ref{quasi10})--(\ref{quasi20}), assuming that the charges ${\cal P}$ and ${\cal E}$ depend on $w,\bar w,\rho_0,\rho_1$. 

Next, one considers the following matching conditions at $x=0$
\br
\label{wx1}
\pa_{x} w &=& \pa_t{\bar{w}} + \frac{1}{2}(A +B )   \\
\pa_{x} \bar{w} &=& \pa_t w  + \frac{1}{2}(B -A ).  
\label{bwx1}
\er
In the particular case of the integrable type-I, the sewing functions $A $ and $B $ become
\br
\label{ABe1100}
  A^{\epsilon} =-2 \epsilon\, m_0 \sigma \rho_0\sin w_{+},\,\,
  B^{\epsilon} =-2 \epsilon\, \frac{m_0}{\sigma} \rho_1\sin w_{-},\,\,\,\,\,\epsilon \equiv \pm 1.
\er
In fact, the frozen version of the B\"acklund transformation equations (\ref{bacvm1})-(\ref{bacvm2}) can be written as (\ref{wx1})-(\ref{bwx1}) with two different sets of sewing functions $(A^{\epsilon}, B^{\epsilon})$ (\ref{ABe1100}) since $\epsilon \equiv \pm 1$. Let us emphasize that in the integrable type-I case one must consider the same background fields $(\rho_0, \rho_1)$ in the whole space-time, i.e. $\bar{\rho}_0 = \rho_0$,\,$\bar{\rho}_1 = \rho_1$. The sign $\epsilon$ can be absorbed by redefining $\sigma\to\epsilon\sigma$; henceforth, we omit the superscript $\epsilon$ from $A^\epsilon$ and $B^\epsilon$. 

Before proceeding, several remarks are in order.
\begin{enumerate}
\item In both the type-I and type-II cases, the defect charges
$P^{(1)}_D$ and $E^{(1)}_D$ are independent of the deformation
fields. They depend on $w$ and $\bar w$ and, in the type-II
case, also on the auxiliary field $\L$.

\item The right-hand sides of the type-I and type-II
B\"acklund equations depend on $w$ and $\bar w$ without
their $\xi$- or $\eta$-derivatives. The type-II system
additionally involves $\L$ and its derivatives.

\item The frozen type-I B\"acklund equations
\eqref{bacvm1}--\eqref{bacvm2} can be expressed as the sewing
conditions \eqref{wx1}--\eqref{bwx1}, with sewing functions
$(A^{\epsilon},B^{\epsilon})$ in (\ref{ABe1100}) depending on
$\{w,\bar w,\rho_0,\rho_1\}$.

\item Under the matching-background condition \eqref{rhodif}, $\bar\rho_j=\rho_j$ for $j=0,1$, the type-I and type-II B\"acklund transformations derived above are applicable. We therefore restrict the applications below to this setting.

\item The existence of a local defect functional
$Q^{(1)}_D$ requires compatibility of its prescribed partial
derivatives. In particular,
\br
\label{compat0}
\frac{\partial^2 Q^{(1)}_D}{\partial w\,\partial\bar w}
=
\frac{\partial^2 Q^{(1)}_D}{\partial\bar w\,\partial w}.
\er
For type-II defects, the corresponding mixed-derivative
conditions involving $\L$ must also be satisfied. Since
these first-order defect charges are independent of the
background fields, no additional mixed-derivative
conditions involving those fields arise.
\end{enumerate} 

In what follows, we restrict our analysis to the type-I sewing conditions \eqref{wx1}--\eqref{bwx1} and their parameter deformations. We consider sewing functions $(A, B)$ that depend on the fields $\{w,\bar{w},\rho_0,\rho_1\}$ and on parameters deforming those of the type-I B\"acklund transformation. Although the integrable analytical construction includes both type-I and type-II defects, the parameter-deformation analysis, quasi-conservation results, and numerical simulations presented below are restricted to type-I defects. Extending these investigations to type-II defects, including the evolution of their localized auxiliary field, lies beyond the scope of this work and is left for future study.
 
Replacing (\ref{wx1})-(\ref{bwx1}) into (\ref{P0}) one has  
\begin{align}
\frac{dP^{(1)}_{\mbox{\tiny vm}}}{dt}
=&\frac14\Big\{ \Big[\frac{2}{\rho_0} A-\frac{2}{\rho_1} B\Big]\dot w
 +\Big[\frac{2}{\rho_0} A+\frac{2}{\rho_1} B\Big]\dot{\bar w}
 + \rho_+ (AB)
\nonumber\\[0.4em]
&+2m_0^2\Big[
(\rho_0+\rho_1)(\cos 2w- \cos 2\bar w)
\Big]
\Big\}_{x=0}.
\label{eq:dP-substituted0}
\end{align}
Under the matching-background condition (\ref{rhopm}), substituting
(\ref{wx1})--(\ref{bwx1}) into (\ref{P0}) cancels the terms quadratic
in the time derivatives, leaving only terms linear in these derivatives in (\ref{eq:dP-substituted0}).   

Taking into account $P^{(1)}_D = P^{(1)}_D(w,\bar{w})$ one has 
\begin{equation}
    \dot P^{(1)}_D=\frac{\partial P^{(1)}_D}{\partial w}\dot w
    +\frac{\partial P^{(1)}_D}{\partial \bar w}\dot{\bar w}.
    \label{eq:PD-time-derivative0}
\end{equation}  
The terms linear in $\dot w$ and $\dot{\bar w}$ in (\ref{eq:dP-substituted0}) can be absorbed into a defect contribution $P^{(1)}_D$ by imposing
\begin{equation}
    \frac{\partial P^{(1)}_D}{\partial w}
    =-\frac14\Big[\frac{2}{\rho_0} A-\frac{2}{\rho_1} B\Big],
    \label{eq:PD-w0}
\end{equation}
\begin{equation}
    \frac{\partial P^{(1)}_D}{\partial \bar w}
    =-\frac14\Big[\frac{2}{\rho_0} A+\frac{2}{\rho_1} B\Big].
    \label{eq:PD-wbar0}
\end{equation}
The Eqs. (\ref{eq:PD-w0})-(\ref{eq:PD-wbar0}) together with the compatibility condition $\frac{\partial P^{(1)}_D}{\partial w \partial \bar w}=\frac{\partial P^{(1)}_D}{\partial \bar w \partial w}$ provides 
\br
\label{compat11}
\frac{1}{\rho_0} A_{\bar{w}} - \frac{1}{\rho_1} B_{\bar{w}}=\frac{1}{\rho_0} A_w+\frac{1}{\rho_1} B_w.
\er

Using \eqref{eq:PD-w0}--\eqref{eq:PD-wbar0}, Eq.~\eqref{eq:dP-substituted0} can be written in the form (\ref{quasi10})
where
\begin{align}
\mathcal A_1^{(1)}
=&\Bigg[\frac1{4} (\rho_+) (AB) 
 +\frac{m_0^2}{2} 
(\rho_0+\rho_1)(\cos 2w- \cos 2\bar w)
\Bigg]_{x=0}.
\label{eq:A1-general0}
\end{align}

For matching backgrounds, $\bar{\rho}_j=\rho_j$, the type-I defect contribution \eqref{pd1}, with $P^{(1)}_D$ independent of the deformation fields, yields $\mathcal A_1^{(1)}=0$ when $A$ and $B$ are given by \eqref{wx1}--\eqref{bwx1} and \eqref{ABe1100}, consistently with the frozen B\"acklund equations \eqref{bacvm1}--\eqref{bacvm2}. More general choices of $A$ and $B$, including parameter deformations, may produce a nonvanishing anomaly in the balance law \eqref{quasi10}. If the compatibility condition \eqref{compat11} fails, however, the existence of a local defect charge cannot be assumed. An example of parameter deformation is examined below.

Similarly, replacing (\ref{wx1})-(\ref{bwx1}) into (\ref{E0}) one has  
\begin{align}
\frac{dE^{(1)}_{\mbox{\tiny vm}}}{dt}
=&\frac14\Big\{ \Big[\frac{2}{\rho_0} A+\frac{2}{\rho_1} B\Big]\dot w
 +\Big[\frac{2}{\rho_0} A-\frac{2}{\rho_1} B\Big]\dot{\bar w}
 + \rho_- (AB)
\nonumber\\[0.4em]
&+2m_0^2\Big[
(\rho_1-\rho_0)(\cos 2w- \cos 2\bar w)
\Big]
\Big\}_{x=0}.
\label{eq:dP-substituted}
\end{align}
Under the matching-background condition (\ref{rhopm}), substituting
(\ref{wx1})--(\ref{bwx1}) into (\ref{E0}) cancels the terms quadratic
in the time derivatives, leaving the linear terms displayed in
(\ref{eq:dP-substituted}).

Considering $E^{(1)}_D = E^{(1)}_D(w,\bar{w} )$ one can write
\begin{equation}
    \dot E^{(1)}_D=\frac{\partial E^{(1)}_D}{\partial w}\dot w
    +\frac{\partial E^{(1)}_D}{\partial \bar w}\dot{\bar w}.
    \label{eq:PD-time-derivative1}
\end{equation}
The terms linear in $\dot w$ and $\dot{\bar w}$ in (\ref{eq:dP-substituted}) can be absorbed into a defect contribution $E^{(1)}_D$ by imposing
\begin{equation}
    \frac{\partial E^{(1)}_D}{\partial w}
    =-\frac14\Big[\frac{2}{\rho_0} A+\frac{2}{\rho_1} B\Big],
    \label{eq:PD-w1}
\end{equation}
\begin{equation}
    \frac{\partial E^{(1)}_D}{\partial \bar w}
    =-\frac14\Big[\frac{2}{\rho_0} A-\frac{2}{\rho_1} B\Big].
    \label{eq:PD-wbar1}
\end{equation}
The Eqs. (\ref{eq:PD-w1})-(\ref{eq:PD-wbar1}) together with the compatibility condition $\frac{\partial E^{(1)}_D}{\partial w \partial \bar w}=\frac{\partial E^{(1)}_D}{\partial \bar w \partial w}$ provides 
\br
\label{compat22}
\frac{1}{\rho_0} A_{\bar w}+\frac{1}{\rho_1} B_{\bar w} =\frac{1}{\rho_0} A_w-\frac{1}{\rho_1}B_w.
\er

Using \eqref{eq:PD-w1}--\eqref{eq:PD-wbar1}, Eq.~\eqref{eq:dP-substituted} can be written in the form (\ref{quasi20}) with
\begin{align}
\mathcal A_2^{(1)}
=&\Bigg[\frac1{4} (\rho_-) (AB) 
 +\frac{m_0^2}{2} 
(\rho_1-\rho_0)(\cos 2w- \cos 2\bar w)
\Bigg]_{x=0}.
\label{eq:A1-general}
\end{align}
 
Notice that the type-I defect contribution (\ref{ed1}), for which $E^{(1)}_D$ is independent of the deformation fields and $\bar{\rho}_j= \rho_j$, together with the expressions for $A$ and $B$ in (\ref{wx1})--(\ref{bwx1}) and (\ref{ABe1100}) consistent with the frozen B\"acklund transformations (\ref{bacvm1})--(\ref{bacvm2}), leads to $\mathcal A_2^{(1)}=0$.  

In summary, one has that the sawing fields $A,B$ satisfy the mixed-derivative conditions (\ref{compat11}) and (\ref{compat22}) which ensure local compatibility in field space, i.e. from (\ref{compat11}) and (\ref{compat22}) one has  
\br
\label{ABcompat}
A_w = A_{\bar{w}},\,\,\,\,B_w = -B_{\bar{w}}. 
\er
In the following subsection, we illustrate how a controlled deformation of the defect sewing parameters generates a non-vanishing anomaly. 

\subsection{An example: matching condition generating anomaly terms}

Having established the deformed defect conditions and the corresponding quasi-conservation laws, we now examine their action on explicit nonlinear excitations. One-soliton configurations provide the simplest setting for analyzing the physical effects of the defect. Matching the soliton solutions across $x=0$ determines the constraints on their spectral parameters and the defect-induced phase and position shifts, while also identifying the conditions under which the purely transmitting character of the integrable defect is preserved. The analysis further reveals critical regimes associated with soliton absorption, emission, and changes of the outgoing topological sector. We therefore solve the defect sewing conditions for one-soliton configurations on the two half-lines and classify the resulting transmission, absorption, emission, and soliton--antisoliton conversion processes.
 
For simplicity, we consider identical variable-mass backgrounds, $\bar\rho_0=\rho_0$ and $\bar\rho_1=\rho_1$, and introduce a non-integrable defect through the deformations of the frozen B\"acklund parameter. Consider the symmetric deformation of the type-I defect functions
\br
\label{ABe11}
  A=-2 m_0 \sigma (1+\epsilon_A)\rho_0\sin w_+,
  \qquad
  B=-\frac{2 m_0}{\sigma} (1+\epsilon_B)\rho_1\sin w_- .
\er
Note that these fields $A,B$ satisfy the mixed-derivative conditions (\ref{ABcompat}). So, we allowed the functions $A$ and $B$, together with the defect contributions $P^{(1)}_D$ and $E^{(1)}_D$, to depart from their integrable forms through the nonzero deformation parameters $\epsilon_A$ and $\epsilon_B$. The integrable type-I defect functions (\ref{ABe1100}) are recovered in the limit $\epsilon_A=\epsilon_B=0$.

Then, the momentum-sector conditions for the defect contribution are
\begin{equation}
  \frac{\partial P^{(1)}_D}{\partial w}
  =
  -\frac12\left(\frac{A}{\rho_0}-\frac{B}{\rho_1}\right),
  \qquad
  \frac{\partial P^{(1)}_D}{\partial\bar w}
  =
  -\frac12\left(\frac{A}{\rho_0}+\frac{B}{\rho_1}\right).
\end{equation}
 
Therefore, the corresponding defect momentum contribution becomes
\begin{equation}
  P_D^{(1)}
  =
  - m_0\sigma (1+\epsilon_A)\cos w_+ +
  \frac{m_0}{\sigma}(1+\epsilon_B)\cos w_-.
\end{equation}

Notice that they satisfy the consistency condition $\frac{\partial^2 P_D}{\partial\bar w \partial w } = \frac{\partial^2 P_D}{\partial w \partial \bar w }$ in (\ref{compat11}) and the undeformed type-I result is recovered in the limit $
  \epsilon_A=\epsilon_B=0$.

So, for the equal-background case and for a defect contribution to $P_D$ independent of the deformation fields, the momentum-sector anomaly reduces to
\begin{equation}
\label{anomaly11}
  \mathcal A_1^{(1)}
  =-\frac{
  m_0^2}{2}(\rho_0+\rho_1)
  \left(\a \b -1 \right)
  \left(\cos 2w-\cos 2\bar w\right)
  \bigg|_{x=0},\,\,\,\,\, \a \equiv 1+ \epsilon_A,\,\,\b \equiv 1+ \epsilon_B.
\end{equation}
Thus the anomaly is controlled by the symmetric deformation parameters and vanishes in the undeformed type-I limit $\a \b =1$.

In the energy sector, the equal-background conditions that absorb the terms linear in
$\dot w$ and $\dot{\bar w}$ are
\begin{equation}
  \frac{\partial E^{(1)}_D}{\partial w}
  =-\frac12\left(\frac{A}{\rho_0}+\frac{B}{\rho_1}\right),
  \qquad
  \frac{\partial E^{(1)}_D}{\partial \bar w}
  =-\frac12\left(\frac{A}{\rho_0}-\frac{B}{\rho_1}\right).
  \label{EDconditions}
\end{equation}
 A defect contribution is therefore 
\begin{equation}   
  E_D^{(1)}
  =-m_0\left[
  \sigma(1+\eps_A)\cos(w+\bar w)
  +\frac{1+\eps_B}{\sigma}\cos(w-\bar w)
  \right].
\end{equation} 
Similarly,  they satisfy the consistency condition $\frac{\partial^2 E_D^{(1)}}{\partial\bar w \partial w } = \frac{\partial^2 E_D^{(1)}}{\partial w \partial \bar w }$ in (\ref{compat22}).

So, for equal backgrounds and for $E_D^{(1)}$ independent of the deformation fields, the energy-sector anomaly is 
\begin{equation}
  {\cal A}_2^{(1)}
  =-\frac{m_0^2}{2}(\rho_1-\rho_0)
  \left(\a \b -1\right)
  \left(\cos 2w-\cos 2\bar w\right)
  \bigg|_{x=0},
  \label{A2cos}
\end{equation}
where $\a$ and $\b$ are defined in (\ref{anomaly11}).

The anomalies $\mathcal{A}_j^{(1)},\,j=1,2$ above vanish generically when the deformation parameters obey
\br
\label{defectparam}
\a \b =1 \,\,\rightarrow \,\,\epsilon_B = -\frac{\epsilon_A}{1+ \epsilon_A}.
\er
Under this constraint, both the momentum- and energy-type quasi-conservation laws (\ref{quasi10})-(\ref{quasi20}) become exact, independently of the values of $\rho_0, \rho_1, w$ and $ \bar{w}$.  The undeformed integrable limit $\epsilon_A=\epsilon_B=0$ is included as a special case. In addition, the momentum- and energy-type anomalies vanish identically
for the background satisfying $\rho_0 = \mp\rho_1$, respectively.

Consider the variable-mass sine-Gordon solutions on the two regions separated by a defect at $x=0$:
\[
 w_L(x,t)=w(x,t),\qquad x<0,
 \qquad
 w_R(x,t)=\bar w(x,t),\qquad x>0.
\]
We impose the equal-background condition $\bar{\rho}_j = \rho_j,\,j=0,1$. So, the defect equations (\ref{wx1})-(\ref{bwx1}) can be written as
\br
\label{defeqs11}
 \partial_\xi(w_L-w_R)=A,
 \qquad
 \partial_\eta(w_L+w_R)=-B.
\er
Let us consider the one-soliton solutions. Define the variable-mass light-cone primitives
\begin{equation}
 X(\xi)=\int^{\xi}\rho_0(s)\,ds,
 \qquad
 Y(\eta)=\int^{\eta}\rho_1(s)\,ds.
\end{equation}
On each side take a one-kink solution \cite{Kundu2007}
\begin{equation}
\label{kink1}
 w_j=2\arctan E_j,
 \qquad
 E_j=e^{\Theta_j},
 \qquad
 j=L,R,
\end{equation}
with phases
\begin{equation}
 \Theta_j=2m_0\left(p_jX-\frac{1}{p_j}Y\right)+\delta_j.
\end{equation}

In fact, defining $X_\xi=\rho_0$, $Y_\eta=\rho_1$ and $w_j=2\arctan\exp\Theta_j$ with $\Theta_j=2m_0(p_jX-p_j^{-1}Y)+\delta_j$, we obtain $\partial_\xi w_j=2m_0p_j\rho_0\sin w_j$ and $\partial_\eta w_j=-2m_0p_j^{-1}\rho_1\sin w_j$. These imply $\partial_\xi\partial_\eta w_j=-2m_0^2\rho_0\rho_1\sin 2w_j$, verifying the vmSG field equation (\ref{eq:intro-vmsg}) exactly. 

Each soliton satisfies (\ref{bacvm1})-(\ref{bacvm2}) with the B\"acklund parameter taken as $\s = p_j$
\begin{equation}
 \partial_\xi w_j=2 m_0 p_j\rho_0\sin w_j,
 \qquad
 \partial_\eta w_j=-\frac{2 m_0}{p_j}\rho_1\sin w_j.
\end{equation}
Next, let us analize the sewing conditions. Substituting (\ref{kink1}) into the defect equations (\ref{defeqs11}), at $x=0$, and taking into account (\ref{ABe11}) provide us
\begin{equation}
 p_L\sin w_L-p_R\sin w_R
 =-\sigma \alpha\sin(w_L+w_R),
\label{eq:left-matching}
\end{equation}
\begin{equation}
 \frac{1}{p_L}\sin w_L+\frac{1}{p_R}\sin w_R
 =-\frac{\beta}{\sigma }\sin(w_L-w_R).
\label{eq:right-matching}
\end{equation}

Next, we find the relationships between the soliton spectral parameters. For soliton transmission one must assume the phase difference to be a constant. At $x=0$, where $\xi=\eta=t/2$, the phase difference obeys
\begin{equation}
 \frac{d}{dt}(\Theta_R-\Theta_L)
 =m_0\left[(p_R-p_L)\rho_0
 -\left(\frac{1}{p_R}-\frac{1}{p_L}\right)\rho_1\right]_{\xi=\eta=t/2}.
\end{equation}
For arbitrary independent background functions $\rho_0$ and $\rho_1$, this quantity can vanish for all $t$ only if
\begin{equation}
p_R=p_L\equiv p.
\end{equation}
Thus the defect preserves the soliton spectral parameter (and hence its rapidity); it can only produce a phase or position shift.

Next, examine the transmission factor and its compatibility with the matching conditions incorporating $\a$ and $\b$. Pure one-soliton transmission requires the two exponentials at the defect to differ only by a constant factor. With $p_L=p_R=p$, the first matching Eq.~\eqref{eq:left-matching} becomes
\begin{equation}
 p(\sin w_L-\sin w_R)
 =-\sigma \alpha\sin(w_L+w_R).
\end{equation}
Using $\tan(w_L/2)=E_L$ and $\tan(w_R/2)=E_R$, and setting $E_R=z_A E_L$, one obtains
\begin{equation}
 p(1-z_A)=-\sigma \alpha(1+z_A),
\end{equation}
and therefore
\begin{equation}
 z_A=\frac{p+\sigma \alpha}{p-\sigma \alpha}.
\end{equation}
The second matching condition, Eq.~\eqref{eq:right-matching}, by setting $E_R=z_B E_L$ yields independently
\begin{equation}
 z_B=\frac{\beta p+\sigma }{\beta p-\sigma }.
\end{equation}
Equating the two expressions, i.e.  $z_A = z_B$, gives
\begin{equation}
 2p\sigma (\alpha\beta-1)=0.
\end{equation}
For a nontrivial soliton and a nonzero defect parameter one gets the relationship (\ref{defectparam}). This is the same condition that makes the defect anomalies vanish.

Define the effective defect parameter
\begin{equation}
 \sigma_{eff}
 =\sigma (1+\epsilon_A)
 =\frac{\sigma }{1+\epsilon_B} . \label{seff1}
\end{equation}
Then the deformed matching functions recover the standard type-I form,
\begin{equation}
 A=-2m_0\sigma_{\mathrm{eff}}\rho_0\sin(w_L+w_R),
 \qquad
 B=-\frac{2m_0}{\sigma_{\mathrm{eff}}}\rho_1\sin(w_L-w_R),
\end{equation}
and the transmission factor is
\begin{equation}
 z=\frac{p+\sigma_{\mathrm{eff}}}{p-\sigma_{\mathrm{eff}}}.\label{seff2}
\end{equation}

Next, let us show the relation between the constant phases for soliton-soliton and soliton-(anti-)soliton transmission. 

{\bf Soliton-soliton transmission}

Since $E_R=zE_L$, the phase constants satisfy, for $z>0$,
\begin{equation}
 \delta_R-\delta_L
 =\ln z
 =\ln\left(\frac{p+\sigma_{\mathrm{eff}}}
 {p-\sigma_{\mathrm{eff}}}\right) .
\end{equation}
Hence
\br
 w_L(x,t)&=&2\arctan\!\left[\exp\!\left\{
 2m_0\left[pX(\xi)-p^{-1}Y(\eta)\right]+\delta_L
 \right\}\right], \label{wL}\\
 w_R(x,t)&=&2\arctan\!\left[
 z\exp\!\left\{
 2m_0\left[pX(\xi)-p^{-1}Y(\eta)\right]+\delta_L
 \right\}\right].\label{wR}
\er                                                                         {\bf Soliton-(anti-)soliton transmission}                                                            
For real $p$ and $\s_{eff}$, a soliton is converted into an antisoliton when the transmission factor is negative, i.e.  $z< 0$. Then, using Eq.~\eqref{seff2}, this condition becomes
 
\begin{equation}
 (p+\s_{eff})(p-\s_{eff})<0\,\,\,
\Rightarrow \,\,\,
 p^2-\s_{eff}^{\,2}<0.
\end{equation}
Hence the soliton-to-antisoliton conversion condition is
\begin{equation}
|\s_{eff}|>|p|.
 \label{eq:conversion-condition}
\end{equation}
  
Note that the critical values
$
 p=\pm\s_{eff} $
must be excluded from regular transmission. Indeed, one of these values gives $z=0$, whereas the other makes $z$ divergent. They correspond to limiting absorption or emission configurations rather than to a regular transmitted soliton. For $z=0$ the transmitted profile disappears, so the incoming kink is absorbed by the defect. No kink or antikink emerges on the right. At the pole $|z| \rightarrow \infty$, the regular transmission solution becomes singular. An emission configuration is obtained by keeping the outgoing soliton phase finite while sending the incoming phase to the asymptotic past. 

For a type-I defect there is no independent auxiliary field. Thus, the emission limit should not be viewed as the decay of an independently excited degree of freedom. Instead, the defect stores energy and topological information through the field discontinuity at $x=0$ and the associated localized defect contribution determined by the sewing conditions. Accordingly, $|z| \rightarrow \infty$ is naturally interpreted as the limiting, time-reversed B\"acklund counterpart of absorption, in which this stored defect content is converted into an outgoing soliton. In Appendix \ref{app:emission-absorption} we present a more complete
demonstration of absorption, emission, and topological-charge transfer. 

On the conversion branch one may write
$
 z=-|z|.
$
Then the $x>0$ solution  becomes
\begin{align}
 w_R(x,t)
 &=2\arctan\!\left(-|z|e^{\Theta_R}\right)\\
 &=-2\arctan\!\left(e^{\Theta_R+\ln|z|}\right).
 \label{eq:real-antisoliton}
\end{align}
Thus the transmitted field is a real antisoliton whose phase is shifted by
$ \Delta\Theta=\ln|z|.$

In summary, the  branch $z>0$ describes soliton--soliton transmission, whereas
$z<0$, equivalently $|\s_{eff}|>|p|$, describes soliton--antisoliton
conversion.

The incoming and outgoing topological charges are defined by
\begin{equation}
 Q_j=\frac{w_j(+\infty,t)-w_j(-\infty,t)}{\pi}, \,\,\, j = L, R.
 \label{eq:topological-charge}
\end{equation}
The bulk topological charge is defined by the field variation on $x<0$ and $x>0$ separately, so that a discontinuity at $x=0$ is not included in the bulk contribution.

For the branches in Eqs.~ \eqref{wL}-\eqref{wR}  and \eqref{eq:real-antisoliton}, one obtains, respectively 
\begin{equation}
 Q_L=+1,
 \qquad
 Q_R=+1,\,\,\,\,\,\, \Delta Q_{\mathrm{bulk}}=Q_R-Q_L=0,\,\,\,\,\,\, z>0.
 \label{eq:charge-reversal0}
\end{equation}
and
\begin{equation}
 Q_L=+1,
 \qquad
 Q_R=-1,\,\,\,\,\,\, \Delta Q_{\mathrm{bulk}}=Q_R-Q_L=-2,\,\,\,\,\,\, z<0.
 \label{eq:charge-reversal1}
\end{equation}
 
For the actual two-half-line system, let us define 
\br
\label{Qtoptot}
Q_{\rm bulk}(t) = \frac{Q_{\rm L}(0^{-},t)-Q_{\rm L}(-\infty,t)}{\pi} + \frac{Q_{\rm R}(+\infty,t)-Q_{\rm R}(0^{+},t)}{\pi},
\er
and
\br
Q_{\rm D}(t) = \frac{Q_{\rm R}(0^{+},t)-Q_{\rm L}(0^{-},t)}{\pi}
\er 
Then
\br
Q_{\rm top}(t)= Q_{\rm bulk}(t)+Q_{\rm D}(t)
\er 
is constant when the asymptotic vacua at $x= \pm \infty$ are fixed. Therefore
\br
\D Q_{\rm bulk} = - \D Q_{\rm D}.
\er
The relation with the $Q_L$ and $Q_R$ above follows in the asymptotic scattering regime. In fact, in the asymptotic scattering regime they coincide, respectively, with the physical bulk charge at $t \rightarrow - \infty$ and $t \rightarrow + \infty$, respectively. Hence $\D Q_{\rm bulk} = Q_{\rm bulk}(+\infty)-Q_{\rm bulk}(-\infty) = Q_R - Q_L$. Since 
$Q_{\rm top}= Q_{\rm bulk}+Q_{\rm D}$ is conserved, one obtains $\D Q_{\rm bulk} = - \D Q_{\rm D}$. Thus, for kink--antikink conversion, $Q_L =1$ and $Q_R = -1$ imply  $\D Q_{\rm bulk} = -2$, $\D Q_{\rm D} = +2$, and $\D Q_{\rm top}=0$.  

During kink–antikink conversion, the change in bulk topological charge is compensated by the defect contribution, so the exchanged topological information is encoded in the field discontinuity at $x=0$, rather than in an independent defect degree of freedom. Since $p_R=p_L$, the outgoing antisoliton retains the incoming soliton’s rapidity and velocity, while only its topological orientation and position are changed.

\subsubsection{First order charges: Parity and time-integrated vanishing anomalies}

In the equal-background case considered above, the two defect
anomalies (\ref{anomaly11}) and (\ref{A2cos}) can be written in the form
\begin{align}
\mathcal A_1^{(1)}(t)
&=- \frac{m_0^2}{2} \Delta_{\epsilon}\,
  \bigl[\rho_0(0,t)+\rho_1(0,t)\bigr]C(t),
\label{eq:A1}
\\
\mathcal A_2^{(1)}(t)
&=-\frac{m_0^2}{2} \Delta_{\epsilon}\,
  \bigl[\rho_1(0,t)-\rho_0(0,t)\bigr]C(t),\,\,\,\, \Delta_{\epsilon}
=\epsilon_A+\epsilon_B+\epsilon_A\epsilon_B,
\label{eq:A2}
\end{align}
where
\begin{equation}
C(t)
\equiv
\left.
\bigl(\cos (2 w_L)-\cos (2 w_R)\bigr)
\right|_{x=0}.
\label{eq:Cdef}
\end{equation}
Assume that the field-dependent factor has definite time-reversal parity,
\begin{equation}
C(-t)=\eta_C C(t),
\qquad \eta_C=\pm1.
\label{eq:Cparity}
\end{equation}
For both anomalies to be odd under time reversal,
\begin{equation}
\mathcal A_j^{(1)}(-t)=-\mathcal A_j^{(1)}(t),
\qquad j=1,2,
\label{eq:oddA}
\end{equation}
the two background fields must satisfy
\begin{equation}
\rho_j(0,-t)=-\eta_C\rho_j(0,t),
\qquad j=0,1.
\label{eq:generalrho}
\end{equation}
Thus, both variable-mass fields must possess the same time-reversal
parity at the defect.

Let us assume 
\begin{equation}
C(-t)=-C(t),
\qquad \eta_C=-1.
\label{eq:Cparity1}
\end{equation}
Equation~\eqref{eq:generalrho} then becomes
\begin{equation}
\rho_0(0,-t)=\rho_0(0,t),
\qquad
\rho_1(0,-t)=\rho_1(0,t).
\label{eq:evenrho}
\end{equation}
Hence both background fields must be even functions of time about the
collision time. Under these conditions, the combinations
$\rho_0+\rho_1$ and $\rho_1-\rho_0$ are even, while $C(t)$ is odd, and
therefore
\begin{equation}
\label{anodd}
\mathcal A_j^{(1)}(-t)=-\mathcal A_j^{(1)}(t),
\qquad j=1,2.
\end{equation}
It follows immediately that the integrated anomalies vanish over every
time-symmetric interval,
\begin{equation}
\lim_{T\to\infty} \int_{-T}^{T}\mathcal A_j^{(1)}(t)\,dt=0,
\qquad j=1,2,
\label{eq:vanishingintegral}
\end{equation}
which yields asymptotic conservation of the corresponding modified
charges. So, from (\ref{quasi10})-(\ref{quasi20}) one has
\br
\label{asympt1}
{\cal P} (T\to\infty) = {\cal P} (T\to -\infty),\,\,\,\,\,\,\,\mbox{and}\,\,\,\,\,\,\,{\cal E} (T\to\infty) = {\cal E} (T\to -\infty).
\er

Next, let us discuss  the parity symmetries in both soliton-soliton and solito-antisoliton transmission processes. So, at $x=0$, let us define
\begin{equation}
 F(t)=2m_0\left[
 pX\!\left(\frac{t}{2}\right)
 -\frac{1}{p}Y\!\left(\frac{t}{2}\right)
 \right],
 \label{eq:F}
\end{equation}
where the primitives are normalized by $X(0)=Y(0)=0$.  If the
variable-mass backgrounds are even about the collision time (\ref{eq:evenrho}) 
then $X$ and $Y$ are odd along the defect trajectory and hence
\begin{equation}
 F(-t)=-F(t).
 \label{eq:Fodd}
\end{equation}
It is convenient to introduce the operator $K$ such that 
\begin{equation}
 K[u]=2\arctan e^u,
 \qquad
 K[-u]=\pi-K[u].
 \label{eq:Kidentity}
\end{equation}

{\bf Parity conditions for soliton-soliton transmission: $z>0$}

For the positive-transmission branch, write
\begin{equation}
 w_L(0,t)=K\bigl[F(t)+\delta_L\bigr],
 \qquad
 w_R(0,t)=K\bigl[F(t)+\delta_R\bigr],
 \label{eq:ssfields}
\end{equation}
with
\begin{equation}
 \delta_R-\delta_L=\ln z.
 \label{eq:ssphasedifference}
\end{equation}
For the defect-centered soliton symmetry one has
\begin{equation}
w_L(0,-t)=\pi-w_R(0,t),
\qquad
w_R(0,-t)=\pi-w_L(0,t).
\label{eq:solsym}
\end{equation}
The first Eq. (\ref{eq:solsym}), using Eqs.~\eqref{eq:Fodd} and \eqref{eq:Kidentity}, becomes
\begin{align}
 K\bigl[-F(t)+\delta_L\bigr]
 &=\pi-K\bigl[F(t)+\delta_R\bigr] \\
 &=K\bigl[-F(t)-\delta_R\bigr].
\end{align}
Since $K$ is one-to-one, this equality holds precisely when
\begin{equation}
 \delta_L=-\delta_R.
 \label{eq:center}
\end{equation}
The second relation in (\ref{eq:solsym}) gives the same condition.  Combining
Eqs.~\eqref{eq:ssphasedifference} and \eqref{eq:center}, one finds the
centered phases
\begin{equation}
 \delta_L=-\frac{1}{2}\ln z,
 \qquad
 \delta_R=\frac{1}{2}\ln z.
 \label{eq:sscentered}
\end{equation}
Therefore, under the above conditions (\ref{eq:solsym}) and (\ref{eq:center}), on the soliton--soliton branch one has that the parity symmetry (\ref{eq:Cdef}) is satisfied.

{\bf Parity conditions for soliton-antisoliton transmission: $z<0$}

On the conversion branch, write $z=-|z|$ and represent the real
transmitted antisoliton as
\begin{equation}
 w_R(0,t)=-K\bigl[F(t)+\delta_R\bigr],
 \qquad
 \delta_R-\delta_L=\ln|z|,
 \label{eq:safields}
\end{equation}
while the incoming field remains
\begin{equation}
 w_L(0,t)=K\bigl[F(t)+\delta_L\bigr].
 \label{eq:saleft}
\end{equation}
 
The correct branch-sensitive transformation is obtained again by
choosing the centered phases $\delta_L=-\delta_R$.  One then finds
\begin{align}
 w_L(0,-t)
 &=K\bigl[-F(t)-\delta_R\bigr] \\
 &=\pi-K\bigl[F(t)+\delta_R\bigr]
 =\pi+w_R(0,t),
 \label{eq:sasym1}
\end{align}
and
\begin{align}
 w_R(0,-t)
 &=-K\bigl[-F(t)+\delta_R\bigr] \\
 &=K\bigl[F(t)-\delta_R\bigr]-\pi
 =w_L(0,t)-\pi.
 \label{eq:sasym2}
\end{align}
Hence the soliton--antisoliton branch obeys
\begin{equation}
 w_L(0,-t)=\pi+w_R(0,t),
 \qquad
 w_R(0,-t)=w_L(0,t)-\pi.
 \label{eq:sasymmetry}
\end{equation}

{\bf Parity of the anomaly factor}

The field-dependent factor entering the two defect anomalies is
\begin{equation}
 C(t)=\cos\bigl(2w_L(0,t)\bigr)
      -\cos\bigl(2w_R(0,t)\bigr).
 \label{eq:C}
\end{equation}
For the soliton--soliton branch, Eq.~\eqref{eq:solsym} directly gives
\begin{equation}
 \cos\bigl(2w_L(0,-t)\bigr)
 =\cos\bigl(2w_R(0,t)\bigr),
\end{equation}
and the analogous relation with $L$ and $R$ interchanged. 

The same identities follow from Eq.~\eqref{eq:sasymmetry} on the
soliton--antisoliton branch, because the cosine is invariant under shifts
by integer multiples of $2\pi$.  Therefore, in both cases, one has the symmetry (\ref{eq:Cparity1}).
 
When the backgrounds satisfy Eq.~\eqref{eq:evenrho}, the combinations
$\rho_0+\rho_1$ and $\rho_1-\rho_0$ are even.  Consequently, the two
defect anomalies satisfy the parity symmetry (\ref{anodd}) for both transmission branches, and their integrals vanish on any symmetric time interval.
 
\subsubsection{Third order charges: Parity and time-integrated non-vanishing anomalies}
\label{subsec:third-order-defect-parity}

The third-order charges derived in Subsection~\ref{3order} admit analogous balance
relations in the presence of the deformed type-I defect. We retain the
equal-background condition $\bar\rho_0=\rho_0$, $\bar\rho_1=\rho_1$ and
introduce
\begin{equation}
 \sigma_A\equiv\sigma \alpha,
 \qquad
 \sigma_B\equiv\frac{\sigma }{\beta},
 \qquad
 \Delta_{\epsilon}\equiv\frac{\sigma_A}{\sigma_B}-1
 =\alpha\beta-1
 =\epsilon_A+\epsilon_B+\epsilon_A\epsilon_B,
 \label{eq:third-order-effective-parameters}
\end{equation}
with $w_{\pm}=w\pm\bar w$. In terms of the covariant derivatives introduced
in Eq.~(\ref{eq:vmSG-covariant-derivatives}), let
\begin{equation}
 u=D_0w,
 \qquad \bar u=D_0\bar w,
 \qquad v=D_1w,
 \qquad \bar v=D_1\bar w,
 \qquad u_1=D_0u,
 \qquad v_1=D_1v.
 \label{eq:third-order-covariant-fields}
\end{equation}
Considering the deformed sewing functions in Eq.~(\ref{ABe11}), 
the Eqs.~(\ref{wx1})-(\ref{bwx1}) become
\begin{equation}
 u-\bar u=-2m_0\sigma_A\sin w_+,
 \qquad
 v+\bar v=\frac{2m_0}{\sigma_B}\sin w_-.
 \label{eq:third-order-covariant-sewing}
\end{equation}
The third-order bulk charges, defect terms and balance equations analogues of Eqs.~(\ref{quasi10})-(\ref{quasi20}) are
\begin{equation}
 \frac{d}{dt}\left(P^{(3)}_{\mbox{\tiny vm}}+P_D^{(3)}\right)=\mathcal A_1^{(3)},
 \qquad
 \frac{d}{dt}\left(E^{(3)}_{\mbox{\tiny vm}}+E_D^{(3)}\right)=\mathcal A_2^{(3)},
 \label{eq:third-order-PE-balance}
\end{equation}
where
\begin{equation}
 \mathcal A_1^{(3)}
 =-\frac{\Delta_{\epsilon}}{2}
 \left(R_-^{(3)}+R_+^{(3)}\right)_{x=0},
 \qquad
 \mathcal A_2^{(3)}
 =-\frac{\Delta_{\epsilon}}{2}
 \left(R_-^{(3)}-R_+^{(3)}\right)_{x=0}.
 \label{eq:third-order-anomalies}
\end{equation}
The bulk charge expressions  $P^{(3)}_{\mbox{\tiny vm}}$ and $E^{(3)}_{\mbox{\tiny vm}}$ are provided in (\ref{eq:vmSG-P3})-(\ref{eq:vmSG-E3}). The defect contributions $P_D^{(3)}$ and $E_D^{(3)}$ and the expressions for $R_-^{(3)}$ and $R_+^{(3)}$ are provided in the Appendix \ref{app:thirdorder}.
 
Thus, the same closure obstruction $\Delta_{\epsilon}=\alpha\beta-1$ that
 governs the first-order anomalies also controls the third-order balance
equations, since $\a \b =1 $ implies the vanishing of the corresponding anomalies without
requiring any additional parity condition.

For clarity, let us factorize the residuals at the defect as
\begin{equation}
 R_-^{(3)}(t)
 =\rho_1(0,t)\sin w_-(0,t)\,\mathcal C_-^{(3)}(t),
 \qquad
 R_+^{(3)}(t)
 =\rho_0(0,t)\sin w_+(0,t)\,\mathcal C_+^{(3)}(t),
 \label{eq:third-order-C-factorization}
\end{equation}
where the quantities $\mathcal C_\pm^{(3)}(t)$ are provided in (\ref{eq:C-third-minus})-(\ref{eq:C-third-plus}). They are local,
time-dependent functions at $x=0$. Their subscripts label the negative-grade
($D_0$-chiral) and positive-grade ($D_1$-chiral) sectors, respectively, and
do not by themselves specify time-reflection parity. The cubic trigonometric
terms originate from the cubic defect-potential pieces, the terms linear in
$u$ or $v$ come from mixed derivative--trigonometric contributions, and the
terms involving $u^2,u_1,v^2,v_1$ encode the higher-derivative content of the
third-order bulk charges.

Assume, as in Eq.~(\ref{eq:evenrho}), that $\rho_0(0,t)$ and $\rho_1(0,t)$ are even under $t\rightarrow -t$. For the centered {\bf soliton--soliton} branch, Eq.~(\ref{eq:solsym})
implies
\begin{equation}
 w_+(0,-t)=2\pi-w_+(0,t),
 \qquad
 w_-(0,-t)=w_-(0,t).
 \label{eq:third-order-parity-ss}
\end{equation}
A convenient sufficient realization of the required parity is
\begin{equation}
 u(-t)=-u(t),
 \qquad u_1(-t)=-u_1(t),
 \qquad v(-t)=v(t),
 \qquad v_1(-t)=v_1(t),
 \label{eq:third-order-derivative-parity-ss}
\end{equation}
all quantities being evaluated at $x=0$. Note that the soliton solutions (\ref{kink1}) do not meet these parity conditions separately in the left and right sectors.

Equations~\eqref{eq:C-third-minus}
--\eqref{eq:C-third-plus} then give
\begin{equation}
 \mathcal C_-^{(3)}(-t)=-\mathcal C_-^{(3)}(t),
 \qquad
 \mathcal C_+^{(3)}(-t)=+\mathcal C_+^{(3)}(t),
 \qquad (z>0),
 \label{eq:third-order-C-parity-ss}
\end{equation}
so both $R_-^{(3)}$ and $R_+^{(3)}$ are odd.

For the centered {\bf soliton--antisoliton} branch, Eq.~(\ref{eq:sasymmetry}) instead gives
\begin{equation}
 w_+(0,-t)=w_+(0,t),
 \qquad
 w_-(0,-t)=2\pi-w_-(0,t).
 \label{eq:third-order-parity-sa}
\end{equation}

In this branch, a corresponding sufficient realization is
\begin{equation}
 u(-t)=u(t),
 \qquad u_1(-t)=u_1(t),
 \qquad v(-t)=-v(t),
 \qquad v_1(-t)=-v_1(t).
 \label{eq:third-order-derivative-parity-sa}
\end{equation}
Note again that the soliton solutions (\ref{kink1}) do not satisfy these parity conditions separately in the left and right sectors.
So, equations~\eqref{eq:C-third-minus}
--\eqref{eq:C-third-plus} then give
\begin{equation}
 \mathcal C_-^{(3)}(-t)=+\mathcal C_-^{(3)}(t),
 \qquad
 \mathcal C_+^{(3)}(-t)=-\mathcal C_+^{(3)}(t),
 \qquad (z<0).
 \label{eq:third-order-C-parity-sa}
\end{equation}
Again, the complete products in~\eqref{eq:third-order-C-factorization} are odd. Under either set of branch-dependent parity conditions,
\begin{equation}
 \mathcal A_j^{(3)}(-t)=-\mathcal A_j^{(3)}(t),
 \qquad j=1,2,
 \label{eq:third-order-anomaly-parity}
\end{equation}
and therefore
\begin{equation}
 \lim_{T\to\infty}\int_{-T}^{T}\mathcal A_j^{(3)}(t)\,dt=0,
 \qquad j=1,2.
 \label{eq:third-order-integrated-anomaly}
\end{equation}
Hence, the modified third-order charges would be asymptotically conserved if these additional parity conditions were dynamically realized; however, they are not satisfied by the one-soliton configurations (\ref{kink1}) considered here.

The parity assignments (\ref{eq:third-order-derivative-parity-ss}) and (\ref{eq:third-order-derivative-parity-sa}) are only sufficient conditions for odd third-order anomalies and are not simultaneously realized by the one-soliton configurations (\ref{kink1}). Hence, unlike at first order, asymptotic cancellation cannot be inferred from soliton parity alone and must be assessed dynamically, as we examine numerically below.

\section{Numerical simulations}
\label{sec:background}

The numerical simulations in this section are restricted to type-I defects, including their integrable and deformed sewing conditions. Type-II defect dynamics are left for future work. Although the preceding analytical results determine the exact soliton--soliton and soliton--antisoliton transmission branches, a numerical treatment is required both to verify their dynamical consistency and to explore the non-integrable regime beyond the exact one-soliton ansatz. We therefore employ an explicit second-order finite-difference scheme adapted to the two-half-line geometry. The variable-mass sine-Gordon equation is evolved independently on $x<0$ and $x>0$, using centered spatial differences and a leapfrog time discretization, whereas the independent defect values $w_L(0,t)$ and $w_R(0,t)$ are updated directly from the discretized sewing conditions. No bulk stencil crosses $x=0$; consequently, all interaction between the two regions is mediated exclusively by the defect equations. For the closure-compatible case $ \alpha \beta  =1$, 
the two sewing relations define a common effective defect parameter and the simulation reproduces the analytically predicted purely transmitted branches, up to the expected discretization error. By contrast, when $\alpha \beta \neq 1$, the two deformed sewing equations cannot be simultaneously reduced to a single frozen Bäcklund transformation. The resulting incompatibility prevents an exactly phase-shifted one-soliton configuration from satisfying both matching conditions and dynamically excites additional degrees of freedom. Numerically, the incident soliton remains predominantly transmitted, but a small reflected component and an small residual disturbances are emitted from the defect region. The finite-difference evolution thus provides a direct means of measuring defect-induced reflection and radiation and of distinguishing these physical non-integrable effects from numerical artifacts through grid-refinement.

We consider the profiles
\begin{align}
 \rho_0(\xi)
 &=
 1+\epsilon_0\sech^2\!\left[\kappa_0(\xi-\xi_0)\right],
 \label{eq:rho0}
 \\
 \rho_1(\eta)
 &=
 1+\epsilon_1\sech^2\!\left[\kappa_1(\eta-\eta_0)\right],
 \label{eq:rho1}
\end{align}
with
\begin{equation}
 \epsilon_0>-1,
 \qquad
 \epsilon_1>-1,
 \qquad
 \kappa_0>0,
 \qquad
 \kappa_1>0.
 \label{eq:background-conditions}
\end{equation}
The restrictions in \eqref{eq:background-conditions} ensure that
$\rho_0$ and $\rho_1$ remain strictly positive. The profiles (\ref{eq:rho0})–(\ref{eq:rho1}) are smooth localized representatives of the factorized backgrounds required by the integrable vmSG model. They are intended as an idealized testbed, with $\rho_0$ and $\rho_{1}$ representing localized spatiotemporal modulations of the medium. The $\sech^2$ form is convenient because it approaches the homogeneous background asymptotically while introducing a controlled localized inhomogeneity.

The bulk fields are
\begin{equation}
 w_L(x,t),\qquad x<0,
 \qquad\qquad
 w_R(x,t),\qquad x>0,
\end{equation}
with independent limiting values at the defect, $
 w_L(0^-,t)\neq w_R(0^+,t)$ in general.

Let us discuss the continuous bulk and defect equations. The variable-mass sine-Gordon equation on each half-line is
\begin{equation}
 \partial_t^2w_s-\partial_x^2w_s
 +2m_0^2
 \rho_0\!\left(\frac{t+x}{2}\right)
 \rho_1\!\left(\frac{t-x}{2}\right)
 \sin(2w_s)=0,
 \qquad s=L,R.
 \label{eq:bulk-continuous}
\end{equation}

For the one-soliton transmission sector, impose the closure relation
\begin{equation}
 \alpha\beta=1,
 \qquad
 \alpha=1+\epsilon_A,
 \qquad
 \beta=1+\epsilon_B,
 \label{eq:alpha-beta}
\end{equation}
and define the effective B\"acklund parameter
\begin{equation}
 \sigma_{\mathrm{eff}}
 =\sigma \alpha
 =\frac{\sigma }{\beta}.
 \label{eq:sigmaeff}
\end{equation}
The defect functions are then
\begin{align}
 A(t)
 &=-2m_0\sigma_{\mathrm{eff}}
 \rho_0\!\left(\frac{t}{2}\right)
 \sin\!\left[w_L(0,t)+w_R(0,t)\right],
 \label{eq:A-continuous}
 \\
 B(t)
 &=-\frac{2m_0}{\sigma_{\mathrm{eff}}}
 \rho_1\!\left(\frac{t}{2}\right)
 \sin\!\left[w_L(0,t)-w_R(0,t)\right].
 \label{eq:B-continuous}
\end{align}
At $x=0$, the type-I sewing conditions are
\begin{align}
 \left.\partial_xw_L\right|_{0^-}
 &=\left.\partial_tw_R\right|_{0^+}
 +\frac{A+B}{2},
 \label{eq:sewing1}
 \\
 \left.\partial_xw_R\right|_{0^+}
 &=\left.\partial_tw_L\right|_{0^-}
 +\frac{B-A}{2}.
 \label{eq:sewing2}
\end{align}
In Appendix \ref{app:defectrajetory}, we show that a coordinate transformation removes the explicit background factors from the full B\"acklund equations. For a defect fixed at $x=0$, the frozen B\"acklund equations must instead be imposed in the original $(x,t)$ coordinates, where these factors remain explicit as $\rho_0\!\left(\frac{t}{2}\right)$ and $\rho_1\!\left(\frac{t}{2}\right)$.

Next, we discuss the exact localized soliton phases. Normalize the primitives by $X(0)=Y(0)=0$. One has the backgrounds
\eqref{eq:rho0}--\eqref{eq:rho1},
\begin{align}
 X(\xi)
 ={}&\xi+\frac{\epsilon_0}{\kappa_0}
 \left\{
 \tanh[\kappa_0(\xi-\xi_0)]
 +\tanh(\kappa_0\xi_0)
 \right\},
 \label{eq:X}
 \\
 Y(\eta)
 ={}&\eta+\frac{\epsilon_1}{\kappa_1}
 \left\{
 \tanh[\kappa_1(\eta-\eta_0)]
 +\tanh(\kappa_1\eta_0)
 \right\}.
 \label{eq:Y}
\end{align}
Next, introduce the common phase
\begin{equation}
 \Phi(x,t)
 =2m_0\left[
 pX\!\left(\frac{t+x}{2}\right)
 -\frac{1}{p}Y\!\left(\frac{t-x}{2}\right)
 \right].
 \label{eq:Phi}
\end{equation}
The transmission factor is
\begin{equation}
 z=\frac{p+\sigma_{\mathrm{eff}}}
         {p-\sigma_{\mathrm{eff}}},
 \qquad
 p\neq\pm\sigma_{\mathrm{eff}}.
 \label{eq:z}
\end{equation}
It is useful to define
\begin{equation}
 K[u]=2\arctan(e^u).
 \label{eq:K}
\end{equation}

For the {\bf soliton--soliton} branch one has $ z>0$, then the exact profiles are
\begin{align}
 w_L^{\mathrm{ex}}(x,t)
 &=K\!\left[\Phi(x,t)+\delta_L\right],
 \label{eq:exact-left-soliton}
 \\
 w_R^{\mathrm{ex}}(x,t)
 &=K\!\left[\Phi(x,t)+\delta_R\right],
 \qquad
 \delta_R-\delta_L=\ln z.
 \label{eq:exact-right-soliton}
\end{align}
The defect preserves the topological orientation and produces a phase,
or equivalently a position, shift.

For the {\bf soliton-antisoliton} branch one has $ z<0$, let us write $z=-|z|$. A real representative of the transmitted exact antisoliton is
\begin{align}
 w_L^{\mathrm{ex}}(x,t)
 &=K\!\left[\Phi(x,t)+\delta_L\right],
 \label{eq:exact-left-conversion}
 \\
 w_R^{\mathrm{ex}}(x,t)
 &=-K\!\left[\Phi(x,t)+\delta_R\right],
 \qquad
 \delta_R-\delta_L=\ln|z|.
 \label{eq:exact-right-antisoliton}
\end{align}
Equivalently,
\begin{equation}
 w_R^{\mathrm{ex}}(x,t)
 =2\arctan\!\left[
 z\,e^{\Phi(x,t)+\delta_L}
 \right].
 \label{eq:exact-z-unified}
\end{equation}
The right-hand field carries the opposite topological orientation. The
spectral parameter $p$, and therefore the asymptotic propagation speed,
is unchanged.

Both branches can be represented compactly as
\br
 w_L^{\mathrm{ex}}(x,t)
 &=& K\!\left[\Phi(x,t)+\delta_L\right],
 \label{eq:unified-left}
 \\
 w_R^{\mathrm{ex}}(x,t)
 &=&\sgn(z) K\!\left[\Phi(x,t)+\delta_R\right],
 \qquad
 \delta_R=\delta_L+\ln|z|.
 \label{eq:unified-right}
\er

A phase choice centered with respect to the defect is
\begin{equation}
 \delta_L=-\frac{1}{2}\ln|z|,
 \qquad
 \delta_R=+\frac{1}{2}\ln|z|.
 \label{eq:centered-phases}
\end{equation}

Fig. 1 shows kink–kink transmission through the defect. The incoming kink propagates along the left half-line, reaches the defect at $x=0$, and subsequently continues along the right half-line. Top figure: Integrable sewing conditions satisfying $\alpha\beta=1$. The kink crosses the defect as an essentially pure transmitted excitation, following the analytically predicted trajectory without reflection or radiation. Bottom figure: Non-integrable deformed sewing conditions with $\alpha\beta \neq 1$. The evolution is still dominated by a transmitted kink, but the mismatch between the two sewing relations produces a weak reflected component and small-amplitude radiation emitted from the defect region during the crossing.

Fig. 2 shows soliton–antisoliton transmission through the defect. The incoming positive-orientation soliton propagates along the left half-line, crosses the defect at $x=0$, and emerges on the right half-line as a negative-orientation antisoliton. Top figure: Integrable sewing conditions satisfying $\alpha\beta=1$. The defect produces a clean topological conversion, and the transmitted antisoliton follows the analytically predicted trajectory without radiation. Bottom figure: Non-integrable deformed sewing conditions with $\alpha\beta \neq 1$. The evolution remains dominated by the transmitted antisoliton, but the two matching equations impose incompatible phase shifts, generating small oscillatory components and weak radiation localized near the defect during the crossing.

The kink--kink transmission processes through the defect at $x=0$ are illustrated in Fig.~1 and investigated numerically in Figs.~3--11 and 13--14. Figure~1 provides an overview of the transmission dynamics in the integrable $\alpha\beta=1$ and non-integrable $\alpha\beta\neq 1$ regimes. Figures~3--11 examine the numerical evolution, convergence properties, post-interaction residuals, and lowest-order defect anomalies. Figures~13--14 focus on the third-order momentum- and energy-type anomalies and their cumulative time integrals. In the integrable regime, the transmitted kink preserves its topological orientation and soliton parameter, undergoing only a phase shift. For $\alpha\beta\neq1$, transmission remains dominant, although small residual disturbances may arise. The lowest-order integrated anomalies vanish within numerical accuracy for the scattering configurations considered, whereas the third-order integrated anomalies are generally nonzero. These results illustrate the hierarchy-dependent character of asymptotic quasi-conservation in the non-integrable defect theory..

The kink--antikink transmission processes through the defect at $x=0$ are illustrated in Figs.~2, 12, and 15--16. Figure~2 presents the field evolution in both the integrable $\alpha\beta=1$ and non-integrable $\alpha\beta\neq 1$ regimes. In the integrable case, the incoming kink is converted into a transmitted antikink, preserving its soliton parameter and asymptotic velocity while undergoing a phase shift. The associated change in bulk topological charge is compensated by the defect. In the non-integrable case, kink--antikink conversion remains dominant, with small residual disturbances. Figure~12 examines the lowest-order momentum- and energy-type anomalies, whose cumulative time integrals vanish within numerical accuracy for the configurations considered. Figures~15--16 investigate the corresponding third-order momentum- and energy-type anomalies and their cumulative integrals, including mesh-refinement comparisons. The nonvanishing integrated third-order anomalies indicate that the asymptotic quasi-conservation observed at the lowest order does not generally extend to higher-order charges. These results demonstrate the hierarchy-dependent character of quasi-conservation in the deformed defect model. 
 
Fig. 3 shows space--time evolution of the numerical kink $w(x,t)$
transmitted through the integrable defect at $x=0$, with
$\epsilon_A=0.25$, $\epsilon_B=-0.20$, $\alpha\beta=1$,
$\sigma=0.16$, and $p_{\rm in}=0.5$.
The kink connects $w=0$ to $w=\pi$, undergoing a phase shift
while preserving its soliton parameter and asymptotic velocity
$v=0.6$. Surface height and color represent $w$, with the
two half-lines plotted separately to preserve independent
defect traces.

Fig. 4 shows the convergence analysis of the numerical scheme described
in Table~2, evaluated at $t=-10$ with $\Delta t=0.2\,\Delta x$.
The maximum error $\|E\|_{\infty}$ and the root-mean-square
(RMS) error $\|E\|_2$, where
$E_j=w_{\mathrm{num}}(x_j,t)-w_{\mathrm{exact}}(x_j,t)$,
are computed over both half-lines and plotted against the
spatial resolution $\Delta x$. The error norms are defined by
\[
\|E\|_{\infty}=\max_{1\leq j\leq N}|E_j|,
\qquad
\|E\|_2=
\left(\frac{1}{N}\sum_{j=1}^{N}|E_j|^2\right)^{1/2},
\]
where $N$ denotes the total number of sampled grid points
on both half-lines. The dotted reference line represents
the expected second-order scaling,
$\|E\|\propto(\Delta x)^2$, and nearly overlaps the
$\|E\|_{\infty}$ curve. Both error norms decrease
quadratically under mesh refinement, confirming the
second-order accuracy of the numerical scheme in this
benchmark.

\begin{figure}
\centering
\label{fig1}
\includegraphics[width=1.5cm,scale=2, angle=0,height=6cm]{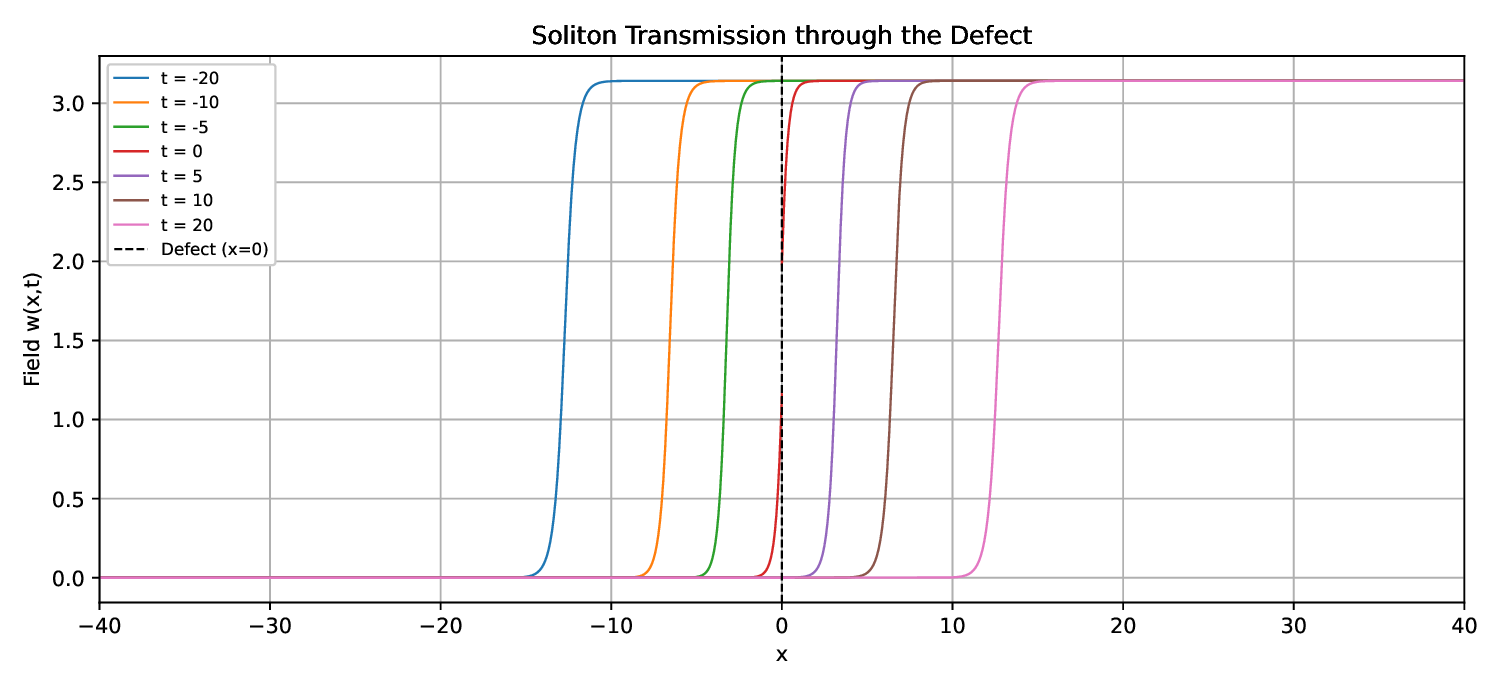}
\includegraphics[width=1.5cm,scale=2, angle=0,height=6cm]{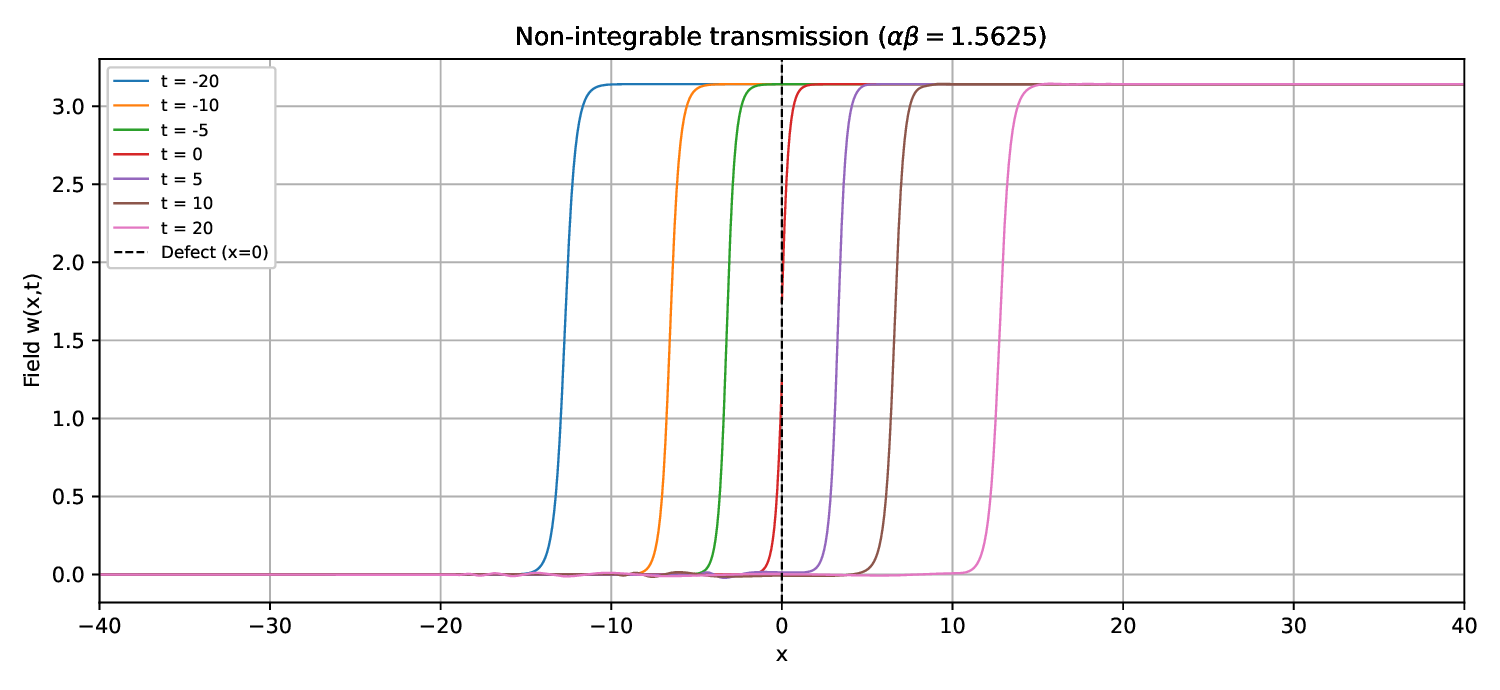}  
\parbox{6in}{\caption{(color online) The kink-kink transmission for $m_0 =1; \epsilon_0 = 0.3; \epsilon_1= 0.5; \kappa_0 = 0.5; \kappa_1=0.7, \s=0.16$.  The kink propagates from the left half-line, crosses the defect at $x=0$, and continues on the right. Top Fig.: integrable case with $\epsilon_A = 0.25, \epsilon_B = -0.20,\,\a \b =1, p_{in}=0.5, \xi_0=\eta_0=0$ with asymptotic velocity $v = (1-p_{in}^2)/(1+p_{in}^2)=0.6$. Bottom: Numerical kink transmission through the non-integrable defect at $x=0$, with $\epsilon_A=\epsilon_B=0.25$,
 $\alpha=\beta=1.25$, and $\alpha\beta=1.5625$, $p_{in}=0.5$. The profiles are shown at the times
 indicated in the legend, using $\Delta x=0.0025$ and
 $\Delta t=0.0005$. The kink propagates from the left to
 the right half-line, leaving small oscillatory disturbances.
 The dashed vertical line marks the defect; the two field
 traces at $x=0^\pm$ are evolved independently.}}
\end{figure}

\begin{figure}
\centering
\label{fig2}
\includegraphics[width=1.5cm,scale=2, angle=0,height=7.5cm]{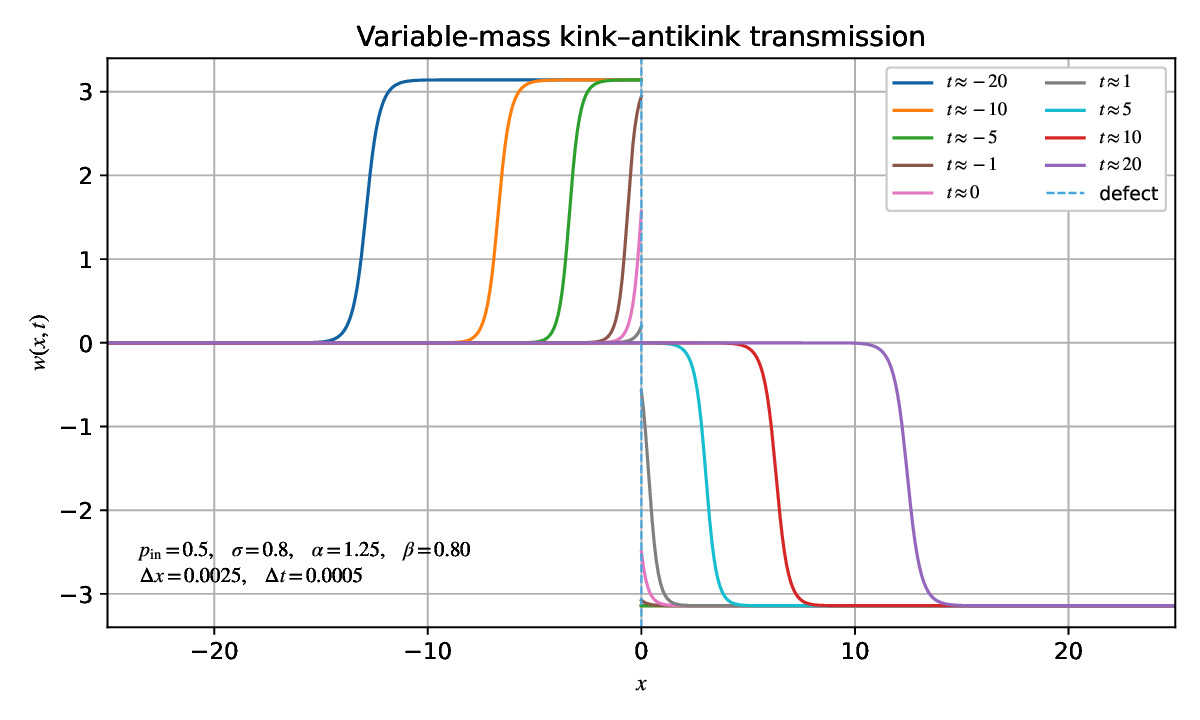}
\includegraphics[width=1.5cm,scale=2, angle=0,height=7.5cm]{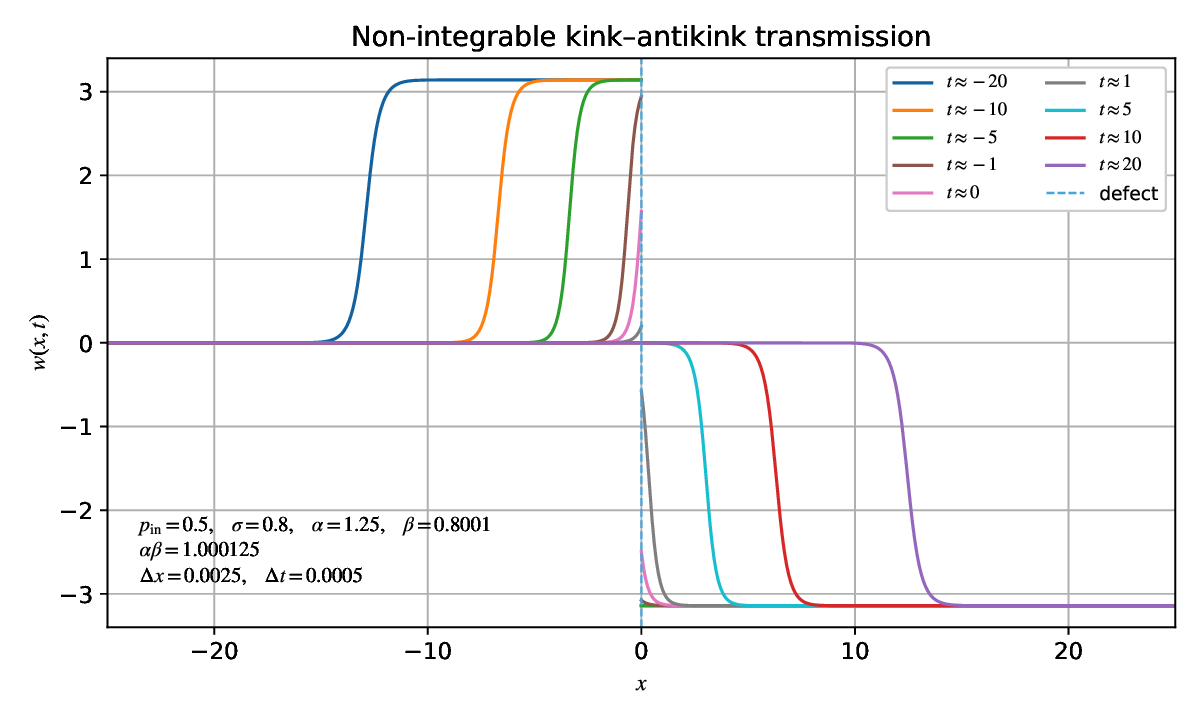}  
\parbox{6in}{\caption{(color online) Top: Numerical kink–antikink transmission in the integrable case, with $m_0=1$, $p_{\mathrm{in}}=0.5$, $\sigma=0.8$, $\alpha=1.25$, and $\beta=0.80$, satisfying $\alpha\beta=1$. Profiles show the incoming kink, its interaction with the defect, and the outgoing antikink $t\approx5,10,20$. The dashed line marks $x=0$. The simulation uses $\Delta x=0.0025$ and $\Delta t=0.0005$. Bottom: Numerical kink–antikink transmission for $m_0=1$, $p_{\mathrm{in}}=0.5$, $\sigma=0.8$, $\alpha=1.25$, and $\beta=0.8001$, with $\alpha\beta=1.000125\neq1$. Profiles show the incoming kink, its interaction with the defect, and the outgoing antikink at $t\approx5,10,20$. The dashed line marks $x=0$. The simulation uses $\Delta x=0.0025$ and $\Delta t=0.0005$.}}
\end{figure}

\begin{figure}
\centering
\label{fig3}
\includegraphics[width=1.5cm,scale=2, angle=0,height=7cm]{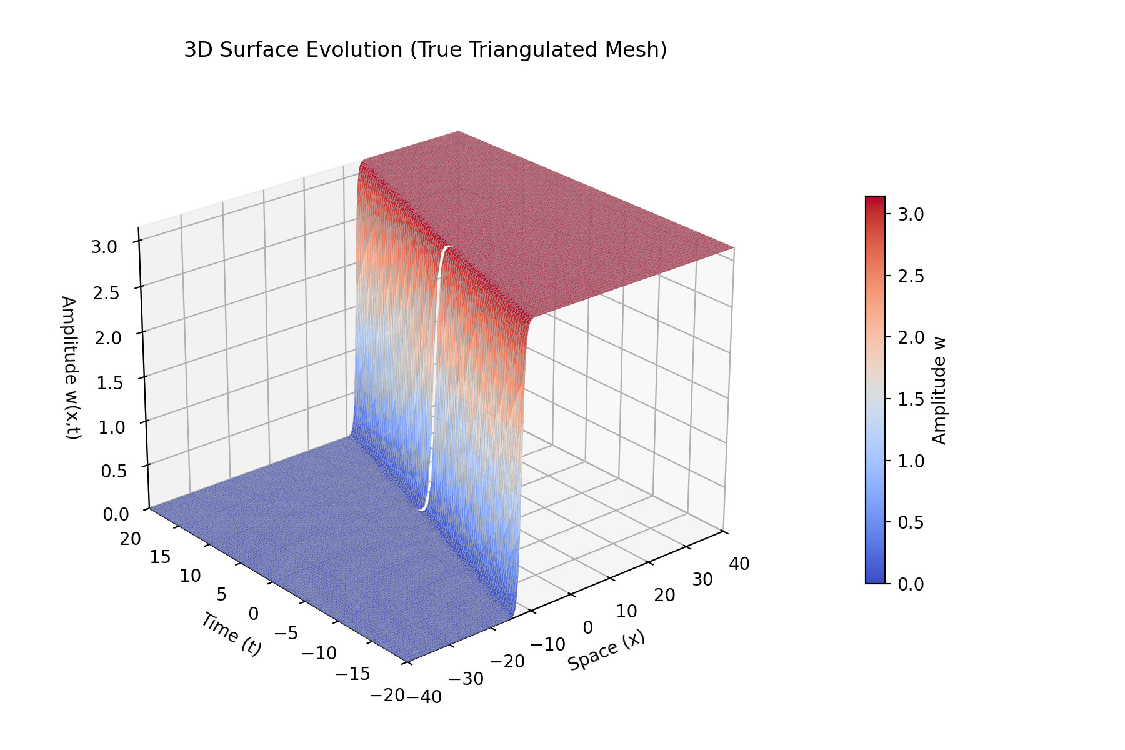}
\parbox{6in}{\caption{(color online) Space--time evolution of the numerical field $w(x,t)$
during kink transmission through the integrable defect at
$x=0$, with $\epsilon_A=0.25$, $\epsilon_B=-0.20$,
$\alpha\beta=1$, $\sigma=0.16$, and $p_{in}=0.5$.
The incoming kink propagates from the left half-line to the
right half-line, connecting the vacua $w=0$ and $w=\pi$.
The defect produces a phase shift while preserving the
soliton parameter and its asymptotic velocity $v=0.6$.
The surface height and color both represent the field value.
The two half-line surfaces are plotted separately to retain
the independent field traces at the defect.}}
\end{figure}

\begin{figure}
\centering
\label{fig4}
\includegraphics[width=1.5cm,scale=2, angle=0,height=7cm]{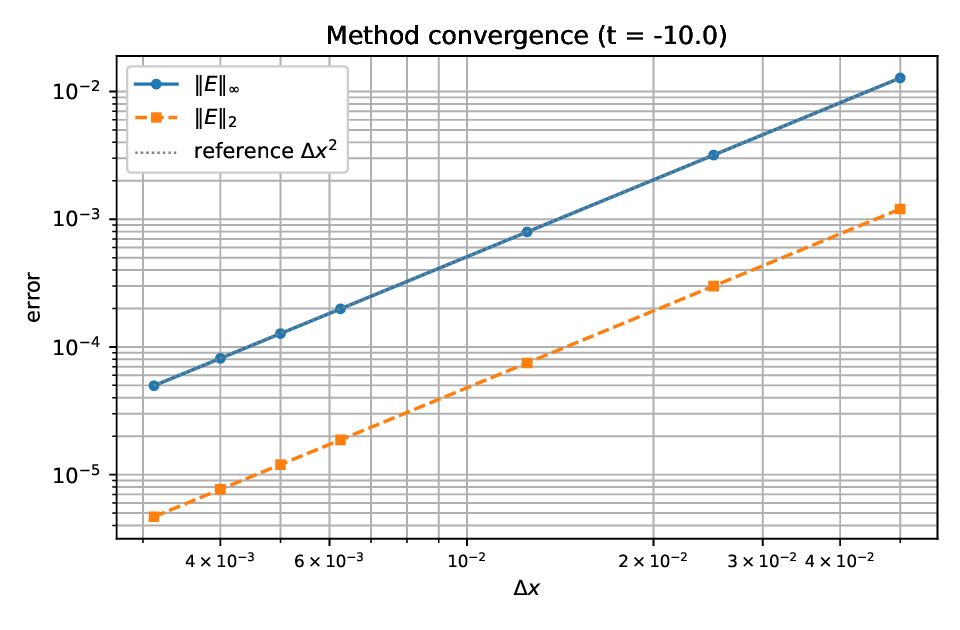}
\parbox{6in}{\caption{(color online) Convergence of the numerical scheme described in Table 2 at $t=-10$, with $\Delta t=0.2\,\Delta x$.
  The maximum error $\|E\|_{\infty}$ and root-mean-square error
  $\|E\|_2$, where $E=w^{\mathrm{num}}-w^{\mathrm{exact}}$,
  are evaluated over both half-lines and plotted against the spatial
  step $\Delta x$. The dotted reference line is proportional to
  $(\Delta x)^2$ (it almost exactly overlaps the blue $\|E\|_\infty$ curve). Both error measures exhibit second-order convergence.
  Here, $\|E\|_2$ denotes the normalized discrete norm,
  $\bigl(N^{-1}\sum_{j=1}^{N}|E_j|^2\bigr)^{1/2}$, consistent with
  the RMS calculation used for this figure; $N$ is the total number
  of sampled values over both half-lines.}}
\end{figure}

\begin{figure}
\centering
\label{fig5}
\includegraphics[width=1.5cm,scale=2, angle=0,height=5.5cm]{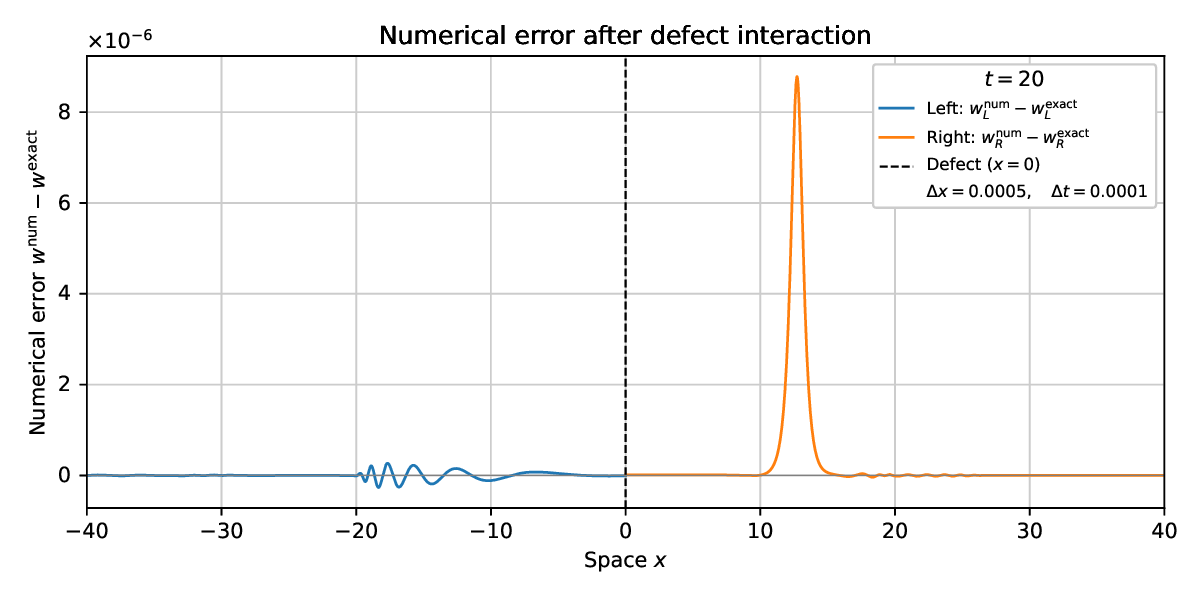}
\includegraphics[width=1.5cm,scale=2, angle=0,height=5.5cm]{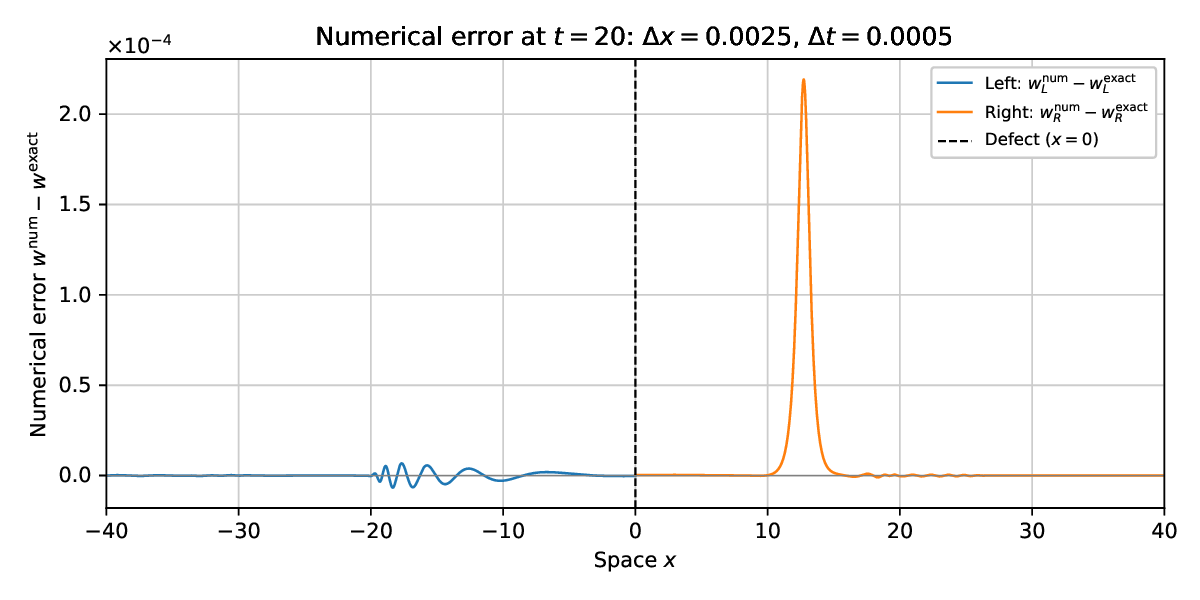}
\includegraphics[width=1.5cm,scale=2, angle=0,height=5.5cm]{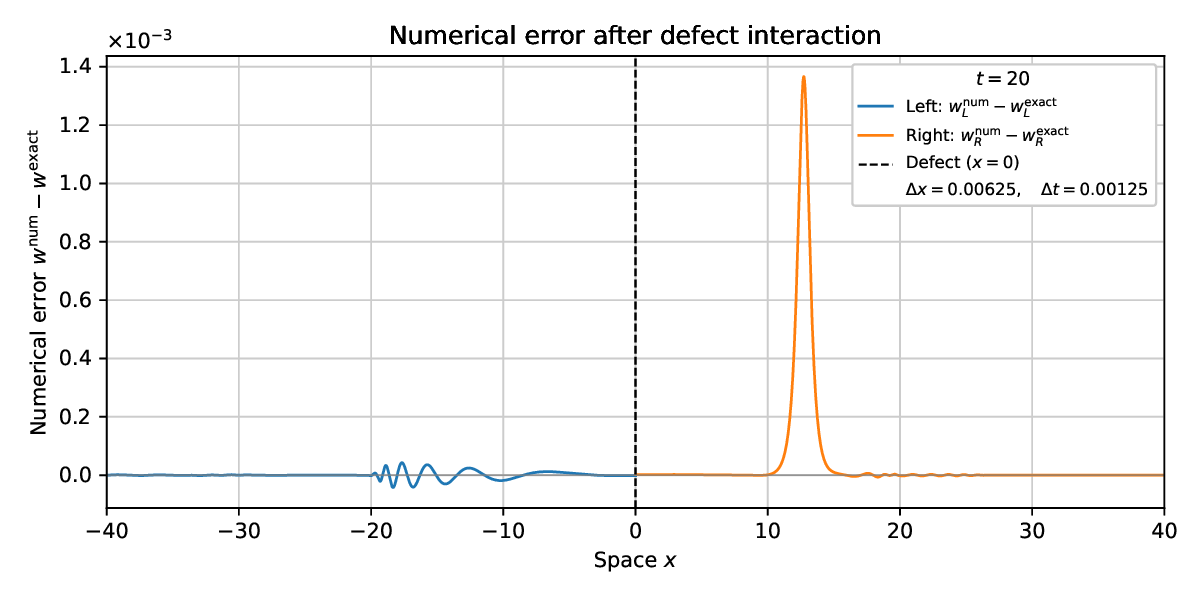}
\parbox{6in}{\caption{(color online)  Numerical error $w^{\mathrm{num}}-w^{\mathrm{exact}}$
at $t=20$ after soliton transmission through the integrable
defect, with $\epsilon_A=0.25$, $\epsilon_B=-0.20$, and
$\alpha\beta=1$. Panels (a), (b), and (c) correspond to
$\Delta x=0.0005$, $0.0025$, and $0.00625$, respectively,
with $\Delta t=0.2\,\Delta x$ in each case.
The blue and orange curves show the errors on the left and
right half-lines, respectively; the dashed vertical line
marks the defect at $x=0$. The reference field is the exact
transmitted one-soliton solution, without refitting its
parameters. The maximum absolute errors are approximately
$8.78\times10^{-6}$, $2.19\times10^{-4}$, and
$1.37\times10^{-3}$, respectively, consistent with
second-order error reduction under mesh refinement.
These residuals measure numerical error in the integrable
benchmark.}}
\end{figure}
\[\]

\begin{minipage}{15cm}
\centering
\begin{tabular}{rrrrrr}
\toprule
$\Delta x$ & $\Delta t$ & $\|E\|_\infty$ & $\|E\|_2$ & $q_\infty$ & $q_2$ \\
\midrule
0.0500000 & 0.0100000 & $1.2818\times10^{-2}$ & $1.2070\times10^{-3}$ & --- & --- \\
0.0250000 & 0.0050000 & $3.1997\times10^{-3}$ & $3.0084\times10^{-4}$ & 2.0022 & 2.0044 \\
0.0125000 & 0.0025000 & $7.9915\times10^{-4}$ & $7.5163\times10^{-5}$ & 2.0014 & 2.0009 \\
0.0062500 & 0.0012500 & $1.9974\times10^{-4}$ & $1.8789\times10^{-5}$ & 2.0003 & 2.0001 \\
0.0050000 & 0.0010000 & $1.2783\times10^{-4}$ & $1.2025\times10^{-5}$ & 2.0001 & 2.0000 \\
0.0041667 & 0.0008333 & $8.8772\times10^{-5}$ & $8.3507\times10^{-6}$ & 2.0001 & 2.0000 \\
0.0035714 & 0.0007143 & $6.5220\times10^{-5}$ & $6.1353\times10^{-6}$ & 2.0000 & 2.0000 \\
\bottomrule
\end{tabular}
\captionof{table}{Convergence results in the integrable limit
$\epsilon_A=\epsilon_B=0$ at $t=-10$, with $\Delta t/\Delta x=0.2$. Test validates before the soliton--defect interaction. 
The refinement factors are $r=1,2,4,8,10,12,14$, with $\Delta x=0.05/r$.
Both error measures exhibit second-order convergence.}
\end{minipage}\\

Table 2 demonstrates second-order
convergence in the integrable limit $\epsilon_A=\epsilon_B=0$
at $t=-10$, with $\Delta t/\Delta x=0.2$ held fixed.
Both the maximum error $\|E\|_\infty$ and the normalized
discrete RMS error $\|E\|_2$ decrease approximately as
$(\Delta x)^2$, with empirical orders
$q_\infty\simeq q_2\simeq2$. In particular, halving
$\Delta x$ reduces both errors by approximately a factor
of four. This test validates the expected accuracy before
the soliton--defect interaction; the accuracy of the
post-interaction disturbances is assessed separately
through mesh refinement.

\begin{figure}
\centering
\label{fig6}
\includegraphics[width=1.5cm,scale=2, angle=0,height=5.5cm]{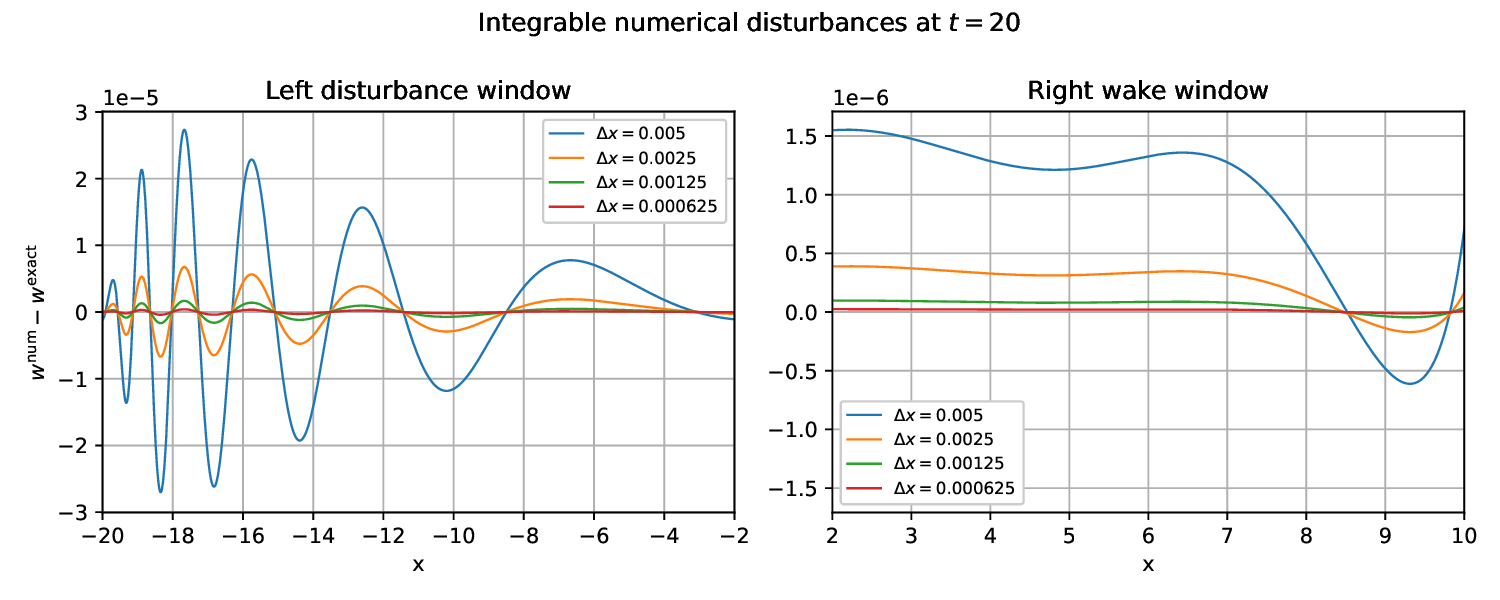}
\includegraphics[width=1.5cm,scale=2, angle=0,height=5cm]{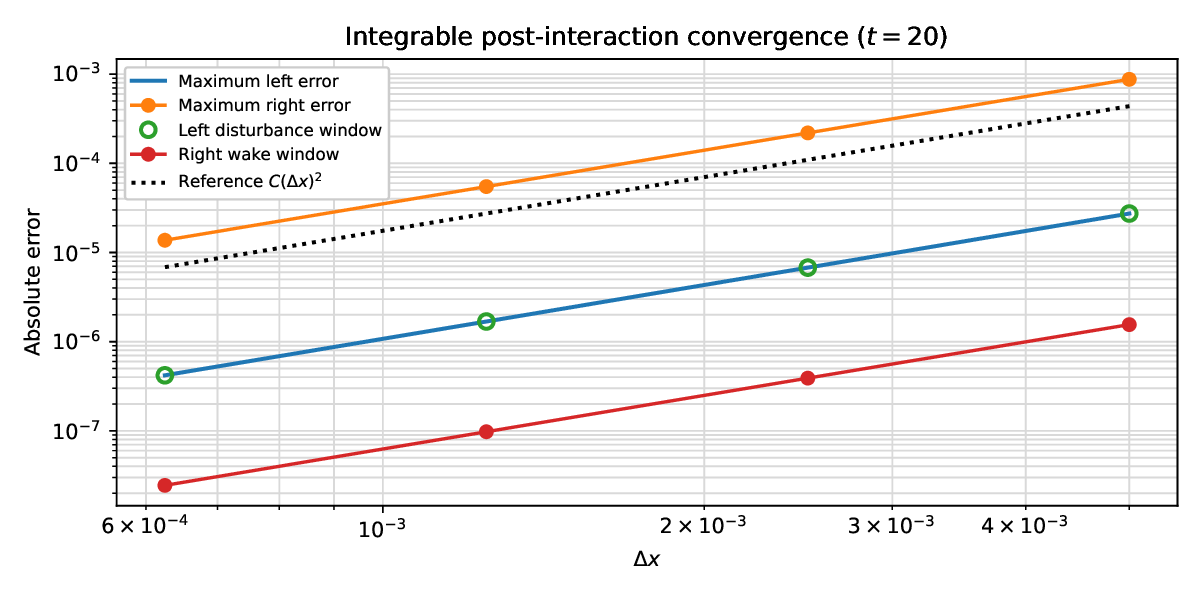}  
\parbox{6in}{\caption{(color online) Top panels: Post-interaction numerical disturbances
$w^{\mathrm{num}}-w^{\mathrm{exact}}$ in the integrable
benchmark at $t=20$. The left and right panels show the
fixed windows $[-20,-2]$ and $[2,10]$, respectively,
excluding the defect and transmitted kink core.
The four curves correspond to
$\Delta x=0.005$, $0.0025$, $0.00125$, and $0.000625$,
with $\Delta t=0.2\,\Delta x$.
The disturbance amplitudes decrease approximately
quadratically under mesh refinement, consistent with
discretization artifacts rather than physical radiation.
The two panels use different vertical scales. Bottom : Post-interaction error convergence in the integrable
benchmark at $t=20$, with $\Delta t=0.2\,\Delta x$.
The blue and orange curves show the maximum absolute errors
$|w^{\mathrm{num}}-w^{\mathrm{exact}}|$ over the left and right
half-lines, respectively. Green markers and the red curve
show the corresponding maxima in $[-20,-2]$ and $[2,10]$.
The green markers coincide with the blue curve.
The dotted reference line indicates second-order scaling,
$(\Delta x)^2$. These values quantify numerical discrepancies,
including phase and shape errors; they are not physical
radiation amplitudes.}}
\end{figure}

Figure~5 presents the numerical error
$E(x,t)=w_{\mathrm{num}}(x,t)-w_{\mathrm{exact}}(x,t)$
at $t=20$, following kink transmission through the
integrable defect ($\epsilon_A=0.25$, $\epsilon_B=-0.20$,
$\alpha\beta=1$). Panels (a)--(c) correspond to
$\Delta x=0.0005$, $0.0025$, and $0.00625$, respectively,
with $\Delta t=0.2\,\Delta x$. The blue and orange curves
represent the errors on the left and right half-lines,
while the dashed vertical line indicates the defect at $x=0$.
The numerical fields are compared directly with the exact
transmitted one-soliton solution, without parameter refitting.
The corresponding maximum absolute errors are
$8.78\times10^{-6}$, $2.19\times10^{-4}$, and
$1.37\times10^{-3}$, respectively, confirming the expected
second-order convergence under mesh refinement.
These residuals quantify numerical discretization errors
in the integrable benchmark and should not be interpreted
as physical radiation.

Figure~6 shows the post-interaction numerical errors in the integrable benchmark at $t=20$, with $\Delta t=0.2\,\Delta x$.
Top: Residual fields $w^{\rm num}-w^{\rm exact}$
in the left and right windows $[-20,-2]$ and $[2,10]$,
respectively, for $\Delta x=0.005$, $0.0025$, $0.00125$,
and $0.000625$. The panels use different vertical scales.
Bottom: Maximum absolute errors over both half-lines
(blue and orange) and the corresponding restricted windows
(green markers and red curve). The green markers overlap
the blue curve. The dotted line indicates the expected
second-order scaling, $(\Delta x)^2$.
The approximately quadratic error reduction confirms
that the residuals, including phase and shape discrepancies,
are numerical artifacts rather than physical radiation.
 
\begin{figure}
\centering
\label{fig7}
\includegraphics[width=1.5cm,scale=2, angle=0,height=4cm]{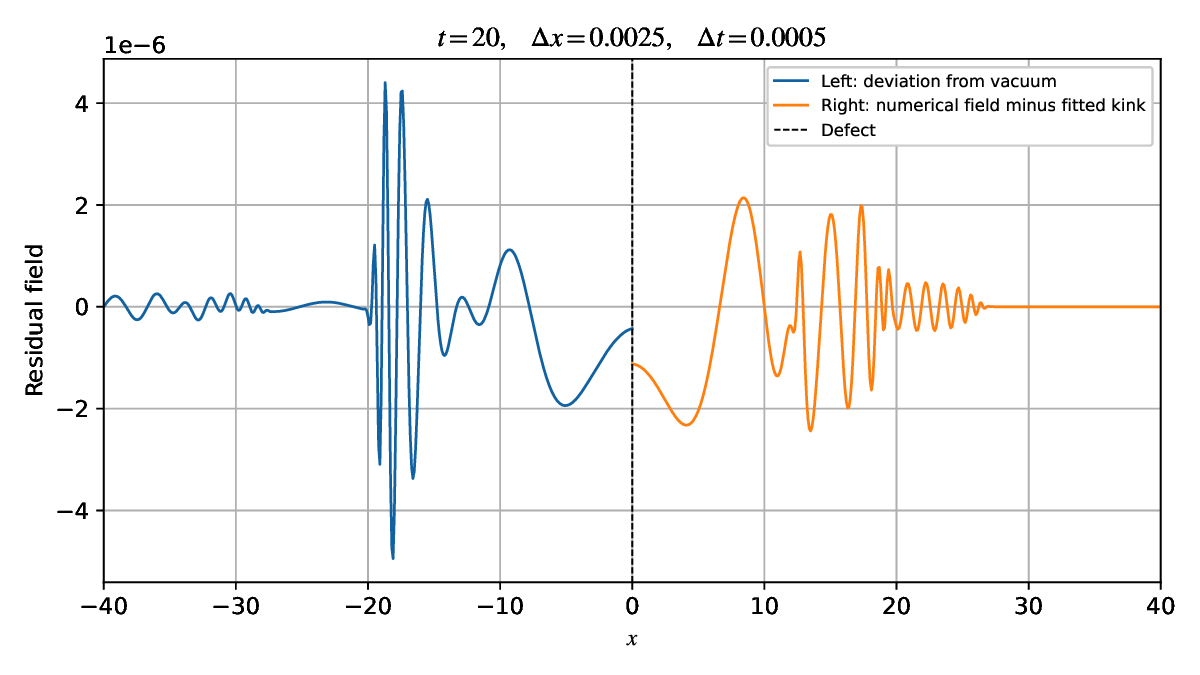}
\includegraphics[width=1.5cm,scale=2, angle=0,height=4cm]{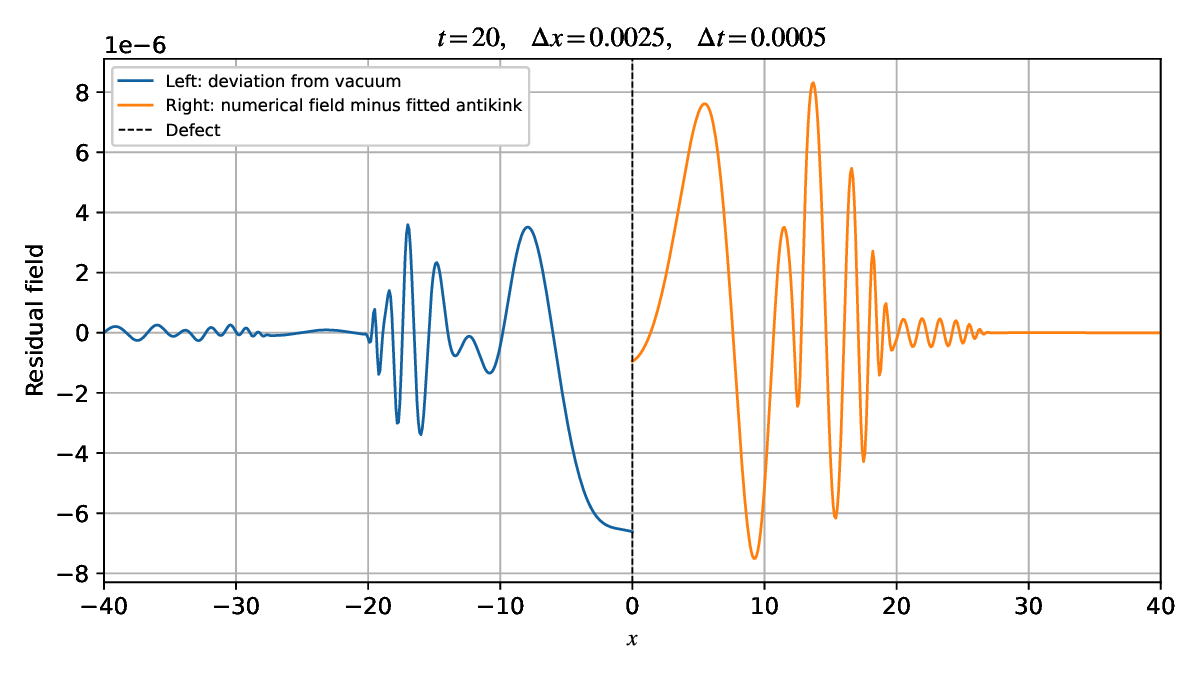}  
\parbox{6in}{\caption{(color online) Left: Field residuals at $t=20$ following non-integrable kink--kink
transmission, with $\alpha=1.25$, $\beta=0.8001$, $\sigma=0.16$, $p_{\mathrm{in}}=0.5$ ($z=7/3>0$), and $\alpha \beta \neq 1$. The curves show the deviation
from the vacuum on the left and from the fitted transmitted kink on
the right; the defect at $x=0$. Computed with
$\Delta x=0.0025$ and $\Delta t=0.0005$, the residuals are of order
$10^{-6}$ and include numerical and fitting errors. Right: Post-transmission field residuals at $t=20$ for the non-integrable case $\alpha=1.25$, $\beta=0.8001$, $\sigma=0.8$, and $p_{\mathrm{in}}=0.5$ for kink-antikink. Blue: left-field deviation from vacuum. Orange: right-field deviation from the antikink fitted over $x\in [11.8,13.1]$. The dashed line marks the defect. The simulation uses $\Delta x=0.0025$ and $\Delta t=0.0005$; residuals include numerical errors and are not established physical radiation amplitudes.}}
\end{figure}

Figure 7 shows the post-transmission field residuals at $t=20$ for
non-integrable soliton--defect interactions with
$\alpha=1.25$, $\beta=0.8001$, $p_{\rm in}=0.5$,
$\Delta x=0.0025$, and $\Delta t=0.0005$.
Left: Kink--kink transmission ($\sigma=0.16$, $z=7/3>0$).
Right: Kink--antikink transmission ($\sigma=0.8$),
with the outgoing antikink fitted over $x\in[11.8,13.1]$.
Blue and orange curves represent deviations from the
left vacuum and fitted right soliton, respectively.
The dashed line marks the defect at $x=0$.
The residuals include numerical and fitting errors
and do not establish physical radiation amplitudes.

Figure 8 shows the momentum- and energy-type anomalies, Eqs.~(4.30)
and (4.33), for the non-integrable kink--kink transmission
shown in Fig.~1 (bottom panel).
Left: Momentum-type anomaly.
Right: Energy-type anomaly.
The insets display the corresponding time integrals,
which vanish within numerical accuracy, supporting
asymptotic quasi-conservation of the lowest-order charges.

Figure 9 shows time-integrated anomalies versus $\alpha\beta$
for non-integrable kink--kink transmission, with
$\alpha=1.25$, $p_{\rm in}=0.5$, and $\sigma=0.16$.
Left: Momentum-type anomaly
$\int_{-20}^{20}\mathcal{A}_1^{(1)}(t)\,dt$.
Right: Energy-type anomaly
$\int_{-20}^{20}\mathcal{A}_2^{(1)}(t)\,dt$.
Results at two spatial resolutions, with
$\Delta t=0.2\,\Delta x$, exhibit approximately quadratic
suppression near the integrable limit $\alpha\beta=1$,
where both anomalies vanish identically.

Figure 10 shows momentum- and energy-type bulk-plus-defect charges
during non-integrable soliton transmission.
Left: Momentum-type charge.
Right: Energy-type charge.
The total quantities $Q=Q_{\rm vmSG}+Q_{D,I}$,
defined in Eqs.~(3.41) and (3.75), recover their
initial values within a numerical accuracy of $10^{-9}$,
supporting asymptotic quasi-conservation.
The parameters are indicated in the panels.

Figure 11 shows the momentum- and energy-type balance-law residuals
for non-integrable kink--kink transmission with
$\alpha=1.25$, $\beta=0.8001$, $p_{\rm in}=0.5$,
and $\sigma=0.16$.
Left: Momentum residual
$R_P(t)=d(P_{\rm vmSG}^{(1)}+P_D^{(1)})/dt
-\mathcal{A}_1^{(1)}(t)$.
Right: Energy residual
$R_E(t)=d(E_{\rm vmSG}^{(1)}+E_D^{(1)})/dt
-\mathcal{A}_2^{(1)}(t)$.
For $\Delta x=0.0025$ and $\Delta t=0.0005$,
the respective maximum residuals are
$4.27\times10^{-4}$ and $3.75\times10^{-4}$,
while their signed time integrals are
$2.008\times10^{-9}$ and $2.45\times10^{-10}$,
consistent with numerical discretization errors.

\begin{figure}
\centering
\label{fig8}
\includegraphics[width=1.5cm,scale=2, angle=0,height=4cm]{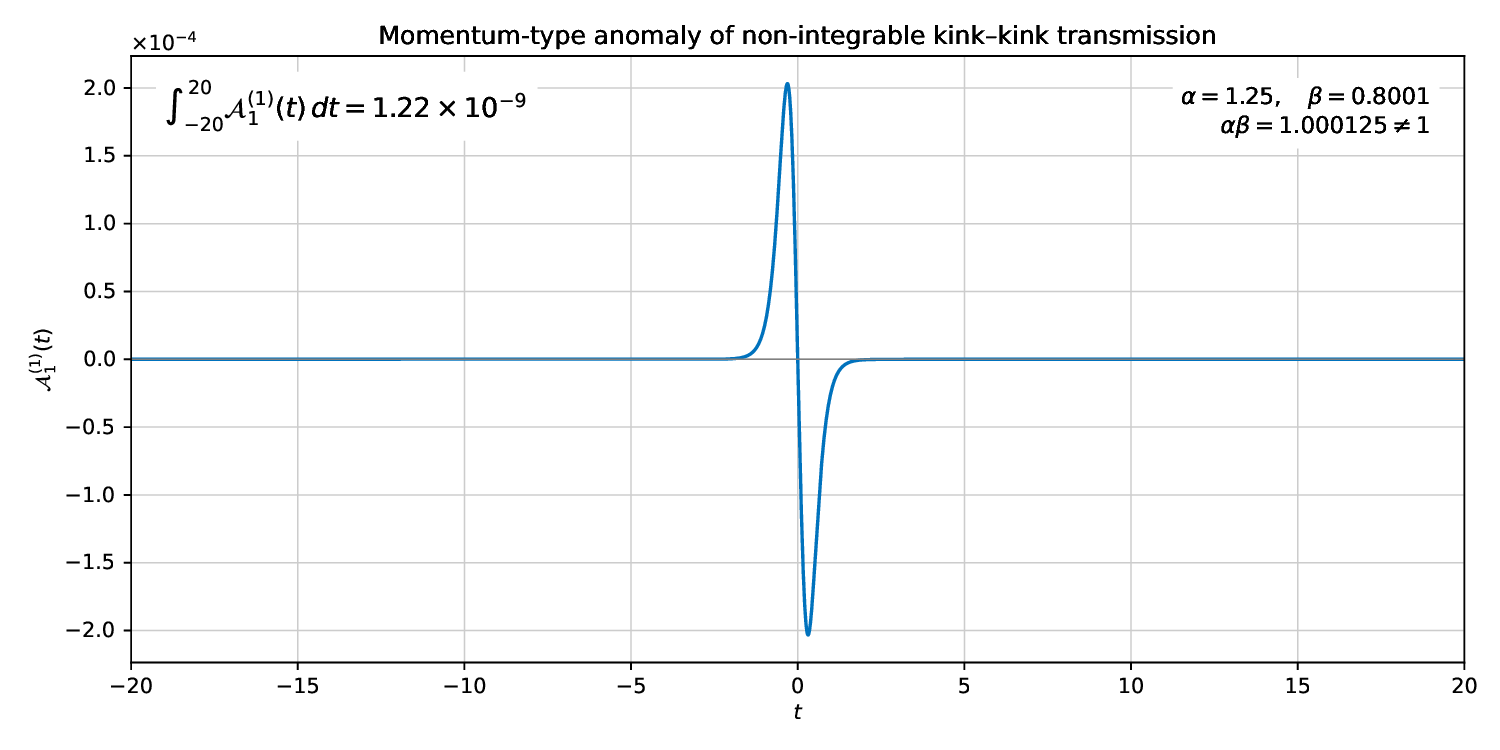}
\includegraphics[width=1.5cm,scale=2, angle=0,height=4cm]{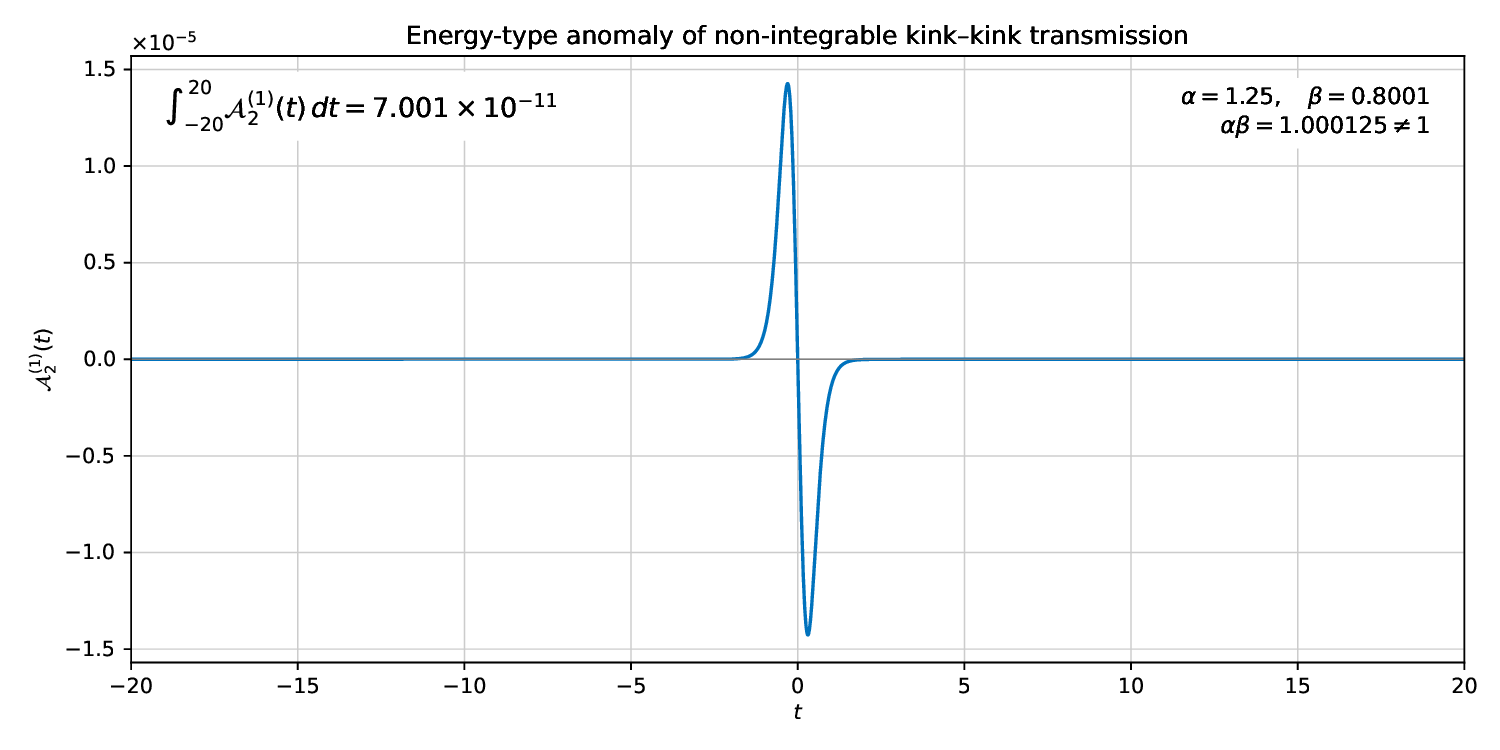}  
\parbox{6in}{\caption{(color online)  Plot of the anomalies (\ref{anomaly11}) and  (\ref{A2cos}) as functions of $t$  associated with the non-integrable kink–kink transmission process of the type shown in the bottom panel of Fig. 1. Left: Momentum-type sector anomaly. The legends show the parameters. Right: Energy-type sector anomaly. The insets in the upper-left corner of each panel display the corresponding time-integrated anomaly, which vanishes within numerical accuracy.}}
\end{figure}

\begin{figure}
\centering
\label{fig9}
\includegraphics[width=1.5cm,scale=2, angle=0,height=4.5cm]{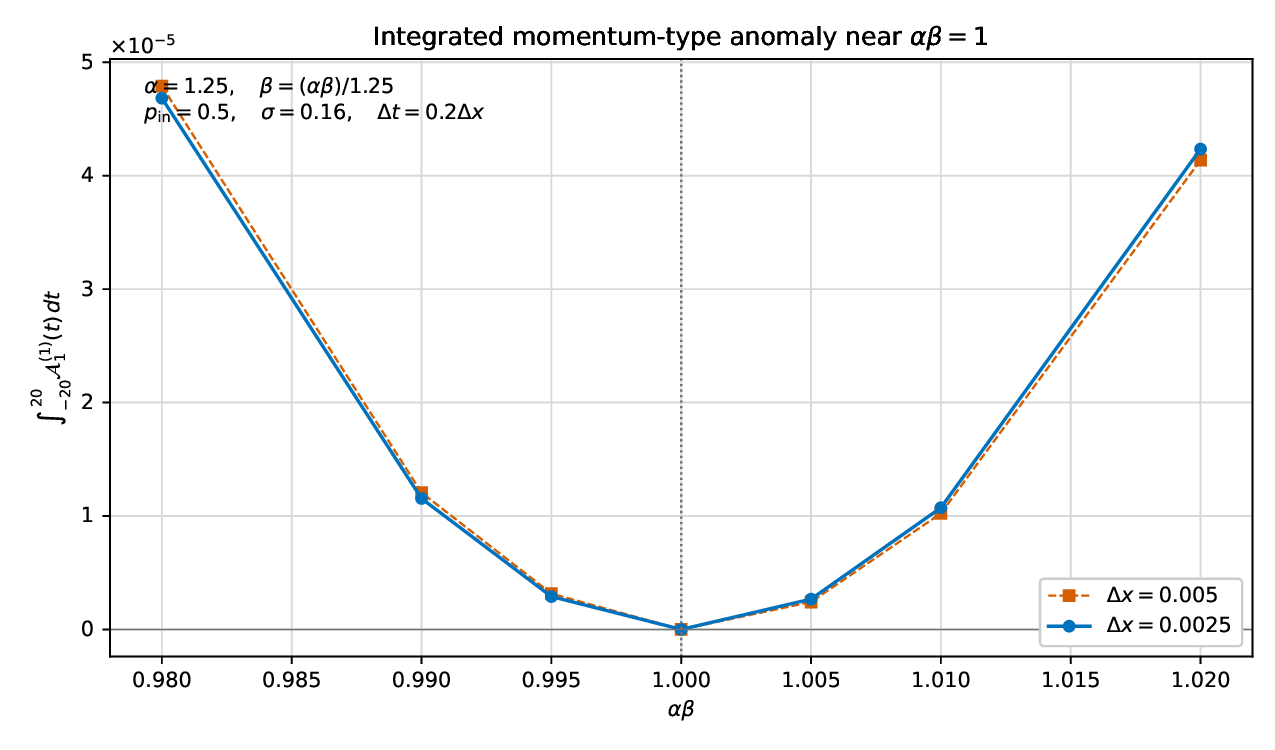}
\includegraphics[width=1.5cm,scale=2, angle=0,height=4.5cm]{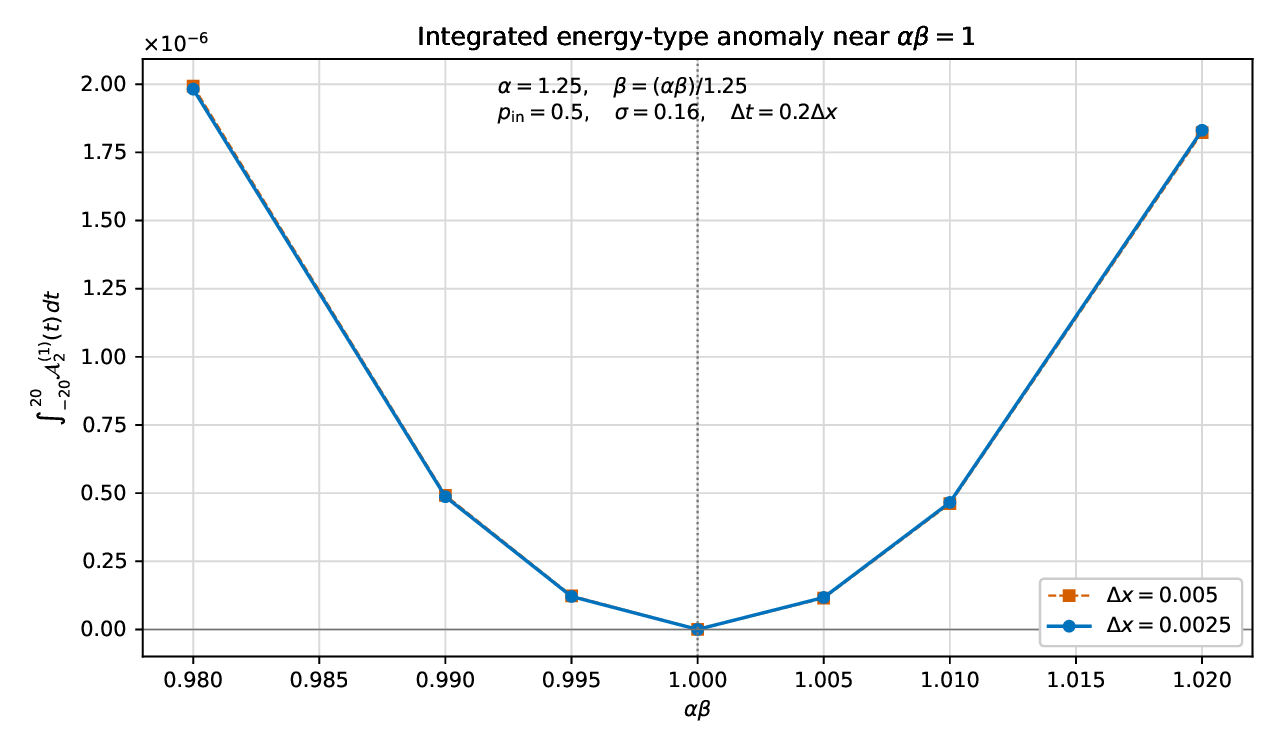}  
\parbox{6in}{\caption{(color online) Left: Time-integrated momentum-type anomaly $\int_{-20}^{20}\mathcal A_1^{(1)}(t)\,dt$\, versus \,$\alpha\beta$ for kink–kink transmission, with $\alpha=1.25$, $p_{\mathrm{in}}=0.5$, and $\sigma=0.16$. Results for two spatial resolutions, with $\Delta t=0.2\Delta x$, show approximately quadratic suppression near $\alpha\beta=1$, where the anomaly vanishes identically. Right: Time-integrated energy-type anomaly $\int_{-20}^{20}\mathcal A_2^{(1)}(t)\,dt$ versus $\alpha\beta$ for kink–kink transmission, with $\alpha=1.25$, $p_{\mathrm{in}}=0.5$, and $\sigma=0.16$. Results for two spatial resolutions, with $\Delta t=0.2\Delta x$, show approximately quadratic suppression near the integrable condition $\alpha\beta=1$, where the anomaly vanishes identically.}}
\end{figure}

\begin{figure}
\centering
\label{fig10}
\includegraphics[width=1.5cm,scale=2, angle=0,height=4.5cm]{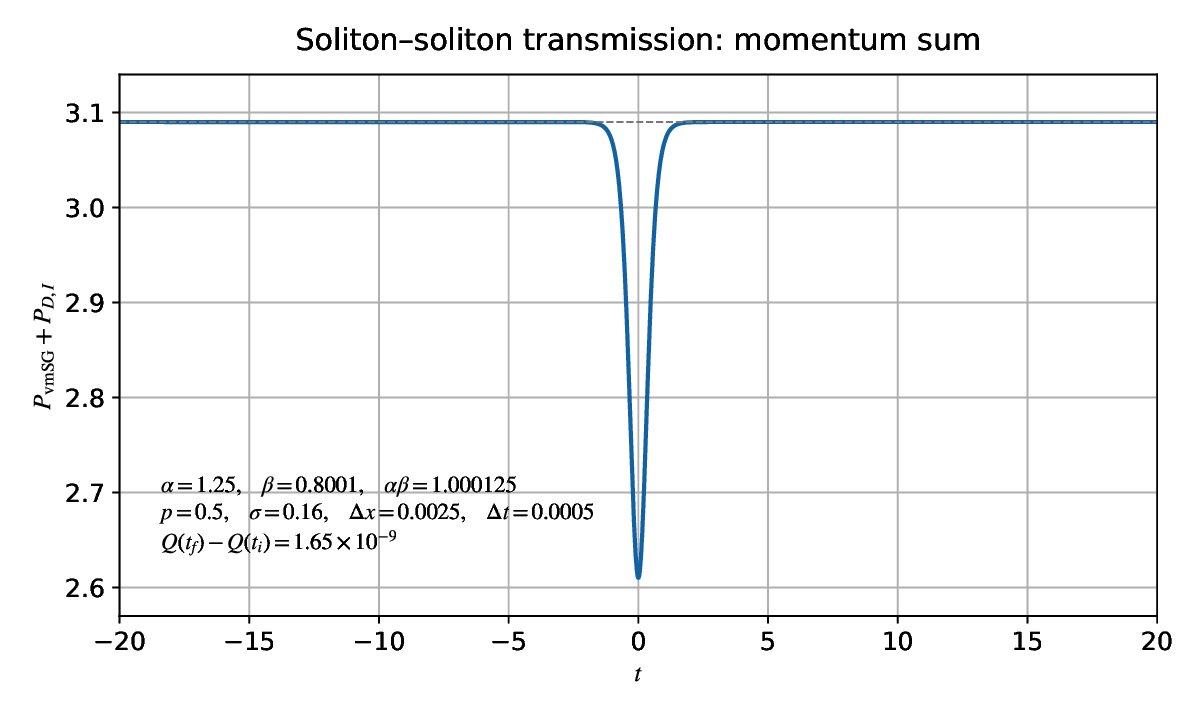}
\includegraphics[width=1.5cm,scale=2, angle=0,height=4.5cm]{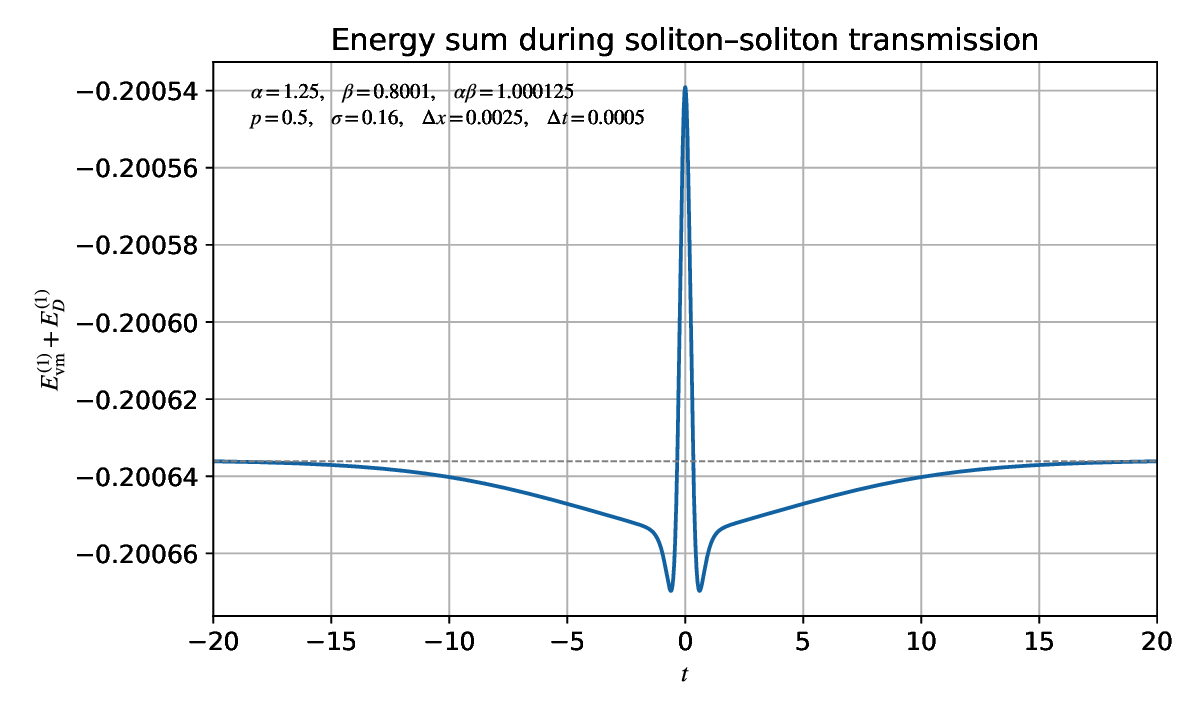}  
\parbox{6in}{\caption{(color online) Nonintegrable soliton transmission and momentum and energy-type charges. Left: Momentum sum. Right: Energy sum. The parameters are displayed in the panels. Note that the sum $Q=Q_{\mathrm{vmSG}}+Q_{D,I}$ from (3.41) and (3.75) returns to its initial value within the accuracy of $10^{-9}$. }}
\end{figure}

Fig. 12 shows the lowest-order momentum- and energy-type anomalies
for non-integrable kink--antikink transmission (Fig.~2),
with $\alpha=1.25$, $\beta=0.8001$, $\sigma=0.8$,
$p_{\rm in}=0.5$, $\Delta x=0.0025$, and $\Delta t=0.0005$.
Left: $\mathcal{A}_1^{(1)}(t)$, with integrated value
$-1.233\times10^{-6}$.
Right: $\mathcal{A}_2^{(1)}(t)$, with integrated value
$7.423\times10^{-7}$.
The nearly canceling lobes support asymptotic
quasi-conservation within numerical accuracy.

Fig. 13 shows the third-order momentum-type anomaly
$\mathcal{A}_1^{(3)}(t)$ for non-integrable kink--kink
transmission with $\alpha=1.25$, $\beta=0.8001$,
$\sigma=0.16$, and $p_{\rm in}=0.5$.
Left: Anomaly profile with integrated value
$4.2253\times10^{-4}$, computed using
$\Delta x=0.0025$ and $\Delta t=0.0005$.
Right: Mesh comparison and cumulative integral,
showing convergence to a nonzero value and hence
non-conservation of the third-order momentum-type charge.

\begin{figure}
\centering
\label{fig11}
\includegraphics[width=1.5cm,scale=2, angle=0,height=4.5cm]{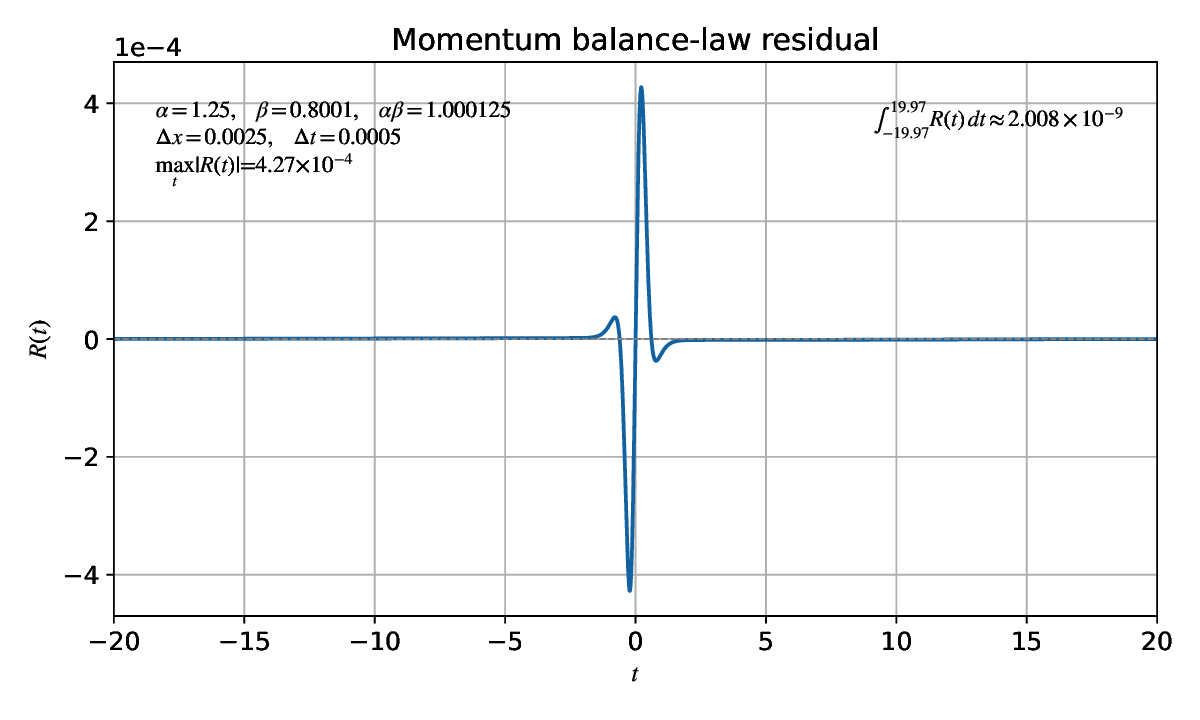}
\includegraphics[width=1.5cm,scale=2, angle=0,height=4.5cm]{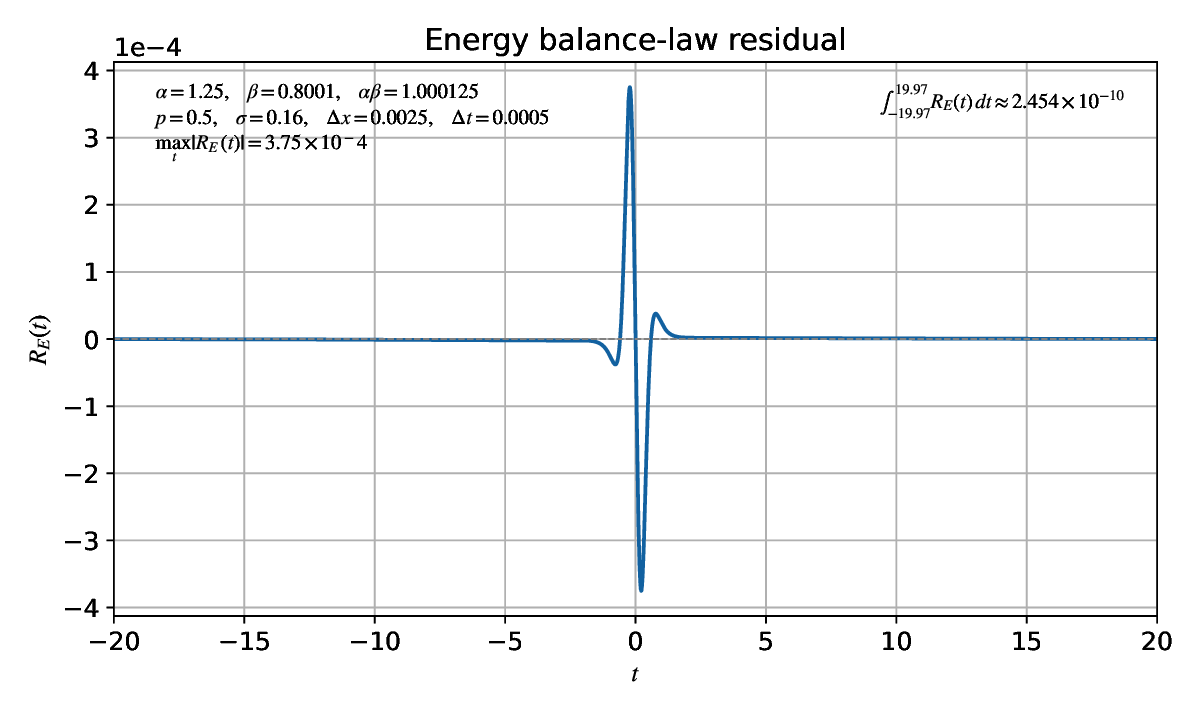}  
\parbox{6in}{\caption{(color online) Left: Momentum balance-law residual $R(t)=d(P_{\mathrm{vmSG}}^{(1)}+P_D^{(1)})/dt-\mathcal A_1^{(1)}$, using the deformed defect contribution (4.29), for soliton–soliton transmission with $\alpha=1.25$, $\beta=0.8001$, $p_{in}=0.5$, and $\sigma=0.16$. At $\Delta x=0.0025$ and $\Delta t=0.0005$, the maximum residual is $4.27\times10^{-4}$, while its signed time integral is $2.008\times10^{-9}$, consistent with numerical discretization error. Right: Energy balance-law residual $R(t)=d(E_{\mathrm{vmSG}}^{(1)}+E_D^{(1)})/dt-\mathcal A_2^{(1)}$, using the deformed defect contribution (4.32), for soliton–soliton transmission with $\alpha=1.25$, $\beta=0.8001$, $p_{in}=0.5$, and $\sigma=0.16$. At $\Delta x=0.0025$ and $\Delta t=0.0005$, the maximum residual is $3.75\times10^{-4}$, while its signed time integral is $2.45\times10^{-10}$, consistent with numerical discretization error. }}
\end{figure}

Fig. 14 provides the third-order energy-type anomaly
$\mathcal{A}_2^{(3)}(t)$ for non-integrable kink--kink
transmission with $\alpha=1.25$, $\beta=0.8001$,
$\sigma=0.16$, and $p_{\rm in}=0.5$.
Left: Anomaly profile with integrated value
$-4.2625\times10^{-4}$, computed at
$\Delta x=0.0025$ and $\Delta t=0.0005$.
Right: Mesh comparison and cumulative integral,
showing convergence to a nonzero value and
non-conservation of the third-order energy-type charge.

Fig 15 shows the third-order momentum-type anomaly
$\mathcal{A}_1^{(3)}(t)$ for non-integrable kink--antikink
transmission with $\alpha=1.25$, $\beta=0.8001$
($\alpha\beta=1.000125$), $\sigma=0.8$, and $p_{\rm in}=0.5$.
Left: Anomaly profile.
Right: Mesh comparison and cumulative time integral,
with $\Delta t=0.2\,\Delta x$, showing convergence to a
nonzero integrated anomaly and non-conservation of the
third-order momentum-type charge.

Finally, Figure 16 shows the third-order energy-type anomaly
$\mathcal{A}_2^{(3)}(t)$ for non-integrable kink--antikink
transmission with $\alpha=1.25$, $\beta=0.8001$
($\alpha\beta=1.000125$), $\sigma=0.8$, and $p_{\rm in}=0.5$.
Left: Anomaly profile with integrated value
$5.87087\times10^{-5}$, computed at
$\Delta x=0.00125$ and $\Delta t=0.00025$.
Right: Mesh comparison and cumulative time integral,
with $\Delta t=0.2\,\Delta x$, showing convergence to a
nonzero integrated anomaly and non-conservation of the
third-order energy-type charge.

\begin{figure}
\centering
\label{fig12}
\includegraphics[width=1.5cm,scale=2, angle=0,height=4cm]{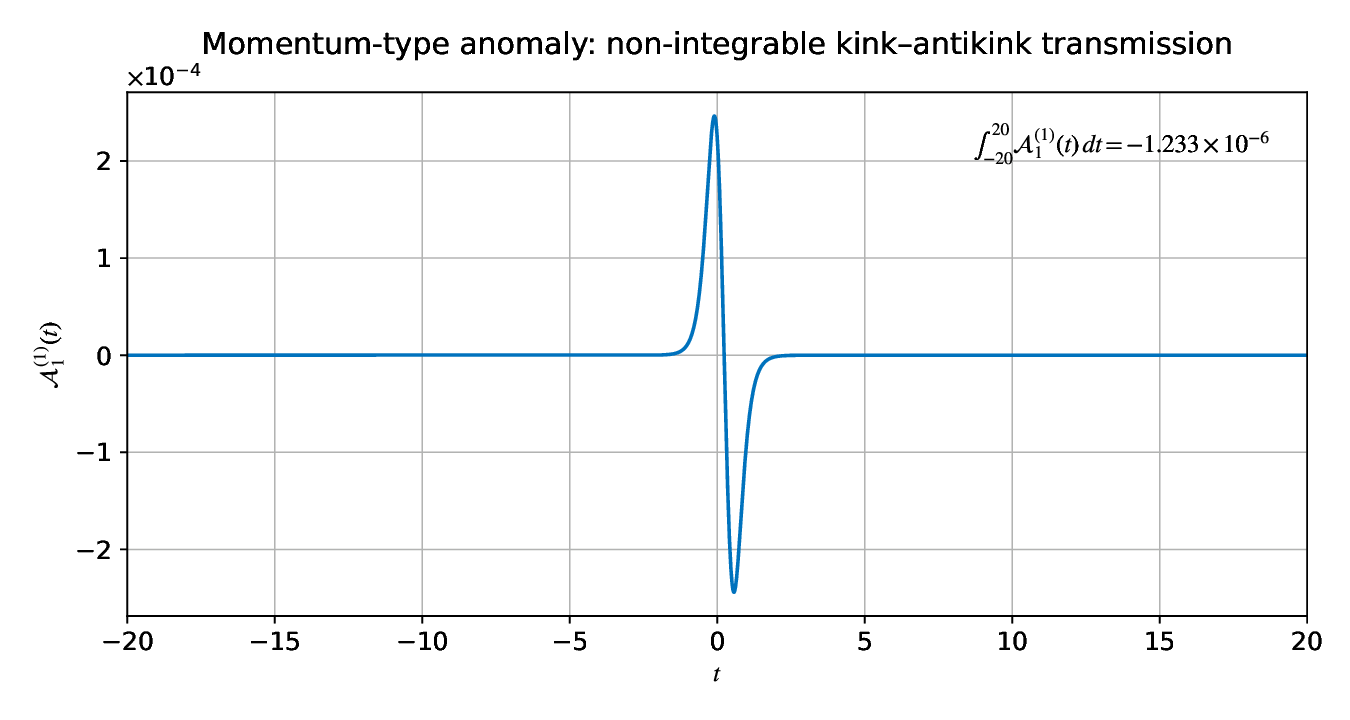}
\includegraphics[width=1.5cm,scale=2, angle=0,height=4cm]{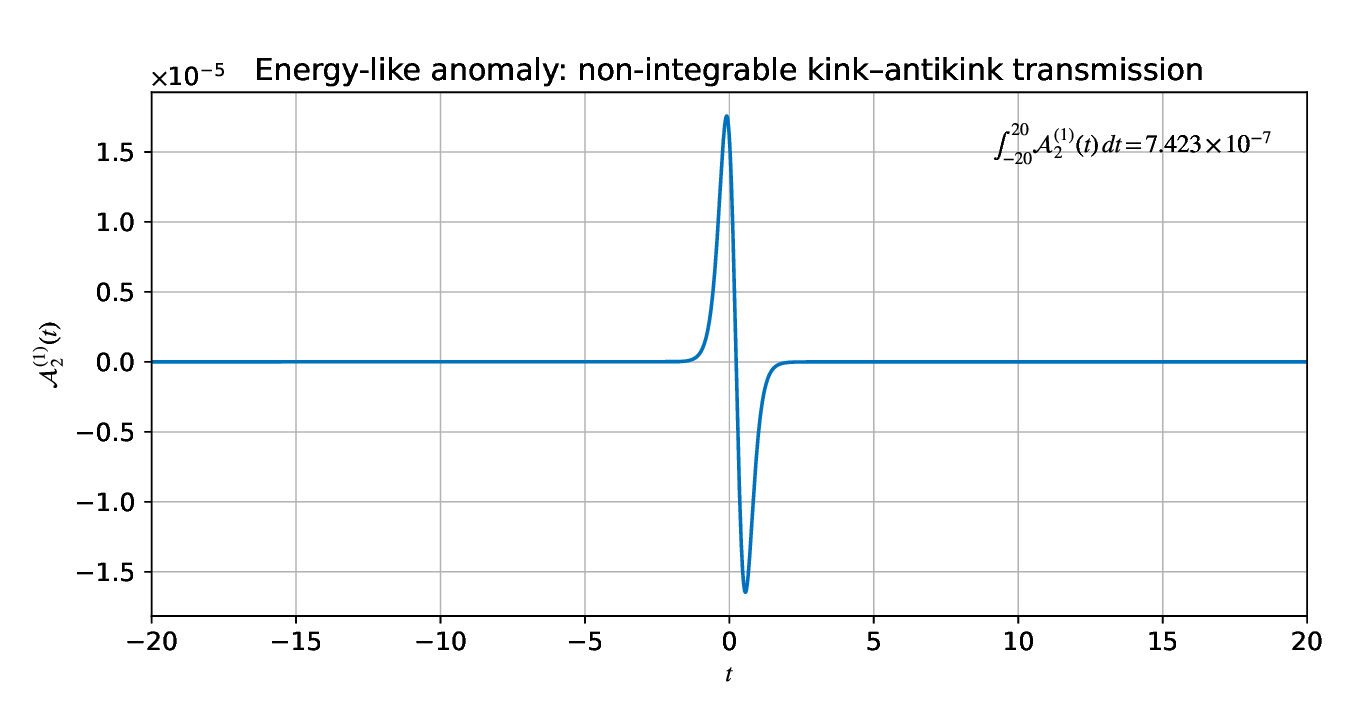}  
\parbox{6in}{\caption{(color online)  Plots of the anomalies (\ref{anomaly11}) and (\ref{A2cos}) as functions of $t$ associated with the non-integrable kink–antikink transmission process shown in the bottom panel of Fig. 2. Left: Momentum-type anomaly $\mathcal{A}^{(1)}_1(t)$ for
non-integrable kink--antikink transmission with $\alpha=1.25$,
$\beta=0.8001$, $\sigma=0.8$, and $p_{\mathrm{in}}=0.5$.
The simulation uses $\Delta x=0.0025$ and $\Delta t=0.0005$.
The positive and negative lobes nearly cancel, yielding the
finite-mesh integral
$\int_{-20}^{20}\mathcal{A}^{(1)}_1(t)\,dt
\simeq -1.233\times10^{-6}$, shown at the top right. Right: Energy-like anomaly $\mathcal{A}^{(1)}_2(t)$ for
non-integrable kink--antikink transmission with $\alpha=1.25$,
$\beta=0.8001$, $\sigma=0.8$, and $p_{\mathrm{in}}=0.5$,
computed using $\Delta x=0.0025$ and $\Delta t=0.0005$.
The positive and negative lobes nearly cancel, yielding
$\int_{-20}^{20}\mathcal{A}^{(1)}_2(t)\,dt
\simeq 7.423\times10^{-7}$, shown at the top right.}}
\end{figure}
\[\]

\begin{figure}
\centering
\label{fig13}
\includegraphics[width=1.5cm,scale=2, angle=0,height=4cm]{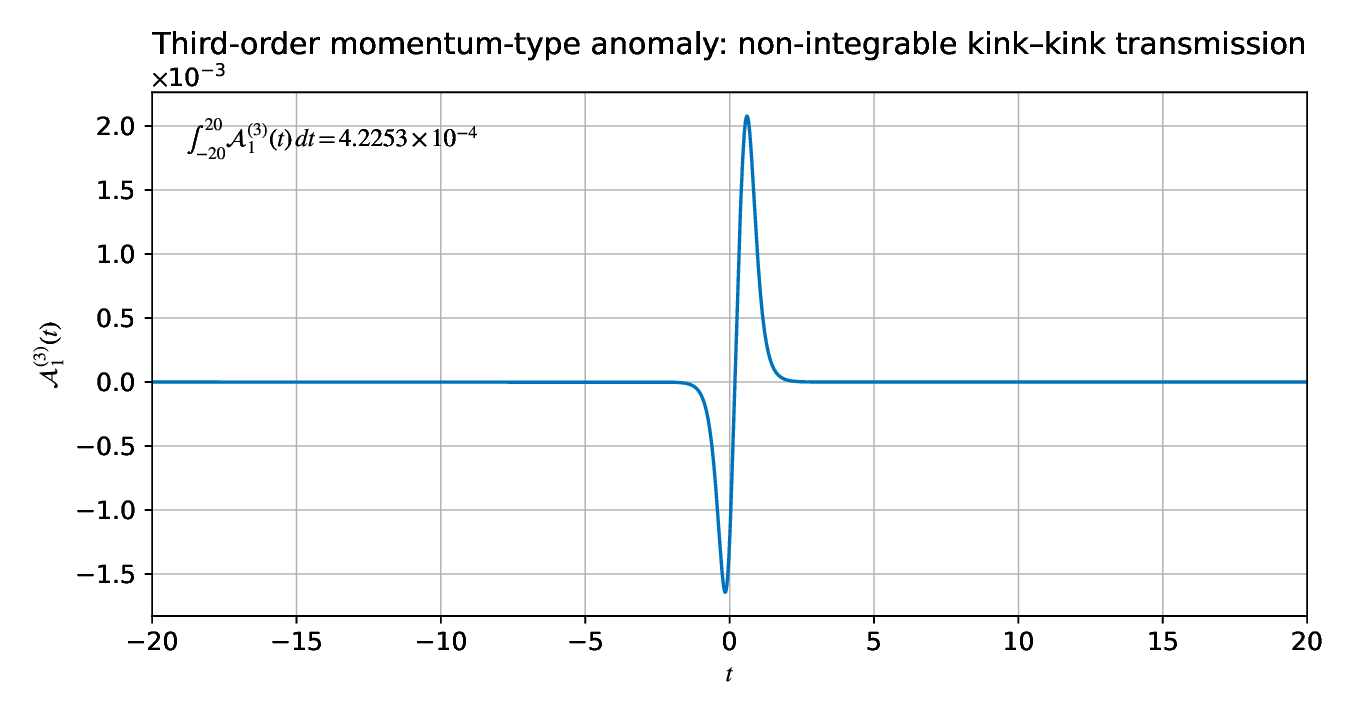}
\includegraphics[width=1.5cm,scale=2, angle=0,height=4.5cm]{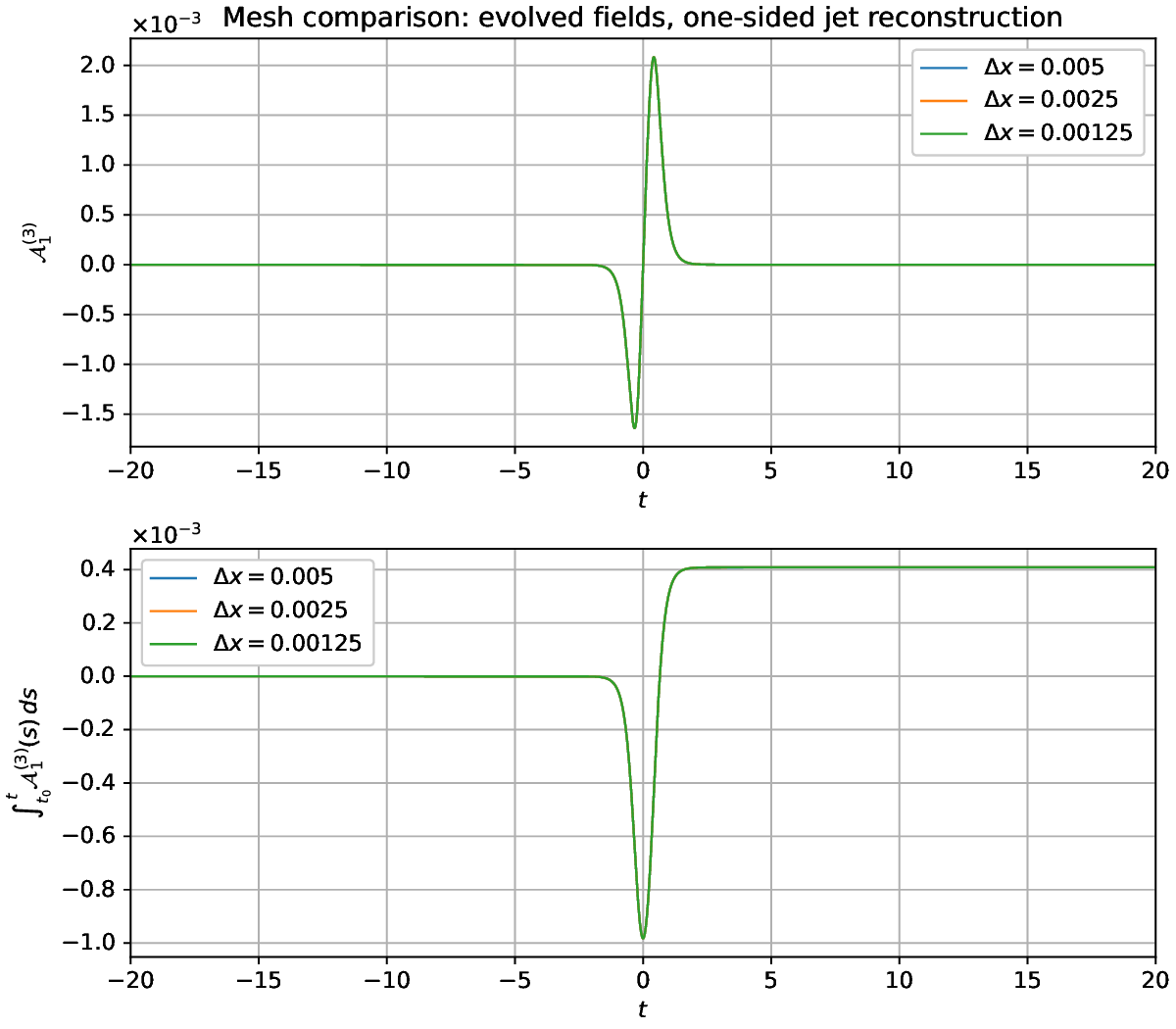}  
\parbox{6in}{\caption{(color online) Left: Plot of the anomaly $\mathcal{A}^{(3)}_1(t)$ in (\ref{eq:third-order-anomalies}) as function of $t$  associated with the non-integrable kink–kink transmission process shown in the bottom panel of Fig. 1. It is used $\alpha=1.25$,
$\beta=0.8001$ ($\alpha\beta=1.000125$), $\sigma=0.16$, and
$p_{\mathrm{in}}=0.5$. Computed with $\Delta x=0.0025$ and
$\Delta t=0.0005$, its time integral is approximately
$4.2253\times10^{-4}$, shown in the panel. Right: Mesh comparison of  
$\mathcal{A}^{(3)}_1(t)$  and its cumulative time integral. The curves nearly overlap. The plot shows mesh convergence to a nonzero integrated anomaly, indicating non-conservation of the third-order energy-type charge.   }}
\end{figure}

\begin{figure}
\centering
\label{fig14}
\includegraphics[width=1.5cm,scale=2, angle=0,height=4cm]{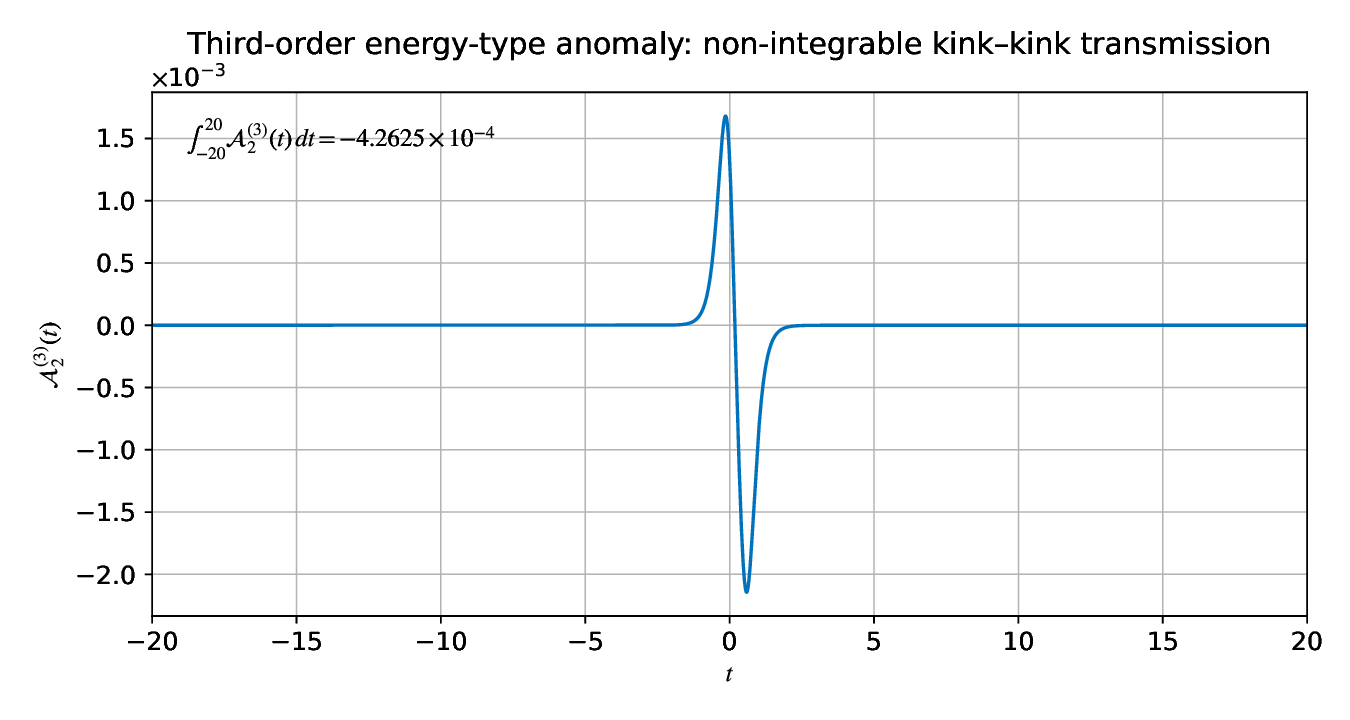}
\includegraphics[width=1.5cm,scale=2, angle=0,height=4.5cm]{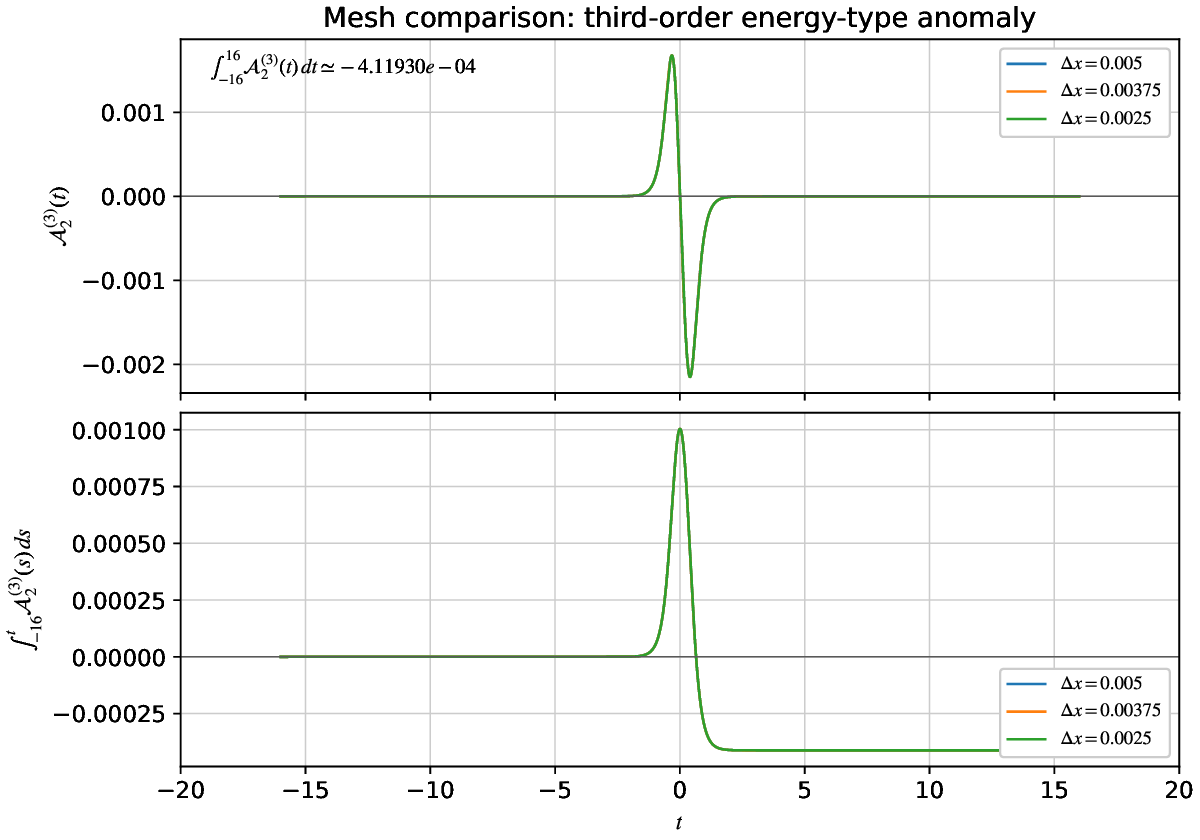}  
\parbox{6in}{\caption{(color online) Left: Plot of the anomaly $\mathcal{A}^{(3)}_2(t)$ in (\ref{eq:third-order-anomalies}) as function of $t$  associated with the non-integrable kink–kink transmission process shown in the bottom panel of Fig. 1. It is used $\alpha=1.25$,
$\beta=0.8001$ ($\alpha\beta=1.000125$), $\sigma=0.16$, and
$p_{\mathrm{in}}=0.5$. Computed with $\Delta x=0.0025$ and
$\Delta t=0.0005$, its time integral is approximately
$-4.2625\times10^{-4}$, shown in the upper-left corner of the panel. Right: Mesh comparison of the anomaly
$\mathcal{A}_{2}^{(3)}(t)$. Right: cumulative integral
$I_{2}^{(3)}(t)=\int_{-12}^{t}\mathcal{A}_{2}^{(3)}(s)\,ds$. The curves nearly overlap (top figure). The curves nearly overlap. The bottom plot shows mesh convergence to a nonzero integrated anomaly, indicating non-conservation of the third-order energy-type charge. }}
\end{figure}

\begin{figure}
\centering
\label{fig15}
\includegraphics[width=1.5cm,scale=2, angle=0,height=4.5cm]{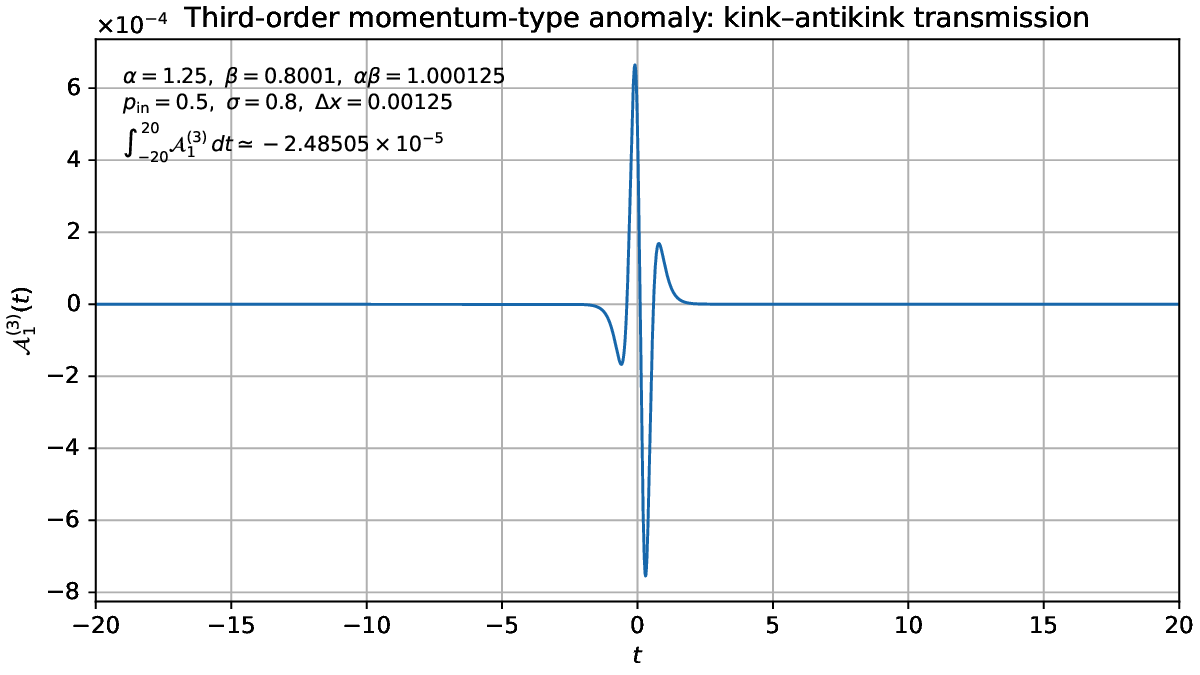}
\includegraphics[width=1.5cm,scale=2, angle=0,height=4.5cm]{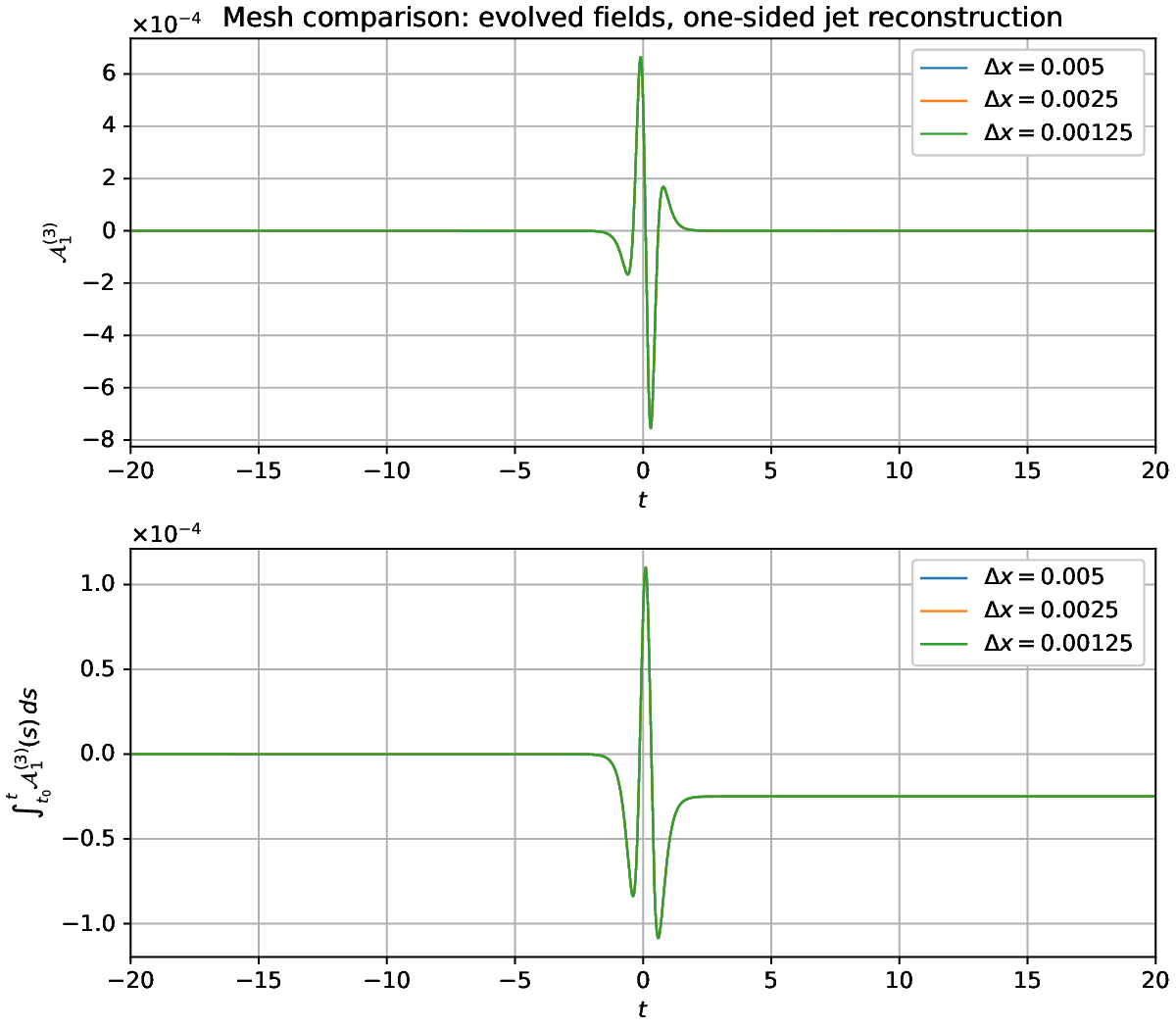}  
\parbox{6in}{\caption{(color online)  Left: Plot of the anomaly $\mathcal{A}^{(3)}_1(t)$ in (\ref{eq:third-order-anomalies}) as function of $t$  associated with the non-integrable kink–antikink transmission process shown in the bottom panel of Fig. 2. Right: Mesh comparison of the third-order momentum-type anomaly
$\mathcal{A}^{(3)}_1(t)$ (top) and its cumulative time integral (bottom)
for kink--antikink transmission, with $\alpha=1.25$, $\beta=0.8001$,
$\alpha\beta=1.000125$, $\sigma=0.8$, and $p_{\rm in}=0.5$.
The three resolutions use $\Delta t=0.2\Delta x$. The curves nearly overlap. The plot shows mesh convergence to a nonzero integrated anomaly, indicating non-conservation of the third-order energy-type charge. }}
\end{figure}

\begin{figure}
\centering
\label{fig16}
\includegraphics[width=1.5cm,scale=2, angle=0,height=4.5cm]{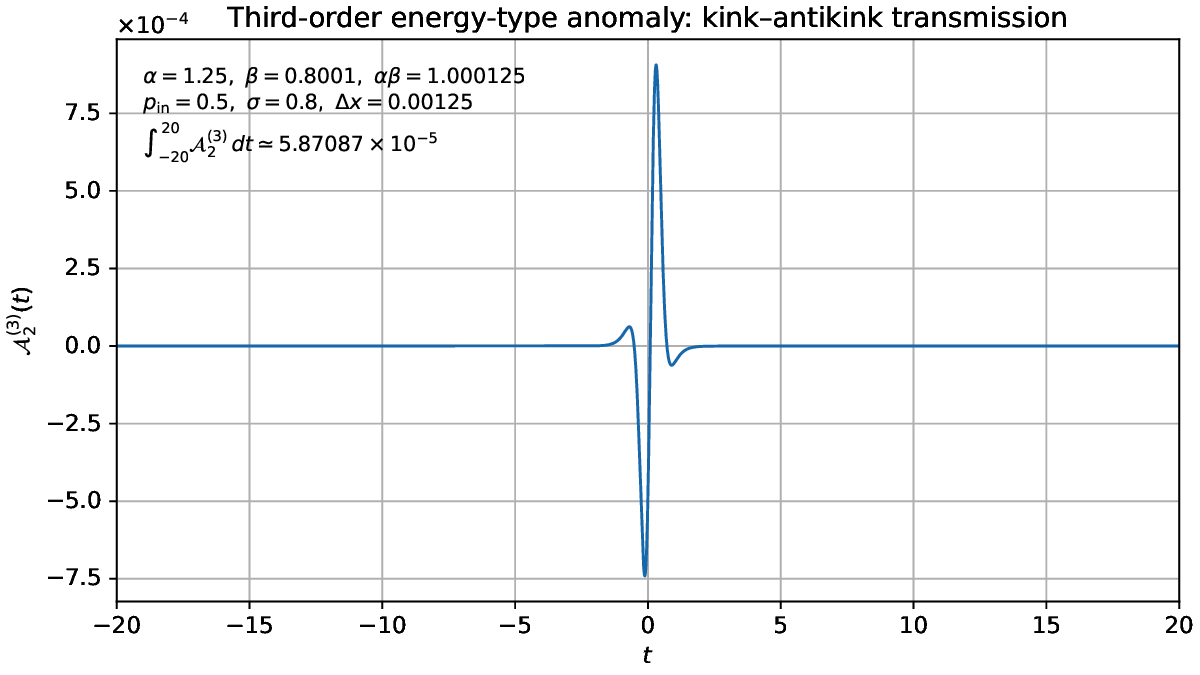}
\includegraphics[width=1.5cm,scale=2, angle=0,height=4.5cm]{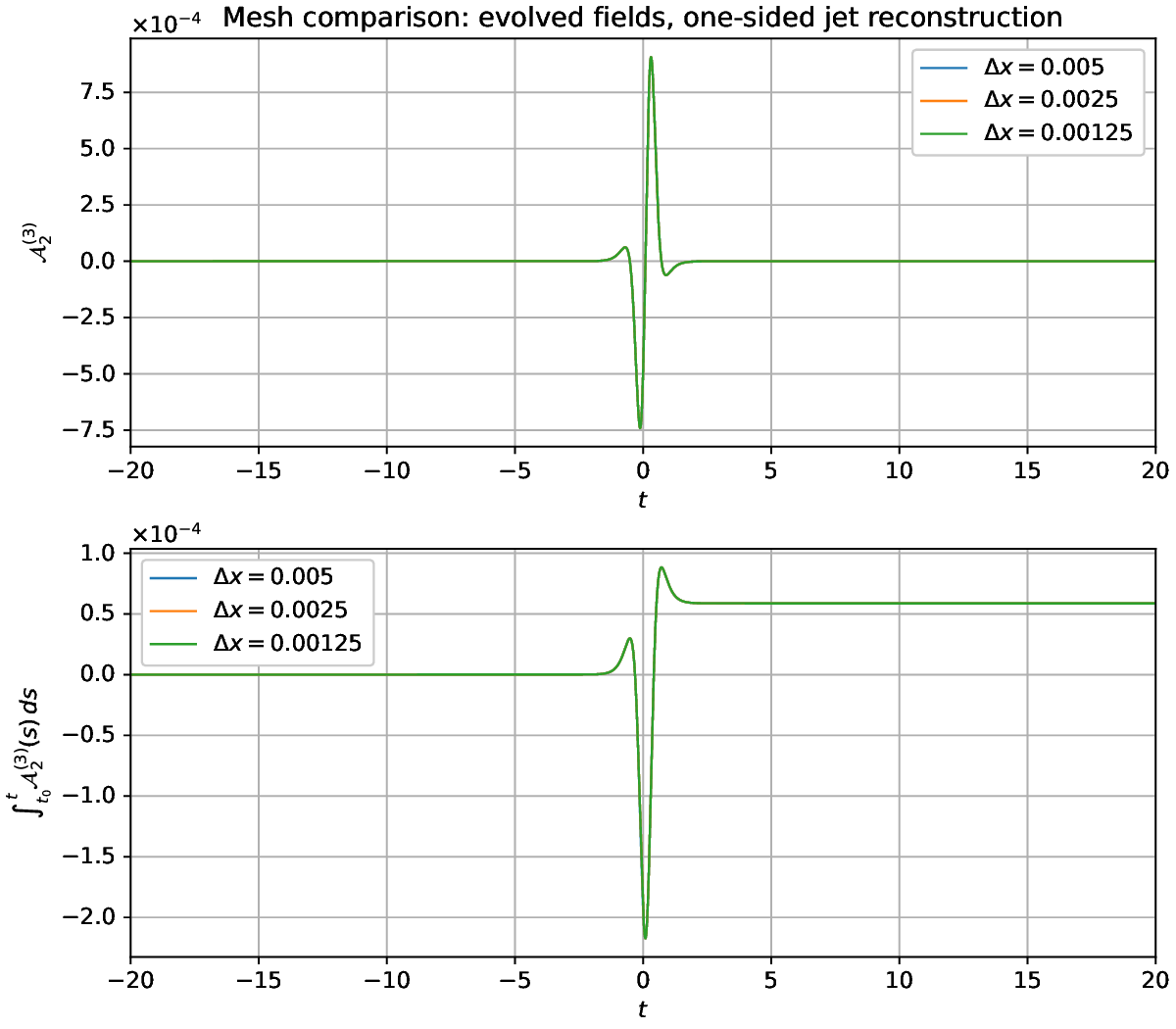}  
\parbox{6in}{\caption{(color online)  Left: Plot of anomaly $\mathcal A_2^{(3)}$ in (\ref{eq:third-order-anomalies}) as function of $t$  associated with the non-integrable kink–antikink transmission process shown in bottom panel of Fig. 2. The inset in the upper-left corner of  panel displays the non-vanishing time-integrated anomaly. It is used $\alpha=1.25$, $\beta=0.8001$, $\alpha\beta=1.000125$,
$\sigma=0.8$, $p_{\rm in}=0.5$, $m_0=1$, $\Delta x=0.00125$ and $\Delta t=0.00025$. The finite-time
integral is approximately $5.87087\times10^{-5}$. Right: Mesh comparison of $\mathcal{A}^{(3)}_2(t)$ (top)
and its cumulative time integral (bottom) for the same
kink--antikink transmission.
The three resolutions use $\Delta t=0.2\Delta x$. The curves nearly overlap. The plot shows mesh convergence to a nonzero integrated anomaly, indicating non-conservation of the third-order energy-type charge. }}
\end{figure}

The numerical results show that quasi-conservation is hierarchy-dependent. For centered configurations, the first-order anomalies are approximately odd and integrate to zero, whereas the third-order anomalies ${\cal A}_j^{(3)}, j=1,2,$ are generally asymmetric and have nonzero integrals, as shown in Figs. 13-16. Hence the corresponding third-order charges are not asymptotically recovered. The present results thus establish low-order, charge-dependent quasi-conservation rather than quasi-integrability of the full hierarchy. 

The above results suggest that quasi-integrability in the variable-mass sine-Gordon model may be achieved by deforming not only the parameters of the type-I sewing conditions, but also their field content. In particular, the sewing functions can be extended to depend on parity-adapted combinations of the background-covariant jets $u,v,u_{1},v_{1},\ldots $. These additional terms can be chosen so that the residuals entering the higher-order balance equations transform oddly under the combined time reflection, field transformation, and interchange of the two half-lines.

Our constructions suggest that quasi-integrability is hierarchy-dependent: parameter deformations may preserve the first-order charges, while higher-order charges require additional jet-dependent terms. It therefore supports the conjecture that a simultaneous deformation of the parameters and field content of a type-I defect can generate a quasi-integrable variable-mass sine-Gordon theory with asymptotically conserved charges. Exact integrability requires all anomalies to vanish identically, whereas quasi-integrability requires only their time-integrated cancellation for parity-symmetric soliton processes. An all-order recursive derivation of the sewing terms and generalized defect contributions lies beyond the scope of this work.

\section{Conclusions and discussion}
\label{sec:conclusions}

We have developed a defect formulation for the variable-mass sine-Gordon model. Starting from its zero-curvature representation, we constructed the type-I and type-II B\"acklund--gauge matrices and derived the associated sewing conditions. The defect matrices preserve the algebraic structure of their homogeneous sine-Gordon counterparts, whereas the B\"acklund equations acquire an explicit dependence on the light-cone backgrounds $\rho_0(\xi)$ and $\rho_1(\eta)$. For the defect constructions considered here, exact compatibility requires suitably matched backgrounds across the two half-lines; in particular, the type-I and II B\"acklund--gauge systems impose $\bar\rho_0=\rho_0$ and $\bar\rho_1=\rho_1$. Consequently, coupling genuinely distinct inhomogeneous media would require a more general defect matrix, additional localized degrees of freedom, or an anomalous extension of the gauge equations, as in the quasi-integrability framework \cite{nuclear1}

The Riccati formulation provides a systematic construction of the bulk and defect contributions to the charge hierarchy. The first nontrivial
combinations define momentum- and energy-type charges which reduce to the
canonical sine-Gordon momentum and energy in the homogeneous limit. For
generic variable-mass backgrounds, however, space-time translation
invariance is absent, and these quantities must be regarded as charges
generated by the integrable structure rather than as ordinary Noether
charges. The second-order bulk combinations are surface terms for localized
configurations, but become nontrivial in the two-half-line theory because
their jumps at $x=0^\pm$ are compensated by the corresponding defect
contributions. The third-order members provide the first higher odd charges with genuinely integrability-generated defect contributions that depend explicitly on the light-cone backgrounds.

A deformation of the type-I sewing conditions was introduced through the
parameters
$
 \alpha=1+\epsilon_A,
 \,
 \beta=1+\epsilon_B .
$
The resulting balance equations contain anomalies localized at the defect.
The condition
$
 \alpha\beta=1
$
simultaneously removes these anomalies and ensures compatibility of the two
sewing equations with an exact one-soliton transmission ansatz. In this
sector the deformation can be absorbed into a single effective B\"acklund
parameter; consequently, it does not define a genuinely new non-integrable
defect, but rather a reparametrization of the integrable type-I defect. When
$\alpha\beta\neq1$, the two sewing equations impose incompatible
transmission factors and no exactly phase-shifted one-soliton profile can
satisfy both relations. The defect is then genuinely non-integrable and the
transmitted soliton must be accompanied by additional small residual disturbances.

The transmission factor separates two distinct branches. Its positive
sector describes kink--kink transmission, while its negative sector
describes kink--antikink conversion. In the latter process, the change in
the bulk topological charge is exchanged with the defect. The limiting
values at which the transmission factor vanishes or diverges correspond to
critical absorption or emission configurations rather than to regular
transmission. These results show that the defect may control not only the
position shift and delay of the excitation, but also its outgoing
topological sector.

For parity-centered kink--kink and kink--antikink configurations, the
first-order momentum- and energy-type anomalies are approximately odd in
time and their integrated values vanish within numerical accuracy. The
corresponding modified charges may therefore vary during the interaction
but recover their incoming values asymptotically. This property is,
however, hierarchy dependent. At third order the cancellation requires
additional parity assignments for the covariant derivatives
$u,u_1,v,$ and $v_1$. These assignments are sufficient conditions for
odd anomalies, but they do not follow in general from the field parity or
from the deformed sewing equations. The numerical third-order anomalies
display asymmetric temporal profiles and finite time integrals for the
non-integrable trajectories considered.

In summary, exact defect integrability is established in the closure-compatible sector $\a \b =1$, whereas the deformed cases studied here exhibit only low-order, charge-dependent asymptotic quasi-conservation. Since the higher-order anomalies do not generically integrate to zero, no infinite quasi-conserved hierarchy is demonstrated. Whether more general correlated bulk–defect deformations can generate an infinite asymptotically conserved subsector remains an open question.

The finite-difference simulations reproduce the integrable branches shown in the top panels of Figs. 1 and 2. Away from the closure condition, the dominant excitation remains transmitted, as seen in the bottom panels of Figs. 1 and 2, while small residual disturbances are generated near the defect (see Fig. 7 for nonintegrable kink-kink transmission). These features provide a dynamical signature of the incompatibility between the two deformed sewing relations. Their small amplitude is consistent with the approximate lowest-order balance, while the parity structure accounts for the vanishing integrated first-order anomalies.

A particularly natural continuation of this work is the construction of a
quasi-integrable variable-mass sine-Gordon theory in which the bulk
potential and the defect sewing conditions are deformed simultaneously.
One may consider a bulk deformation $V_{SG}(w) \rightarrow V_{SG}^{def}(w)$ dressed by the variable-mass backgrounds, together with independent
defect deformations controlled by $\epsilon_A$ and $\epsilon_B$.
The anomalous zero-curvature or Riccati formulation should then lead to
combined balance equations of the form
\[
 \frac{d}{dt}\({\cal Q}^{(n)}+ {\cal Q}^{(n)}_D\)
 =
 \int_{-\infty}^{0}{\cal X}^{(n)}_{L}(x,t)\,dx
 +
 \int_{0}^{+\infty}{\cal X}^{(n)}_{R}(x,t)\,dx
 +
 {\cal A}^{(n)}_{D}(t),
\]
where ${\cal X}^{(n)}_{L,R}$ are bulk anomaly densities and
${\cal A}^{(n)}_{D}$ is the localized defect anomaly. Asymptotic
quasi-conservation could then arise either because the bulk and defect
contributions are separately odd under a suitable space-time parity or
because they compensate each other after integration,
\[
 \int_{-\infty}^{+\infty}dt\,
 \left[
 \int_{-\infty}^{0}{\cal X}^{(n)}_{L}\,dx
 +
 \int_{0}^{+\infty}{\cal X}^{(n)}_{R}\,dx
 +
 {\cal A}^{(n)}_{D}
 \right]
 =0 .
\]
This second possibility would define a genuinely coupled bulk--defect
mechanism of quasi-integrability, in which the defect acts as a localized
reservoir capable of absorbing or compensating the anomalous flux generated
in the deformed bulk.

An important objective would be to determine whether the bulk and defect
deformation parameters can be correlated so that the total anomaly
vanishes for an infinite subsector of the hierarchy, even though neither
the bulk nor the defect is integrable separately. This requires the
construction of generalized anomalous B\"acklund or defect matrices,
recursive derivation of the combined higher-order anomalies, and the
identification of parity conditions satisfied dynamically by the complete
bulk--defect solution.

Other relevant extensions include deformed type-II defects with a dynamical
auxiliary field, multi-soliton and breather scattering, multiple and moving
defects, backgrounds without definite time-reflection parity, and defects
joining different variable-mass media. It would also be useful to develop
structure-preserving numerical schemes that reproduce the combined
bulk--defect balance laws at the discrete level. Finally, the classical
results provide a basis for studying semiclassical and quantum transmission
factors, defect-induced topological conversion, and experimentally relevant
inhomogeneous realizations in Josephson, magnetic, nonlinear-optical, and
effective molecular systems.

\noindent {\bf Acknowledgements}

We thank N. A. de Almeida, G. B. S. Marcelino, L. T. Teixeira, G. K. R. de Souza and A. O. Assun\c c\~ao for useful discussions. HB thanks CNPq for partial financial support. The authors thank UFMT for financial support through Edital $N^o$ 06/2026/PROPESQ.

\appendix

\section{Transformation of the defect trajectory}
\label{app:defectrajetory}

The coordinate transformation to $(X,Y)$ in (\ref{tr01}) makes the local relation with the homogeneous sine-Gordon theory explicit. Nevertheless, the physical defect at $x=0$, where $\xi=\eta=t/2$, is mapped to the background-dependent trajectory $
X_d(t)=X(t/2), Y_d(t)=Y(t/2).
$
Introducing $\tau=X+Y$ and $z=X-Y$, this trajectory becomes
\br
\tau(t)=X(t/2)+Y(t/2),\qquad
z(t)=X(t/2)-Y(t/2),
\er
with velocity
\br
V_d(t)=\frac{dc}{d\tau}
=\frac{\rho_0(t/2)-\rho_1(t/2)}
{\rho_0(t/2)+\rho_1(t/2)},\,\,\,\,\,\,\, z = c(t),
\er
provided the denominator is nonzero. The transformed defect is therefore generally moving; it is stationary when $\rho_0(t/2)=\rho_1(t/2)$.

The coordinate map is locally invertible where $\rho_0\rho_1\ne0$. Parametrization of the induced interface by $\tau$ additionally requires $\rho_0+\rho_1\ne0$ along the defect. If $\rho_0\rho_1>0$, the transformed trajectory is timelike, since $1-V_d^2=4\rho_0\rho_1/(\rho_0+\rho_1)^2>0$. The derivatives in the sewing relations are traces of bulk partial derivatives, whereas $d/d\tau=\partial_\tau+V_d\partial_z$ denotes differentiation along the interface. The curve is fixed by the prescribed backgrounds rather than by an independent defect equation of motion.

Since $\partial_\xi=\rho_0\partial_X$ and $\partial_\eta=\rho_1\partial_Y$, the explicit background factors are locally removed from the B\"acklund equations. However, the resulting sewing relations (\ref{wx1})-(\ref{bwx1}) must be imposed along this moving trajectory.  In the integrable type-I sector, with the common sign absorbed into $\sigma$, one can show explicitly that
\begin{equation*}
\begin{aligned}
(\partial_\tau+\partial_z)(w-\bar w)&=-2m_0\sigma\sin(w+\bar w),\\
(\partial_\tau-\partial_z)(w+\bar w)&=\frac{2m_0}{\sigma}\sin(w-\bar w),
\end{aligned}
\qquad z=c(\tau).
\end{equation*}
These follow from $\partial_\xi=\rho_0\partial_X$, $\partial_\eta=\rho_1\partial_Y$, and $\partial_X=\partial_\tau+\partial_z$, $\partial_Y=\partial_\tau-\partial_z$. They exhibit the homogeneous B\"acklund form while retaining the moving evaluation curve.  

The Riccati variables are auxiliary-component ratios and transform as scalars
\begin{equation}
\Gamma_{lk}(\xi,\eta;\lambda)=\G_{lk}(X,Y;\lambda).
\end{equation}
Dividing manuscript Eqs.\ (\ref{eq3.18})--(\ref{eq3.21}) by $\rho_0$ or $\rho_1$ gives the homogeneous Riccati system. For example,
\begin{align}
\partial_X\G_{21}&=iW_X\G_{21}+\lambda m_0(e^{-iW}+e^{iW}\G_{21}^2),\\
\partial_Y\G_{21}&=-iW_Y\G_{21}+\frac{m_0}{\lambda}(e^{iW}+e^{-iW}\G_{21}^2).
\end{align}
Here and throughout this appendix, subscripts $X$ and $Y$ denote derivatives with respect to the transformed coordinates. In particular,
$
W_X=\partial_X W=\rho_0^{-1}\partial_\xi w,
\,\,
W_Y=\partial_Y W=\rho_1^{-1}\partial_\eta w,
$
where $W(X,Y)=w(\xi,\eta)$. These relations follow directly from the coordinate transformation (\ref{tr01}).

Define the homogeneous currents, retaining the vacuum normalization of manuscript Eqs.\ (\ref{eq3.15})--(\ref{eq3.16}), by
\begin{equation}\label{crossed}
\widehat J_k^X=\frac{J_k^\xi}{\rho_1},\qquad
\widehat J_k^Y=\frac{J_k^\eta}{\rho_0},\qquad k=1,2.
\end{equation}
The crossed factors in \eqref{crossed} are essential. Explicitly,
\begin{align}
\widehat J_1^X&=\frac{i}{2}W_Y+\frac{im_0}{\lambda}(1+ie^{-iW}\G_{21}),&
\widehat J_1^Y&=\frac{i}{2}W_X-i\lambda m_0(1+ie^{iW}\G_{21}),\\
\widehat J_2^X&=-\frac{i}{2}W_Y-\frac{im_0}{\lambda}(1+ie^{iW}\G_{12}),&
\widehat J_2^Y&=-\frac{i}{2}W_X+i\lambda m_0(1+ie^{-iW}\G_{12}).
\end{align}
Because $\partial_\xi\rho_1=\partial_\eta\rho_0=0$, the conservation law Eq.\ (\ref{cl2}) becomes
\begin{equation}
\pa_\xi J_k^{\xi} + \pa_\eta J_k^{\eta}=0 \,\,\, \Rightarrow \,\,\,\rho_0\rho_1(\partial_X\widehat J_k^X+\partial_Y\widehat J_k^Y) =0.
\end{equation}
Set
\begin{equation}
q_k=\frac{\widehat J_k^X+\widehat J_k^Y}{2},\qquad
j_k=\frac{\widehat J_k^X-\widehat J_k^Y}{2},\qquad
\omega_k=q_k\,\dd z-j_k\,\dd\tau
=\widehat J_k^Y\,\dd X-\widehat J_k^X\,\dd Y.
\end{equation}
Then $\partial_\tau q_k+\partial_zj_k=0$ and $\dd\omega_k=0$ in each bulk region. On a physical constant-$t$ slice,
\begin{equation}
\dd X=\frac{\rho_0}{2}\dd x,\qquad \dd Y=-\frac{\rho_1}{2}\dd x,
\end{equation}
so that
\begin{equation}\label{exactdensity}
\left.\omega_k\right|_{t=\mathrm{const}}
=\frac12(J_k^\xi+J_k^\eta)\,\dd x
=\frac12[(\rho_0+\rho_1)q_k-(\rho_0-\rho_1)j_k]\,\dd x. 
\end{equation}
Thus the pullback is exactly the generating density Eq.\ (\ref{eq3.17}); its piecewise integral reproduces Eq.\ (\ref{dch}).

Next, examine the type-I logarithmic defect term and its moving balance. 
Let $W^\pm=W\pm\bar W$. The matrix in Eq.\ (\ref{eq2.10vm}) is unchanged algebraically
\begin{equation}
K_I=\begin{pmatrix}
e^{-iW^-/2}&\sigma\lambda^{-1}e^{-iW^+/2}\\
-\sigma\lambda^{-1}e^{iW^+/2}&e^{iW^-/2}
\end{pmatrix}.
\end{equation}
The transformed sewing equations are
\begin{equation}
W^-_X=-2m_0\sigma\sin W^+,\qquad
W^+_Y=2m_0\sigma^{-1}\sin W^-,\qquad z=c(\tau).
\end{equation}
All interface derivatives below are understood as traces of the respective bulk derivatives. The defect contribution of Eq.\ (\ref{Def12}) is
\begin{align}
D_1&=-\ln\left[e^{-iW^-/2}+\frac{\sigma}{\lambda}e^{-iW^+/2}\G_{21}\right]_{z=c(\tau)},\\
D_2&=-\ln\left[e^{iW^-/2}-\frac{\sigma}{\lambda}e^{iW^+/2}\G_{12}\right]_{z=c(\tau)}.
\end{align}
 
To verify the sign and normalization, let $v$ be the homogeneous auxiliary vector and define $S_k=K_{kk}+K_{kl}\G_{lk}$. The relation $\bar v=K_Iv$ implies $\bar v_k=S_kv_k$ on the interface. Directly from the Lax pair and the currents above,
\begin{equation}
\omega_k=-\dd\ln v_k+\dd F_k,\qquad
F_1=-im_0(\lambda X+\lambda^{-1}Y),\quad F_2=-F_1.
\end{equation}
The equal-background chart makes $F_k$ common to both sides. 

For the interface embedding $\iota(\tau)=(\tau,c(\tau))$, therefore,
\begin{equation}\label{jump}\iota^*(\omega_{k,R}-\omega_{k,L})
=-\dd\ln S_k=\dd D_k.
\end{equation}
Here $\iota^*$ denotes the pullback onto the defect trajectory $\iota(\tau)=(\tau,c(\tau))$. For a differential one-form $\omega=q\,dz-j\,d\tau$, its restriction to the trajectory is
\[
\iota^*\omega=(Vq-j)\,d\tau,
\qquad V=\frac{dc}{d\tau},
\]
where the functions $q$ and $j$ are evaluated at $z=c(\tau)$. This operation expresses the bulk current balance along the moving interface.
Since $\iota^*\omega_{k,a}=(Vq_{k,a}-j_{k,a})\dd\tau$, Eq.\ \eqref{jump} gives
\begin{equation}\label{Dbalance}
\frac{\dd D_k}{\dd\tau}
=j_{k,L}-j_{k,R}-V(q_{k,L}-q_{k,R}). 
\end{equation}
Here $\dd/\dd\tau$ is the total derivative along the moving interface. For ordered domains on a constant-$\tau$ slice, define
\begin{equation}
Q_{k,\mathrm{bulk}}=\int_{-\infty}^{c(\tau)}q_{k,L}\,\dd z
+\int_{c(\tau)}^\infty q_{k,R}\,\dd z.
\end{equation}
With vanishing asymptotic fluxes, differentiation gives
\begin{equation}
\frac{\dd Q_{k,\mathrm{bulk}}}{\dd\tau}
=j_{k,R}-j_{k,L}+V(q_{k,L}-q_{k,R}),\qquad
\frac{\dd}{\dd\tau}(Q_{k,\mathrm{bulk}}+D_k)=0.
\end{equation}
No additional velocity correction to the logarithmic $D_k$ is required. The velocity term follows from evaluating the same defect generating function along the moving curve. Together with \eqref{exactdensity}, this establishes the manuscript's bulk-plus-defect generating charges on the transformed physical slices. Comparing their values with charges on constant-$\tau$ slices also requires compatible endpoints and vanishing flux through the intervening outer boundaries.
 
Let us perform the first-order check and current improvements.
Define
\begin{equation}
u=W_X=\frac{\partial_\xi w}{\rho_0},\quad
v=W_Y=\frac{\partial_\eta w}{\rho_1},\quad U=2m_0^2(1-\cos2W).
\end{equation} 
The homogeneous energy, momentum and stress densities in the manuscript's momentum convention are
\begin{equation}
e=\frac{u^2+v^2}{4}+U,\qquad
p=\frac{u^2-v^2}{4},\qquad s=\frac{u^2+v^2}{4}-U.
\end{equation}
The bulk equation gives $\partial_\tau e-\partial_zp=0$ and $\partial_\tau p-\partial_zs=0$. Thus $\omega_E=e\dd z+p\dd\tau$ and $\omega_P=p\dd z+s\dd\tau$. Their pullbacks are
\begin{align}
\mathcal E_{\mathrm{vm}}&=\frac{\rho_0u^2+\rho_1v^2}{4}
+m_0^2(\rho_0+\rho_1)(1-\cos2w),\\
\mathcal P_{\mathrm{vm}}&=\frac{\rho_0u^2-\rho_1v^2}{4}
-m_0^2(\rho_0-\rho_1)(1-\cos2w).
\end{align}
Substituting $\partial_\xi w=w_t+w_x$ and $\partial_\eta w=w_t-w_x$ reproduces Eqs.\ (\ref{e1}) and (\ref{p1}), respectively, without discarding a total derivative. The same linear combinations of defect coefficients give Eqs.\ (\ref{pd1})--(\ref{ed1}):
\begin{align}
P_{D,I}^{(1)}&=-m_0[\sigma\cos W^+-\sigma^{-1}\cos W^-],\\
E_{D,I}^{(1)}&=-m_0[\sigma\cos W^++\sigma^{-1}\cos W^-].
\end{align}
For these conventions the corresponding moving balances read
\begin{align}
\frac{\dd E_{D,I}^{(1)}}{\dd\tau}&=p_R-p_L-V(e_L-e_R),\\
\frac{\dd P_{D,I}^{(1)}}{\dd\tau}&=s_R-s_L-V(p_L-p_R).
\end{align}
These follow either from \eqref{Dbalance} coefficient by coefficient or directly from the transformed sewing equations.

A constant-$t$ slice generally maps to a tilted curve in the transformed $(\tau,z)$ coordinates. Charges defined on this curve and on a constant-$\tau$ slice are integrals of the same conserved one-form $\omega=q\,dz-j\,d\tau$, supplemented by the defect contribution at the respective interface intersections. In the integrable sector, bulk current conservation and the interface identity $\iota^*(\omega_R-\omega_L)=d_DD$ imply, by Stokes’ theorem, that the difference between the total charges is determined entirely by outer-boundary fluxes. The charges therefore coincide provided the coordinate transformation is regular throughout the comparison region, the integrals converge, and these boundary contributions vanish. Their bulk parts may differ when the slices intersect the defect at different events, with the difference compensated by the defect term. This establishes slicing independence of the total conserved charge under the stated conditions; equivalence of the associated Hamiltonian structures requires a separate analysis.

The additional contribution of our analysis concerns the deformed sewing conditions. These reduce to an effective B\"acklund transformation when $\alpha\beta=1$. For $\alpha\beta\neq1$, the two sewing equations involve incompatible B\"acklund parameters and generate localized anomalies. This deformation, and the resulting charge-dependent quasi-conservation, is not implied by the integrable pullback construction. The exact logarithmic balance proved here assumes the integrable type-I sewing conditions 
and does not assert its validity for $\alpha \beta \neq 1$.

The transformed trajectory is of the same geometric character as the prescribed moving interfaces considered in \cite{xia}: the B\"acklund relations are imposed along a time-dependent curve. In our setting, however, this curve is induced by the variable-mass backgrounds through a coordinate transformation of a fixed physical defect. The correspondence concerns the moving-interface formulation; equivalence of the Hamiltonian structures requires a separate analysis.

\section{Third order charge, defect contribution and anomalies}
\label{app:thirdorder}
The localized third order momentum- and energy-type contributions are
\begin{equation}
 P_D^{(3)}=-\frac{m_0}{2}
 \left(\Phi_-^{(3)}+\Phi_+^{(3)}\right)_{x=0},
 \qquad
 E_D^{(3)}=-\frac{m_0}{2}
 \left(\Phi_-^{(3)}-\Phi_+^{(3)}\right)_{x=0},
 \label{eq:third-order-defect-contributions}
\end{equation}
with
\begin{align}
 \Phi_-^{(3)}={}&
 -\frac{2}{3}\sigma_A^3\cos(3w_+)
 +\frac{2\sigma_A^2}{m_0}\,u\sin(2w_+)
 +\frac{\sigma_A}{m_0^2}
 \left(u^2\cos w_+ +u_1\sin w_+\right),
 \label{eq:Phi-third-minus}
 \\
 \Phi_+^{(3)}={}&
 \frac{2}{3\sigma_B^3}\cos(3w_-)
 +\frac{2v}{m_0\sigma_B^2}\sin(2w_-)
 -\frac{1}{m_0^2\sigma_B}
 \left(v^2\cos w_- +v_1\sin w_-\right).
 \label{eq:Phi-third-plus}
\end{align}
Using the sewing relations~\eqref{eq:third-order-covariant-sewing} and their
$D_0$- and $D_1$-derivatives, the remaining non-exact chiral terms are
\begin{align}
 R_-^{(3)}={}&\rho_1\sin w_-
 \Bigl[
 2m_0^2\sigma_A^2\sin(3w_+)
 +4m_0\sigma_Au\cos(2w_+)
 -u^2\sin w_+
 +u_1\cos w_+
 \Bigr],
 \label{eq:R-third-minus}
 \\
 R_+^{(3)}={}&\rho_0\sin w_+
 \Bigl[
 v_1\cos w_-
 -v^2\sin w_-
 -\frac{4m_0}{\sigma_B}v\cos(2w_-)
 +\frac{2m_0^2}{\sigma_B^2}\sin(3w_-)
 \Bigr].
 \label{eq:R-third-plus}
\end{align}
The next expressions are the components of  the anomalies
\begin{align}
 \mathcal C_-^{(3)}(t)={}&
 \Bigl[
 2m_0^2\sigma_A^2\sin(3w_+)
 +4m_0\sigma_Au\cos(2w_+)
 -u^2\sin w_+
 +u_1\cos w_+
 \Bigr]_{x=0},
 \label{eq:C-third-minus}
 \\
 \mathcal C_+^{(3)}(t)={}&
 \Bigl[
 v_1\cos w_-
 -v^2\sin w_-
 -\frac{4m_0}{\sigma_B}v\cos(2w_-)
 +\frac{2m_0^2}{\sigma_B^2}\sin(3w_-)
 \Bigr]_{x=0}.
 \label{eq:C-third-plus}
\end{align}

\section{Type-II reality conditions: A sufficient real reduction}
\label{app:reality}

The type-II contributions in Eqs. (\ref{PII})–(\ref{EII}) and (\ref{P'II})–(\ref{E'II}) to the charges $P_{D,\mathrm{II}}^{(1)}, E_{D,\mathrm{II}}^{(1)}$ are generally complex when $w,\bar w,\Lambda$ are real and $a,b,c$ are arbitrary. Appropriate reality conditions on $a,b,c,c_0$ and $\Lambda$ must therefore be specified to ensure real-valued physical charges and consistent real defect dynamics.

Let
\begin{equation}
 p=w^+=w+\bar w,\qquad q=w^-=w-\bar w,\qquad
 S(q)=e^{\ii q}+e^{-\ii q}+c_0=2\cos q+c_0.
\end{equation}
We take $w,\bar w$ and the matched backgrounds $\bar\rho_j=\rho_j$ to be real, and assume $m_0,a,b,c,c_0\in\mathbb R$ with $abc\ne0$. The auxiliary field in (\ref{eq2.14}) and (\ref{eq2.14vm}) is complex on the real physical sector
\begin{equation}
 \Lambda=\ell+\ii\chi,\qquad \ell,\chi\in\mathbb R,
 \qquad e^{2\chi}=-\frac{ac}{b^2}S(q)>0.
 \label{reality}
\end{equation}
Equivalently,
\begin{equation}
 \chi(q)=\frac12\log\!\left[-\frac{ac}{b^2}(2\cos q+c_0)\right].
\end{equation}
With these restrictions the lowest-order type-II defect contributions are real
\begin{align}
 P_{D,\mathrm{II}}^{(1)}
 &=m_0e^\chi\left[\frac ba\cos(p-\ell)+\frac bc\cos\ell\right],\\
 E_{D,\mathrm{II}}^{(1)}
 &=m_0e^\chi\left[\frac ba\cos(p-\ell)-\frac bc\cos\ell\right].
\end{align}
These conditions apply to the homogeneous contributions (\ref{PII})–(\ref{EII}) and the corresponding lowest-order variable-mass contributions (\ref{P'II})–(\ref{E'II}), which retain the same algebraic structure.

The constraint \eqref{reality} is locally preserved by the full type-II B\"acklund equations, as verified below. Thus $\ell$ is the real auxiliary degree of freedom, while $\chi$ is fixed by $q$. A sufficient condition for this auxiliary-field parametrization to be nonsingular for every real $q$ is $|c_0|>2$ and $ac\,c_0<0$. This excludes zeros of $S$; it does not assert global regularity of every dynamical solution.  

Let us analyze the B\"acklund equations (\ref{back11v})--(\ref{back14v}) and (\ref{b12r})-(\ref{L11}). For real matched backgrounds $\rho_0=\rho_0(\xi)$ and $\rho_1=\rho_1(\eta)$, the equations are
\begin{align}
 \ii q_\xi&=\frac{m_0}{a}\rho_0
 \left(be^{-\ii(\Lambda-p)}+\beta_{12}e^{-\ii p/2}\right), \\
 \ii q_\eta&=-\frac{m_0}{c}\rho_1
 \left(\beta_{12}e^{\ii p/2}+be^{-\ii\Lambda}\right), \\
 \ii\Lambda_\xi&=-\frac{m_0c}{b}\rho_0e^{\ii(\Lambda-p)}
 (e^{\ii q}-e^{-\ii q}), \\
 \ii(\Lambda-p)_\eta&=\frac{m_0a}{b}\rho_1e^{\ii\Lambda}
 (e^{\ii q}-e^{-\ii q}),
\end{align}
and
\begin{align} 
 (\beta_{12})_\xi&=-\frac{\ii}{2}p_\xi\beta_{12}
 +cm_0\rho_0e^{\ii p/2}(e^{\ii q}-e^{-\ii q}),\\
 (\beta_{12})_\eta&=\frac{\ii}{2}p_\eta\beta_{12}
 -am_0\rho_1e^{-\ii p/2}(e^{\ii q}-e^{-\ii q}), \\
 \beta_{12}&=\frac{ac}{b}e^{\ii(\Lambda-p/2)}S(q). 
\end{align}
 
On \eqref{reality}, (\ref{L11}) reads
\begin{equation}
 \beta_{12}=\frac{ac}{b}e^{-\chi}S\,e^{\ii(\ell-p/2)}
 =-be^\chi e^{\ii(\ell-p/2)}.\label{beta}
\end{equation}
Although $\beta_{12}$ is generally complex, its modulus and phase are determined by real fields. Substitution into  (\ref{back11v})--(\ref{back12v}) gives
\begin{align}
 q_\xi&=\frac{2m_0b}{a}\rho_0e^\chi\sin(p-\ell),\label{qx}\\
 q_\eta&=\frac{2m_0b}{c}\rho_1e^\chi\sin\ell.\label{qy}
\end{align}
For example, the bracket in (\ref{back11v}) becomes
$be^\chi[e^{\ii(p-\ell)}-e^{-\ii(p-\ell)}]
=2\ii be^\chi\sin(p-\ell)$, as required for real $q_\xi$.

Separating  (\ref{back13v})--(\ref{back14v}) into real and imaginary parts yields
\begin{align}
 \ell_\xi&=-\frac{2m_0c}{b}\rho_0e^{-\chi}\sin q\cos(\ell-p),\label{lx}\\
 (\ell-p)_\eta&=\frac{2m_0a}{b}\rho_1e^{-\chi}\sin q\cos\ell,\label{ly}\\
 \chi_\xi&=-\frac{2m_0c}{b}\rho_0e^{-\chi}\sin q\sin(\ell-p),\label{hx}\\
 \chi_\eta&=\frac{2m_0a}{b}\rho_1e^{-\chi}\sin q\sin\ell.\label{hy}
\end{align}
Equations \eqref{qx}--\eqref{ly}, with $\chi=\chi(q)$, form a real first-order B\"acklund system for $p,q,\ell$.

Next we verify the preservation of the constraint. Differentiating \eqref{reality} on an admissible branch gives
\begin{equation}
 \chi_j=-\frac{\sin q}{S}q_j,\qquad j=\xi,\eta.\label{chain}
\end{equation}
The same constraint implies
\begin{equation}
 \frac{be^\chi}{aS}=-\frac cb e^{-\chi},\qquad
 \frac{be^\chi}{cS}=-\frac ab e^{-\chi}.
\end{equation}
Using \eqref{qx}--\eqref{qy} in \eqref{chain} reproduces exactly \eqref{hx}--\eqref{hy}. Therefore the imaginary parts of (\ref{back13v})--(\ref{back14v}) are differential consequences of the reality constraint. They introduce no second independent real auxiliary variable. 

Next we verify the consistency equations for $\beta_{12}$. Define $D=e^{\ii q}-e^{-\ii q}$, so that $S_j=\ii Dq_j$. Differentiation of (\ref{L11}), followed by (\ref{back11v})--(\ref{back14v}), gives
\begin{align}
 \frac{(\beta_{12})_\xi}{\beta_{12}}
 &=-\frac{\ii}{2}p_\xi+\frac{m_0b}{a}\rho_0
 \frac{D}{S}e^{-\ii(\Lambda-p)},\\
 \frac{(\beta_{12})_\eta}{\beta_{12}}
 &=\frac{\ii}{2}p_\eta-\frac{m_0b}{c}\rho_1
 \frac{D}{S}e^{-\ii\Lambda}.
\end{align}
Multiplying by (\ref{L11}) reproduces (\ref{b12r})-(\ref{b12rr}). Thus, on this nonzero-$S$ branch, those equations are consequences of  (\ref{back11v})--(\ref{back14v}) and  (\ref{L11}), rather than independent evolution laws for another physical degree of freedom.

Finally, we verify the compatibility with the bulk vmSG equations. Set $R_1=2m_0b\rho_0e^\chi/a$ and $R_2=2m_0b\rho_1e^\chi/c$. The real system implies
\begin{align}
 q_\xi=R_1\sin(p-\ell),\quad &q_\eta=R_2\sin\ell,\nonumber\\
 \ell_\xi=\frac{R_1}{S}\sin q\cos(p-\ell),\quad
 &(\ell-p)_\eta=-\frac{R_2}{S}\sin q\cos\ell,\nonumber\\
 \chi_\xi=-\frac{R_1}{S}\sin q\sin(p-\ell),\quad
 &\chi_\eta=-\frac{R_2}{S}\sin q\sin\ell.\nonumber
\end{align}
Because $\partial_\eta\rho_0=\partial_\xi\rho_1=0$ and
$R_1 R_2/S=-4m_0^2\rho_0\rho_1$, differentiation gives
\begin{equation}
 q_{\xi\eta}=\frac{R_1 R_2}{S}\sin q\cos p
 =-4m_0^2\rho_0\rho_1\sin q\cos p.
\end{equation}
Similarly, differentiating $\ell_\xi$ and $(\ell-p)_\eta$, and subtracting their mixed derivatives, gives
\begin{equation}
 p_{\xi\eta}=-4m_0^2\rho_0\rho_1\sin p\cos q.
\end{equation}
Since $w=(p+q)/2$ and $\bar w=(p-q)/2$, both fields satisfy
\begin{equation}
 w_{\xi\eta}+2m_0^2\rho_0\rho_1\sin(2w)=0,
 \qquad
 \bar w_{\xi\eta}+2m_0^2\rho_0\rho_1\sin(2\bar w)=0.
\end{equation}
The real reduction therefore retains the vmSG B\"acklund property. Setting $\rho_0=\rho_1=1$ gives the corresponding homogeneous sine-Gordon reduction.
 
\section{Absorption and emission limits of the type-I defect}
\label{app:emission-absorption}

The critical values $p=\pm\sigma_{\rm eff}$ in (\ref{seff2}) must be excluded from regular transmission, since they correspond to $z=0$ and $|z|\to\infty$, respectively. These limits, however, remain regular at the level of the fields and the sewing conditions. For $z=0$, i.e. $\sigma_{\rm eff}=-p$, the limiting configuration is
\begin{equation}
 w_L(x,t)=2\arctan e^{\Theta_L(x,t)},\qquad
 w_R(x,t)=0,
\end{equation}
so that the incoming kink is absorbed and no soliton emerges on the right. Substitution into the two matching equations gives identities for $\sigma_{\rm eff}=-p$. Conversely, at the pole $|z|\to\infty$, i.e. $\sigma_{\rm eff}=p$, one keeps the outgoing phase
\begin{equation}
\delta_R=\delta_L+\ln|z|
\end{equation}
finite while sending $\delta_L\to-\infty$. The limiting fields are then
\begin{equation}
 w_L(x,t)=0,\qquad
 w_R(x,t)=2\arctan e^{\Theta_R(x,t)},
\label{profile0}
\end{equation}
and the sewing equations are again satisfied identically for $\sigma_{\rm eff}=p$. The emission solution approaches the vacuum configuration $w_L=0,\ w_R=\pi$ as $t\to-\infty$, locally on the two half-lines. This asymptotic configuration carries a nonzero field discontinuity and the corresponding localized defect charge and energy-type contribution. The outgoing phase $\delta_R$ specifies the emission time; at a finite initial time, the exact emission data are obtained by sampling the limiting profile (\ref{profile0}). Thus, the zero and pole of $z$ are singularities of the transmission parametrization, not of the defect conditions.

The corresponding defect bookkeeping makes this interpretation explicit. The following temporal limits assume $p>0$ and $\Theta_j(0,t)\to+\infty$ as $t\to-\infty$, while $\Theta_j(0,t)\to-\infty$ as $t\to+\infty$, corresponding to the right-moving kink sector. Using
\begin{equation}
 Q_D=\frac{w_R(0^+,t)-w_L(0^-,t)}{\pi},\qquad
 E_D^{(1)}=-m_0\left[
 \sigma_{\rm eff}\cos(w_L+w_R)
 +\frac{1}{\sigma_{\rm eff}}\cos(w_L-w_R)
 \right],
\end{equation}
absorption evolves as
\begin{equation}
 (w_L,w_R):(\pi,0)\to(0,0),\qquad
 Q_D:-1\to0,
\end{equation}
with
\begin{equation}
 E_D^{(1)}:
 -m_0\left(p+\frac1p\right)
 \to
 +m_0\left(p+\frac1p\right),
\end{equation}
whereas emission gives
\begin{equation}
 (w_L,w_R):(0,\pi)\to(0,0),\qquad
 Q_D:+1\to0,
\end{equation}
with
\begin{equation}
 E_D^{(1)}:
 +m_0\left(p+\frac1p\right)
 \to
 -m_0\left(p+\frac1p\right).
\end{equation}
Hence $\Delta Q_{\rm bulk}=-\Delta Q_D$, while the change
\begin{equation}
 |\Delta E_D^{(1)}|=2m_0\left(p+p^{-1}\right)
\end{equation}
equals the one-kink energy-type charge, reducing to the ordinary kink energy in the homogeneous limit.

For a type-I defect, emission therefore does not describe the decay of an independent localized degree of freedom, since no auxiliary defect field is present. Its initial data are instead vacuum bulk fields on the two half-lines with a nontrivial discontinuity,
\begin{equation}
 w_L(0^-)=0,\qquad w_R(0^+)=\pi,
\end{equation}
which carries the corresponding localized defect charge and energy.  Emission is thus the time-reversed B\"acklund counterpart of absorption, in which this localized defect contribution is converted into an outgoing kink.

\section{Finite-difference scheme for the bulk and defect evolution}
\label{app:numerical-scheme}

The point defect divides the spatial domain into two independent half-lines,
with fields $w_L(x,t)$ for $x<0$ and $w_R(x,t)$ for $x>0$.
Consequently, a conventional finite-difference stencil cannot be applied
across $x=0$. The bulk equations must instead be evolved separately on
the two half-lines, while the two limiting values
$w_L(0^-,t)$ and $w_R(0^+,t)$ are advanced through a discrete form of
the sewing conditions. The purpose of this appendix is to describe this
coupled bulk--defect algorithm explicitly. The scheme uses centered
second-order differences in space, leapfrog time integration in the bulk,
and a simultaneous second-order update of the two defect values.

Next, let us describe the spatial grid and bulk evolution. Let
\begin{equation}
 t_n=t_0+n\Delta t,
 \qquad
 x_{L,j}=-j\Delta x,
 \qquad
 x_{R,j}=j\Delta x,
 \qquad
 W^n_{s,j}=w_s(x_{s,j},t_n),
 \qquad s=L,R,
 \label{app:grid}
\end{equation}
where $j=0,\ldots,N$. The two values at the defect are treated as
independent dynamical variables,
\begin{equation}
 L^n\equiv W^n_{L,0}=w_L(0^-,t_n),
 \qquad
 R^n\equiv W^n_{R,0}=w_R(0^+,t_n).
 \label{app:defect-values}
\end{equation}
No bulk finite-difference stencil crosses the defect point.

For convenience, define
\begin{equation}
 C^n_{s,j}
 =
 \rho_0\left(\frac{t_n+x_{s,j}}{2}\right)
 \rho_1\left(\frac{t_n-x_{s,j}}{2}\right),
 \qquad
 r=\frac{\Delta t}{\Delta x}.
 \label{app:C-def}
\end{equation}
The bulk equation on either half-line is
\begin{equation}
 \partial_t^2 w_s-\partial_x^2 w_s
 +2m_0^2
 \rho_0\left(\frac{t+x}{2}\right)
 \rho_1\left(\frac{t-x}{2}\right)
 \sin(2w_s)=0,\,\,\,\, s=L,R.
 \label{app:bulk-equation}
\end{equation}
For the interior points $j=1,\ldots,N-1$, centered spatial
differences and leapfrog time stepping yield
\begin{align}
 W^{n+1}_{s,j}
 ={}&
 2W^n_{s,j}-W^{n-1}_{s,j}
 +r^2
 \left(
 W^n_{s,j-1}-2W^n_{s,j}+W^n_{s,j+1}
 \right)
 \nonumber\\
 &\hspace{1cm}
 -2m_0^2(\Delta t)^2
 C^n_{s,j}\sin\left(2W^n_{s,j}\right).
 \label{app:bulk-update}
\end{align}
Thus, the left and right bulk arrays are advanced independently,
using their respective defect values $L^n$ and $R^n$ as the
endpoints adjacent to $x=0$.

The numerical calculations are performed on
\begin{equation}
 t\in[-20,20], \qquad x\in[-40,40],
 \qquad N=\frac{40}{\Delta x},
\end{equation}
where $N$ is the number of spatial intervals on each half-line.
The spatial resolution $\Delta x$ and time step $\Delta t$ vary
between calculations and are specified in the corresponding figures
and convergence tables. For the refined calculations, we use
$\Delta t=0.2\Delta x$.

Now, describe  the defect sewing conditions. At $t=t_n$, introduce the discrete defect functions
\begin{align}
 A^n
 &=
 -2m_0\sigma \alpha\,
 \rho_0\left(\frac{t_n}{2}\right)
 \sin\left(L^n+R^n\right),
 \qquad
 \alpha=1+\epsilon_A,
 \label{app:An}
 \\
 B^n
 &=
 -2m_0\frac{\beta}{\sigma}\,
 \rho_1\left(\frac{t_n}{2}\right)
 \sin\left(L^n-R^n\right),
 \qquad
 \beta=1+\epsilon_B.
 \label{app:Bn}
\end{align}
When $\alpha\beta=1$, the two coefficients can be written in terms
of the common effective B\"acklund parameter
\begin{equation}
 \sigma_{\mathrm{eff}}
 =
 \sigma\alpha
 =
 \frac{\sigma}{\beta}.
 \label{app:sigma-eff}
\end{equation}

Second-order one-sided approximations to the spatial derivatives at
the defect are
\begin{align}
 D_x^-L^n
 &\equiv
 \frac{3L^n-4W^n_{L,1}+W^n_{L,2}}
      {2\Delta x},
 \label{app:left-derivative}
 \\
 D_x^+R^n
 &\equiv
 \frac{-3R^n+4W^n_{R,1}-W^n_{R,2}}
      {2\Delta x}.
 \label{app:right-derivative}
\end{align}
Here $D_x^-L^n$ approximates
$\partial_xw_L(0^-,t_n)$, whereas $D_x^+R^n$ approximates
$\partial_xw_R(0^+,t_n)$.

Using centered time derivatives, the two sewing conditions become
\begin{align}
 D_x^-L^n
 &=
 \frac{R^{n+1}-R^{n-1}}{2\Delta t}
 +\frac{A^n+B^n}{2},
 \label{app:discrete-sewing-1}
 \\
 D_x^+R^n
 &=
 \frac{L^{n+1}-L^{n-1}}{2\Delta t}
 +\frac{B^n-A^n}{2}.
 \label{app:discrete-sewing-2}
\end{align}
Solving these equations for the new defect values gives the explicit
simultaneous update
\begin{align}
 L^{n+1}
 &=
 L^{n-1}
 +2\Delta t
 \left[
 D_x^+R^n-\frac{B^n-A^n}{2}
 \right],
 \label{app:L-update}
 \\
 R^{n+1}
 &=
 R^{n-1}
 +2\Delta t
 \left[
 D_x^-L^n-\frac{A^n+B^n}{2}
 \right].
 \label{app:R-update}
\end{align}

Both components in Eqs. (\ref{app:L-update})-(\ref{app:R-update}), $(L^{n+1},\,R^{n+1})$, must be
computed from the same time level $n$. In particular, the newly
calculated value $L^{n+1}$ must not be used in the evaluation of
$R^{n+1}$, or conversely.

A complete time step is implemented in the following order:
\begin{enumerate}
 \item evaluate the background coefficients $C^n_{s,j}$, the
 bulk right-hand sides, $A^n$, $B^n$, and the two one-sided
 defect derivatives using only levels $n$ and $n-1$;
 \item calculate all bulk values $W^{n+1}_{s,j}$ from
 Eq.~\eqref{app:bulk-update};
 \item calculate $L^{n+1}$ and $R^{n+1}$ simultaneously from
 Eqs.~\eqref{app:L-update}--\eqref{app:R-update};
 \item replace the pair of time levels
 $(n-1,n)$ by $(n,n+1)$.
\end{enumerate}

Let us discuss the outer boundary conditions. At the outer endpoints $x=\pm20$, time-dependent Dirichlet data
are taken from the corresponding reference profiles:
\begin{equation}
 W^n_{L,N}
 =
 w_L^{\mathrm{ref}}(-20,t_n),
 \qquad
 W^n_{R,N}
 =
 w_R^{\mathrm{ref}}(20,t_n).
 \label{app:outer-boundaries}
\end{equation}
For the time interval considered, these values are numerically close
to the appropriate asymptotic vacua. The outer boundaries are also
sufficiently distant from the defect that signals generated near
$x=0$ do not return to the interaction region during the reported
evolution time.

The initialization is performed using the kink solution. Because the leapfrog scheme requires two initial time levels, the
prescribed reference profiles are sampled at
\begin{equation}
 t_0=-20,
 \qquad
 t_1=t_0+\Delta t,
\end{equation}
according to
\begin{equation}
 W^q_{s,j}
 =
 w_s^{\mathrm{ref}}(x_{s,j},t_q),
 \qquad q=0,1.
 \label{app:bulk-initialization}
\end{equation}
The corresponding defect values are initialized as
\begin{equation}
 L^q
 =
 {\cal K}\!\left[\Phi_L(0,t_q)+\delta_L\right],
 \qquad
 R^q
 =
 \varsigma\,
 {\cal K}\!\left[\Phi_R(0,t_q)+\delta_R\right],
 \qquad
 {\cal K}(u)=2\arctan(e^u),
 \label{app:defect-initialization}
\end{equation}
where
\begin{equation}
 \varsigma=
 \begin{cases}
 +1, & \text{kink--kink transmission},\\
 -1, & \text{kink--antikink transmission}.
 \end{cases}
\end{equation}

For the closure-compatible branch, $\Phi_L=\Phi_R=\Phi$, with
\begin{equation}
 \Phi(x,t)
 =
 2m_0
 \left[
 p\,X\left(\frac{t+x}{2}\right)
 -
 p^{-1}Y\left(\frac{t-x}{2}\right)
 \right],
 \label{app:phase}
\end{equation}
and
\begin{align}
 X(\xi)
 &=
 \xi+
 \frac{\epsilon_0}{\kappa_0}
 \left\{
 \tanh[\kappa_0(\xi-\xi_0)]
 +\tanh(\kappa_0\xi_0)
 \right\},
 \label{app:X}
 \\
 Y(\eta)
 &=
 \eta+
 \frac{\epsilon_1}{\kappa_1}
 \left\{
 \tanh[\kappa_1(\eta-\eta_0)]
 +\tanh(\kappa_1\eta_0)
 \right\}.
 \label{app:Y}
\end{align}
The exact transmission factor and centered phase constants are
\begin{equation}
 z=\frac{p+\sigma_{\mathrm{eff}}}
        {p-\sigma_{\mathrm{eff}}},
 \qquad
 \delta_L=-\frac{1}{2}\ln|z|,
 \qquad
 \delta_R=+\frac{1}{2}\ln|z|.
 \label{app:transmission-phase}
\end{equation}

For $\a \b \neq 1$, the two sewing relations yield distinct factors $z_A$ and $z_B$, so no exact transmitted one-soliton profile exists. The initial data therefore comprise an incoming left soliton, specified by $p_{in}$ and its phase, and the appropriate right-hand vacuum, sampled at 
$t_0$ and $t_1 = t_0 + \D t$. No $p_{out}$ is imposed. The independent defect values at $x^{\pm} =0$ are subsequently evolved by the finite-difference sewing equations. Thus, $z_A$ and $z_B$ diagnose the closure mismatch, while $p_{out}$ and the outgoing phase shift are extracted 
{\sl a posteriori} from a late-time fit.

\subsection{One-sided reconstruction for third-order calculations}
\label{app:third-order-reconstruction}

The defect fields are independent one-sided traces,
$L_0^n=w_L(0^-,t_n)$ and $R_0^n=w_R(0^+,t_n)$.
Unbarred quantities denote left traces and barred quantities right traces.
No spatial stencil crosses the defect. For a grid quantity $F_{s,j}^n$,
we use the centered temporal and one-sided endpoint derivatives
\begin{align}
 \delta_tF_{s,j}^n&=\frac{F_{s,j}^{n+1}-F_{s,j}^{n-1}}{2\Delta t},\\
 \delta_x^LF_{L,0}^n&=\frac{3F_{L,0}^n-4F_{L,1}^n+F_{L,2}^n}{2\Delta x},\\
 \delta_x^RF_{R,0}^n&=\frac{-3F_{R,0}^n+4F_{R,1}^n-F_{R,2}^n}{2\Delta x}.
\end{align}
At interior points,
\begin{equation}
 \delta_x^sF_{s,j}^n=\varepsilon_s
 \frac{F_{s,j+1}^n-F_{s,j-1}^n}{2\Delta x},
 \qquad \varepsilon_L=-1,\quad\varepsilon_R=1.
\end{equation}
The background coefficients are sampled separately on each half-line,
\begin{equation}
 \rho_{0,s,j}^n=\rho_0\!\left(\frac{t_n+x_{s,j}}{2}\right),\qquad
 \rho_{1,s,j}^n=\rho_1\!\left(\frac{t_n-x_{s,j}}{2}\right).
\end{equation}
For the anomaly calculations,
the unbarred jets are reconstructed from $L_j^n=W_{L,j}^n$ at
$x_{L,j}=-j\Delta x$. Define
\begin{align}
 T^n&=\delta_tL_0^n,
 &X^n&=\delta_x^LL_0^n,\\
 TT^n&=\frac{L_0^{n+1}-2L_0^n+L_0^{n-1}}{(\Delta t)^2},
 &XX^n&=\frac{2L_0^n-5L_1^n+4L_2^n-L_3^n}{(\Delta x)^2},\\
 TX^n&=\frac{X^{n+1}-X^{n-1}}{2\Delta t}.
\end{align}
Here $T,X,TT,XX,TX$ approximate $w_t,w_x,w_{tt},w_{xx},w_{tx}$,
respectively, at $x=0^-$, with second-order truncation accuracy for
smooth one-sided fields. Put $q_n=t_n/2$ and
$g_j^n=\rho_j(q_n)$, with $g_j^{\prime n}=\rho_j'(q_n)$ denoting
the derivative with respect to the background argument. The implemented
reconstruction is
\begin{align}
 u^n&=\frac{T^n+X^n}{g_0^n},
 &v^n&=\frac{T^n-X^n}{g_1^n},\\
 u_1^n&=\frac{TT^n+2TX^n+XX^n}{(g_0^n)^2}
       -\frac{g_0^{\prime n}(T^n+X^n)}{(g_0^n)^3},\\
 v_1^n&=\frac{TT^n-2TX^n+XX^n}{(g_1^n)^2}
       -\frac{g_1^{\prime n}(T^n-X^n)}{(g_1^n)^3}.
\end{align}
These formulas follow by expanding $u_1=D_0u$ and $v_1=D_1v$,
where $D_0=\rho_0^{-1}(\partial_t+\partial_x)$ and
$D_1=\rho_1^{-1}(\partial_t-\partial_x)$. Background derivatives
are evaluated analytically. For the backgrounds used here,
\begin{equation}
 \rho_j(q)=1+\epsilon_j\operatorname{sech}^2(\kappa_j q),
 \qquad
 \rho_j'(q)=-2\epsilon_j\kappa_j
 \operatorname{sech}^2(\kappa_j q)\tanh(\kappa_j q).
\end{equation}
The expanded formulas are discretized directly; the first-derivative
operators are not applied recursively to numerical $u,v$ arrays.

The third-order anomalies are evaluated using $u,v,u_1,v_1$
reconstructed from the left-hand numerical field, together with both
defect field values. This calculation requires $L_0,\ldots,L_3$ at
neighboring time levels $n-1,n,n+1$. The quantities $u_2,v_2$,
defined by $u_2=D_0u_1$ and $v_2=D_1v_1$ and required by the third-order bulk charge densities, are not evaluated
in this anomaly-only calculation. Diagnostics exclude time endpoints
for which the centered temporal stencils are unavailable.
Subscripts denote covariant derivatives, so $u_2\neq u^2$ and
$v_2\neq v^2$; Appendix~B uses the squares $u^2,v^2$ and the jets
$u_1,v_1$. Evaluating the anomalies alone does not verify the full
third-order bulk-plus-defect charge balance.

\subsection{Outgoing-soliton fitting and empirical uncertainty assessment}
\label{app:outgoing-fit}

The following assessment concerns only the non-integrable kink--kink
transmission with $\alpha=\beta=1.25$, $\sigma=0.16$,
$p_{\mathrm{in}}=0.5$, and $m_0=1$. The fields are evolved on
$x\in[-40,40]$ from $t=-20$ to $t_*=20$, with
$\Delta t=0.2\Delta x$. At $t_*$, the transmitted field is fitted to
\begin{equation}
 w_{\mathrm{sol}}(x,t;p,\delta)=2\arctan\!\left\{
 \exp\!\left[2\left(pX\!\left(\frac{t+x}{2}\right)
 -\frac{1}{p}Y\!\left(\frac{t-x}{2}\right)\right)+\delta\right]
 \right\},
\end{equation}
where $X(q)=q+0.6\tanh(0.5q)$ and
$Y(q)=q+(0.5/0.7)\tanh(0.7q)$. The convention $X(0)=Y(0)=0$
fixes $\delta$ as the additive phase in the exponent.
The connected core $0.05<w_R^{\mathrm{num}}<\pi-0.05$ on the
finest mesh defines, by linear interpolation of the threshold crossings,
$I=[11.3199767494,14.2521493228]$. These physical endpoints are
held fixed throughout the mesh comparison. We minimize
\begin{equation}
 J(p,\delta)=\frac{1}{N_{\mathrm{fit}}}
 \sum_{x_j\in I}
 \left[w_R^{\mathrm{num}}(x_j,t_*)-
 w_{\mathrm{sol}}(x_j,t_*;p,\delta)\right]^2.
\end{equation}
The bounded trust-region reflective least-squares solver uses
$0.1\leq p\leq2$, $-20\leq\delta\leq20$, initial guess
$(0.5,0.31)$, and function, step, and gradient tolerances $10^{-13}$.

\begin{center}
\begin{tabular}{@{}rrrr@{}}
\toprule
$\Delta x$ & $\Delta t$ & $p_{\mathrm{out}}$ & $\delta_{\mathrm{out}}$\\
\midrule
0.005    & 0.001    & 0.499888551 & 0.310089564\\
0.0025   & 0.0005   & 0.499891755 & 0.309349703\\
0.00125  & 0.00025  & 0.499893613 & 0.309090803\\
0.000625 & 0.000125 & 0.499894315 & 0.309009639\\
\bottomrule
\end{tabular}
\end{center}
The finest-mesh core RMS residual is $\sqrt{J_{\min}}=2.92566\times10^{-3}$.
The extra parameter digits document reproducibility rather than physical
precision. The finest-pair changes are
\begin{equation}
 s_{\mathrm{mesh}}(p)=7.02\times10^{-7},\qquad
 s_{\mathrm{mesh}}(\delta)=8.12\times10^{-5}.
\end{equation}
The empirical convergence orders on the successive three-mesh sets are
$(q_p,q_\delta)=(0.786,1.515)$ and $(1.405,1.673)$.
A stable second-order parameter regime is therefore not established;
no Richardson error bound is assigned. Mesh changes also include the
effect of sampling the fixed physical interval at different grid points.

On the finest mesh, fitting-window sensitivity is assessed using the
connected cores $a<w_R^{\mathrm{num}}<\pi-a$:
\begin{center}
\begin{tabular}{@{}rrrrr@{}}
\toprule
$a$ & $x_a$ & $x_b$ & $p_{\mathrm{out}}$ & $\delta_{\mathrm{out}}$\\
\midrule
0.025 & 11.038423 & 14.533253 & 0.499979727 & 0.303089406\\
0.05  & 11.319977 & 14.252149 & 0.499894315 & 0.309009639\\
0.10  & 11.590981 & 13.965665 & 0.499756157 & 0.318668046\\
\bottomrule
\end{tabular}
\end{center}
For either fitted parameter $\theta$, define
$u_{\mathrm{window}}(\theta)=\max_a|\theta_a-\theta_{0.05}|$.
The resulting maximum deviations are
\begin{equation}
 u_{\mathrm{window}}(p)=1.38\times10^{-4},\qquad
 u_{\mathrm{window}}(\delta)=9.66\times10^{-3}.
\end{equation}
Alternative baseline starting guesses $(0.4,-1)$ and $(0.6,1)$ yield
spreads below $3.5\times10^{-10}$ in $p$ and $2.4\times10^{-8}$
in $\delta$.

These mesh and window variations constitute an empirical uncertainty
assessment and are reported separately as deterministic sensitivity
estimates, not statistical confidence intervals or certified error
bounds. Window dependence dominates the measured finest-pair mesh
change. Coherent profile mismatch and possible radiation contributions
preclude interpreting the residuals as independent random errors.
Accordingly, the baseline results are summarized as
$p_{\mathrm{out}}\simeq0.49989$ and $\delta_{\mathrm{out}}\simeq0.309$,
accompanied by the sensitivities above. Since
$|p_{\mathrm{out}}-p_{\mathrm{in}}|\simeq1.06\times10^{-4}$ is smaller
than $u_{\mathrm{window}}(p)$, these fits do not resolve a change from
$p_{\mathrm{in}}=0.5$. They characterize the transmitted soliton at
$t=20$; late-time stability has not been established. Other transmission
branches and parameter sets require separate uncertainty assessments.

\end{document}